%% file: main_draft.tex
\documentclass[letterpaper,11pt]{extarticle}

\input{preamble_article.tex}

\usepackage{gelasio}
\usepackage{csquotes}
\newgeometry{left=0.8in, right=0.8in, bottom=0.8in, top=0.8in}
\savegeometry{default_margins}
\title{\textbf{Salary Matching and Pay Cut Reduction \\ \, \\for Job Seekers with Loss Aversion}}

\author{\, \\
    \begin{minipage}[t]{0.45\textwidth}
        \centering
        \textbf{Ross Chu} \vspace{0.5em} \\ \itshape{U.C. Berkeley} 
    \end{minipage}
    \\[3em]
}

\date{\monthyeardate\today\thanks{
    \noindent I acknowledge financial support through the National Science Foundation Graduate Research Fellowship under Grant DGE 2146752. This work does not represent the views of NSF, UC Berkeley, or Statistics Korea. All statistics reported in this paper represent the analysis sample alone and are not intended to be nationally representative of South Korea. I am grateful to Jungmin Lee for his contributions to early stages of this project, and I thank my dissertation advisors for invaluable guidance and support: Stefano DellaVigna, Enrico Moretti, and Ricardo Perez-Truglia. I also appreciate numerous faculty and students who gave me helpful feedback to improve this paper: Ned Augenblick, John Beshears, Sydnee Caldwell, David Card, Luisa Cefala, Konhee Chang, Kevin Dano, Hillary Hoynes, Jireh Kang, Joon Ha Kim, Kristy Kim, Woojin Kim, Patrick Kline, Jaepil Lee, Jaeyeon Lee, Tammy Lee, Peter Maxted, Nisha Pathak, Simon Quach, Jesse Rothstein, Emmanual Saez, Benjamin Schoefer, Dmitry Taubinsky, Christopher Walters, Danny Yagan, Chamna Yoon, and seminar participants at Berkeley Labor Lunch, Berkeley Psychology \& Economics Workshop, Berkeley Public Finance Seminar, Berkeley Symposium in Labor/Public Finance, and the Korea Institute of Public Finance (KIPF). All errors are my own. 
}}

\begin{document}

\subfile{1_abstract.tex}

\begingroup
\singlespacing

\renewcommand{\contentsname}{Table of Contents}
{\small \singlespacing \tableofcontents}

\thispagestyle{empty}
\endgroup
\clearpage

\setcounter{page}{1}


\subfile{1_introduction.tex}


\subfile{1_literature.tex}

\clearpage

\subfile{2_model.tex}


\subfile{3_data.tex}


\subfile{4_empirics.tex}


\subfile{5_calib.tex}


\subfile{6_results.tex}


\subfile{7_policy.tex}


\subfile{8_conclude.tex}


\clearpage
\addcontentsline{toc}{section}{Supporting Materials}

\subfile{0_figures.tex}

\clearpage

\subfile{0_tables.tex}

\clearpage
\addcontentsline{toc}{subsection}{References}
\printbibliography[title=References]


\appendix
\renewcommand{\thefigure}{A\arabic{figure}}
\renewcommand{\thetable}{A\arabic{table}}
\setcounter{figure}{0}
\setcounter{table}{0}

\newgeometry{left=0.8in, right=0.8in, bottom=0.5in, top=0.3in}

\begingroup
\onehalfspacing

\makeatletter
\begin{titlepage}
    \begin{center}
        \vspace*{\fill}
        {\huge \textbf{Appendix for \\ \, \\ ``\MyTitle''}\par}
        \vspace{1cm}
        {\large \MyAuthor\par}
        \vspace{1cm}
        {\large \monthyeardate\today\par}
        \vspace{1cm}
        {\large \textbf{Contents}\par}
        \vspace{0.5cm}
        \startcontents[appendix]
        \printcontents[appendix]{}{1}{}
        \vspace{\fill}
    \end{center}
\end{titlepage}
\makeatother

\thispagestyle{empty}
\endgroup
\clearpage
\restoregeometry


\subfile{a1_model.tex}

\clearpage

\subfile{a2_estimation.tex}

\clearpage
\addcontentsline{toc}{section}{Appendix Material}

\subfile{0_figures_apx.tex}

\subfile{0_tables_apx.tex}



\end{document}

%% file: preamble_article.tex
\usepackage[utf8]{inputenc}

\usepackage[american]{babel}

\usepackage{kotex}

\usepackage{subfiles}

\usepackage{xr-hyper}

\makeatletter
\@ifclassloaded{subfiles}{
  \ifx\mainfile\undefined\else
    \externaldocument{aux/\subfix{\@nameuse{mainfile}}}%
  \fi
}{}
\makeatother

\usepackage{amsmath}
\usepackage{amsfonts}
\usepackage{bbm}
\usepackage{accents}

\usepackage[toc]{appendix}
\usepackage{chngcntr}

\usepackage{graphicx} 	
\usepackage{color} 		
\usepackage[super]{nth} 
\usepackage{float} 		
\usepackage{ragged2e} 
\usepackage{changepage} 
\usepackage{titletoc} 
\usepackage{authoraftertitle} 
\usepackage{datetime} 
\newdateformat{monthyeardate}{\monthname[\THEMONTH] \THEYEAR}

\usepackage{booktabs}
\usepackage{multirow}
\usepackage{hhline} 
\usepackage{makecell} 

\usepackage[bottom,hang,flushmargin]{footmisc}
\usepackage[margin=1in]{geometry}

\usepackage{setspace}

\usepackage[all]{nowidow}

\renewcommand{\arraystretch}{1.7}

\usepackage[colorinlistoftodos]{todonotes} 
\usepackage{mdframed} 		

\usepackage{hyperref}
\hypersetup{colorlinks=true, linkcolor=cyan, urlcolor=cyan, citecolor=cyan}

\AtBeginDocument{

}

\newcommand{\autorefapx}[1]{\hyperref[#1]{Appendix~\autoref*{#1}}}
\newcommand{\autorefapxsec}[1]{\hyperref[#1]{Appendix~\ref*{#1}}}

\usepackage{xcolor}
\usepackage{soul}
\definecolor{codecolor}{RGB}{0,102,153}
\definecolor{codebg}{RGB}{240,240,240}
\sethlcolor{codebg}

\usepackage{titlesec}

\titleformat{\subsubsection}[block]
  {\normalsize\itshape}
  {\thesubsubsection}
  {1em}
  {\ul}

\usepackage{caption}
\usepackage{placeins}

\usepackage[
    backend=biber,
    sorting=nyt, 
    uniquelist=false, 
    hyperref=true, 
    maxcitenames=2, 
    style=apa,
    doi=false, 
    url=false, 
]{biblatex}

\DeclareFieldFormat[article]{title}{\mkbibquote{#1}} 

\DeclareCiteCommand{\parencite}[\mkbibparens]
  {\usebibmacro{prenote}}
  {\printtext[bibhyperref]{%
    \printnames{labelname}%
    \setunit{\printdelim{nameyeardelim}}%
    \printfield{year}%
    \printfield{extradate}}}
  {\multicitedelim}
  {\usebibmacro{postnote}}

\DeclareCiteCommand{\citeauthor}
  {\usebibmacro{prenote}}
  {\printtext[bibhyperref]{%
    \printnames{labelname}}}
  {\multicitedelim}
  {\usebibmacro{postnote}}

\DeclareCiteCommand{\colorcite}
  {\usebibmacro{prenote}}
  {\printtext[bibhyperref]{%
    \printnames{labelname}%
    \setunit{\printdelim{nameyeardelim}}%
    \printfield{year}}%
  }
  {\multicitedelim}
  {\usebibmacro{postnote}}

%% file: 1_abstract.tex
\newgeometry{left=0.8in, right=0.8in, bottom=0.5in, top=0.3in}
\begingroup
\onehalfspacing

\maketitle

\begin{abstract}
    \noindent This paper examines how loss aversion affects wages offered by employers and accepted by job seekers. 
    I introduce a behavioral search model with monopsonistic firms making wage offers to job seekers who experience steeper disutility from pay cuts than utility from equivalent pay raises.
    Employers strategically reduce pay cuts to avoid offer rejections, 
    and they exactly match offers to current salaries due to corner solutions. 
    Loss aversion makes three predictions on the distribution of salary growth for job switchers, which I empirically test and confirm with administrative data in Korea.
    First, excess mass at zero wage growth is 8.5 times larger than what is expected without loss aversion.
    Second, the density immediately above zero is 8.8\% larger than the density immediately below it.
    Third, the slope of the density below zero is 6.5 times steeper than the slope above it.
    When estimating model parameters with minimum distance on salary growth bins, incorporating loss aversion substantially improves model fit,
    and the marginal value of additional pay is 12\% higher for pay cuts than pay raises in the primary specification.
    For a hypothetical hiring subsidy that raises the value of labor to employers by half of a standard deviation, incorporating loss aversion lowers its pass-through to wages by 18\% (relative to a standard model) due to higher elasticity for pay cuts and salary matches that constrain subsidized wage offers. 
    Somewhat surprisingly, salary history bans do not mitigate these effects as long as employers can imperfectly observe current salaries with noise.
\end{abstract}

\thispagestyle{empty}
\endgroup
\clearpage
\restoregeometry

%% file: 1_introduction.tex
\section{Introduction} \label{sec:intro}

\addcontentsline{toc}{subsection}{Motivation}

	In a seminal paper, \textcite{kahneman_prospect_1979} ignited a rich intersection between psychology and economics by laying out foundations for reference-dependent preferences. Reference-dependence proposes that decision makers are responsive to gains and losses relative to a reference point of comparison, and loss aversion makes disutility from losses steeper than utility from equivalent gains.\footnote{
		In empirical applications, this reference point is either assumed based on context (e.g. zero taxes owed in \cite{rees-jones_quantifying_2018}) or endogenized with rational expectations about future outcomes \parencite{koszegi_model_2006, koszegi_reference-dependent_2007}.
	}
	Reference-dependence gained prominence in economics as empirical research demonstrated its diverse influence on high-stakes decisions.\footnote{
		Examples include tax reporting \parencite{rees-jones_quantifying_2018,jones_loss_2020,dhami_loss_2025}, home transactions \parencite{genesove_loss_2001,andersen_reference_2022}, investment decisions \parencite{benartzi_myopic_1995,barberis_prospect_2001}, corporate mergers \parencite{baker_effect_2012}, and even sports competitions \parencite{pope_is_2011,allen_reference-dependent_2017}.
	}
	Similarly, reference-dependence gained attention as a promising extension to job search models. Recent applications have shown that prior consumption levels can shape search effort during unemployment (\colorcite{dellavigna_reference-dependent_2017}; \citeyear{dellavigna_evidence_2022}), and comparisons with past wages or peer salaries can affect reservation wages during job search \parencite{boheim_great_2011,fu_social_2019}.

	This paper builds on this literature by focusing on three research questions. First, how does loss aversion affect wages offered by employers and accepted by job seekers? Second, what does empirical evidence say about testable predictions of the model? Third, what does loss aversion imply about the pass-through of hiring subsidies to wages? While there is abundant literature on wage rigidity for job stayers, there are new lessons to be learned by paying attention to job switchers. This paper documents three stylized facts about the distribution of wage growth for job switchers: bunching at zero wage growth, discontinuous density at zero, and sharp changes in its curvature at zero. Despite similar patterns holding true in prior studies \parencite[e.g.][]{grigsby_aggregate_2021}, these empirical facts on job switchers have not received much attention in the wage rigidity literature. In this paper, I show that loss aversion can be an intuitive interpretation for these striking properties.

\addcontentsline{toc}{subsection}{Section Summaries}


	I introduce a behavioral job search model with monopsonistic firms making wage offers to job seekers with loss aversion. Wage offers made by firms are relative to current wages held by job seekers, which become their realized wage growth when they accept the offer and become job switchers. Loss aversion makes three predictions on the distribution of wage growth for switchers: 1) bunching at zero wage growth, 2) discontinuous density at zero, and 3) sharp changes in its curvature at zero (\autoref{fig:anomaly_theory}). The key mechanism is that the job seeker's probability of accepting an offer is kinked at their current wage (\autoref{fig:acp_rate}), which implies discontinuous marginal profits for firms. ``Salary matching'' occurs as employers exactly match offers to current wages due to corner solutions, and ``pay cut reduction'' occurs as employers face higher marginal profits and strategically reduce pay cuts to avoid offer rejections (\autoref{fig:optimal_offers}). I initially develop this model in a simple static framework to highlight intuition for key mechanisms, but I show that key predictions remain unchanged in popular search environments like wage bargaining (\autoref{subsec:wage_bargaining}) and wage posting with on-the-job search (\autoref{subsec:dynamic_search}).


	I empirically test these predictions by using administrative data to measure salary growth for job switchers in Korea. Matched employer--employee data include 13.6 million job transitions between 2015 and 2020 for workers registered with the unemploment insurance (UI) agency. For each job switcher, I measure annual salaries by dividing total earnings by the number of months worked during the transition year, and I define salary growth as the difference in logged salaries between new and previous jobs (\autoref{fig:salary_growth_dist}). Since detection of bunching requires new salaries to precisely align with prior salaries, the analysis sample focuses on switchers who work for entire months at hire and separation to limit measurement error (\autoref{tab:conditions}). 


	I use this data to measure bunching, discontinuity, and sharp changes in curvature in the density of salary growth for switchers (\autoref{fig:anomaly_sample}). 6.0\% of switchers are bunched at zero salary growth, which corresponds to an excess mass that is 8.5 times larger than expected levels without loss aversion. The density of salary growth immediately above zero is 8.8\% larger than the density immediately below it, indicating a discontinuity at zero. The slope of the density below zero is 6.5 times steeper than the slope above it, indicating sharp changes in its curvature at zero (\autoref{tab:anomalies}). I rule out several explanations for these ``anomalies'' based on institutional context and supporting evidence from data, including transitions between minimum wage jobs, collective bargaining with unions, self-reporting errors in income tax filings, regulations on over-time pay, sharp changes in demographics, and arbitrary reassignment of firm IDs. Several factors can contribute to the magnitude of bunching, but they cannot explain the discontinuity or curvature break: examples include rounded salaries, current salaries signaling the market value of labor, and wage-centric workers rejecting all pay cuts. Based on simulated distributions from placebo tests, rounded salaries can explain roughly 1\% of switchers bunching at zero, which only partially explains the magnitude observed from data (\autoref{subsec:contributing_factors}).


	I corroborate the behavioral interpretation of anomalies with additional evidence from heterogeneity and similar patterns in prior studies. Anomalies are more pronounced for employed job switchers without employment gaps, which is consistent with current salaries being more salient as reference points for job seekers continuing to receive those paychecks (\autoref{fig:career_gaps}). Bunching is only observed for switchers in the private-sector, but both private and public sectors exhibit discontinuities at zero (\autoref{fig:govt_private}). This is consistent with negotiation on precise salaries in the private-sector, while the public sector only engages in coarse negotiation based on structured pay scales set by the Korean government. Similar anomalies for switchers can be found in prior studies with different contexts and data sources, and prior studies have also found kinked acceptance rates and wage offers bunched at reference points, which are key mechanisms proposed by the model (\autoref{subsec:hetero_prior_studies}).


	Loss aversion is an appealing interpretation for these anomalies because it intuitively justifies why preferences are kinked at prior salaries and aligns with empirical evidence both in this study and in prior studies. Loss aversion does not take a stance on \underline{why} disutility for pay cuts is steeper than pay raises, which makes it an adaptable framework that embodies a wide range of explanations. I use loss aversion to frame several interpretations that can explain all three anomalies, including heuristic rejections of pay cuts, salary benchmarks for bargaining, fairness perceptions regarding pay, and debt obligations held by job seekers (\autoref{subsec:framing}).


	I use minimum distance to estimate loss aversion $(\lambda)$ and location/scale parameters for labor productivity $(\mu_\phi, \sigma_\phi)$, subject to calibrated parameters for non-wage amenities $(\mu_\epsilon, \sigma_\epsilon)$. The minimum distance procedure compares predicted proportions in each salary growth bin against their empirical counterparts, which account for censoring due to job seekers selectively accepting salary offers (\autoref{fig:datagen_process}). Parameters for labor productivity control the location and spread of the salary offer distribution, while loss aversion controls magnitudes of anomalies and the degree of asymmety between densities for pay raises and pay cuts. For identification, I assume that non-wage amenities are equal on average at previous and new jobs, and I match the variance of non-wage amenities in \textcite{lehmann_non-wage_2025} to calibrate the amenity scale parameter (\autoref{tab:calibration_overview}). While these are reasonable assumptions with realistic magnitudes, I report the sensitivity of results to these calibrations throughout the paper.


	Incorporating loss aversion into the search model substantially improves model fit to proportions in salary growth bins (\autoref{fig:model_fit}). Estimated loss aversion is $\hat{\lambda}=1.123 \; (SE=0.0004)$ in the primary specification, which has stable convergence properties and is robust to alternative specifications (\autoref{tab:parameters_main}). This magnitude is smaller than previous studies of reference-dependence, as larger magnitudes would imply substantially less pay cuts that we do not observe for switchers in Korea. The estimated model can reasonably match the discontinuity and curvature break despite not being explicit targets (\autoref{fig:anomaly_comparison}), but predicted bunching is four times larger than the observed magnitude (possibly because measurement error understates true bunching).


	For a hypothetical hiring subsidy that raises the value of labor to employers by half of a standard deviation, incorporating loss aversion into the model lowers its pass-through to wages by 18\% relative to a standard model (\autoref{fig:passthru_lossav}). This primarily occurs through two channels: 1) higher elasticity for pay cuts lowering tax incidence on workers, and 2) subsidized wage offers constrained by salary matching (\autoref{fig:passthru_constrained}). Somewhat surprisingly, these effects do not rely on precise knowledge of current salaries: implications of loss aversion remain unchanged even with salary history bans, as long as employers can imperfectly observe current salaries with noise (\autoref{subsec:salary_history_bans}). This is likely true in real-world settings, where it is common to ask about ``salary expectations'' or ``fair compensation'' at the application and interview stages of the hiring process.

\,

%% file: 1_literature.tex
\addcontentsline{toc}{subsection}{Literature \& Contributions}


	\noindent \textbf{Literature and Contributions}: Wage rigidity is important to labor and macro economists due to its implications on employment fluctuations and financial volatility.\footnote{
        For survey evidence on wage rigidity, see \colorcite{card_does_1997,kahn_evidence_1997,nickell_nominal_2003,dickens_how_2007,kaur_nominal_2019}. \colorcite{dickens_how_2007} presents cross-country evidence with employment registers. \colorcite{grigsby_aggregate_2021} offers evidence from administrative payroll data, and \colorcite{hazell_downward_2024} offers evidence from job postings. See \colorcite{kaur_nominal_2019,ehrlich_wage_2024,faia_cost_2024} for implications of wage rigidity on employment fluctuations and financial volatility. 
    }
    So far, the literature has mostly focused on the ``fair-wage effort hypothesis'' for job stayers.\footnote{
        \noindent See \colorcite{akerlof_fair_1990, akerlof_labor_1982,shapiro_equilibrium_1984,bhaskar_wage_1990,hart_contracts_2008} for a theoretical formulation of the fair-wage effort hypothesis. 
        Field evidence suggests that workers exert less effort after receiving pay cuts and view them as being unfair \parencite{mas_pay_2006,kaur_nominal_2019,cohn_social_2014,koch_can_2021,kujansuu_fairness_2024}. In surveys, managers report being reluctant to reduce wages out of concern for reduced morale \parencite[e.g.][]{bewley_why_1999,quach_wage_2025}. See \colorcite{fongoni_why_nodate,kahneman_fairness_1986,blinder_shred_1990,campbell_reasons_1997,shafir_money_1997} for additional surveys on perceived unfairness of pay cuts.
	}
    However, this explanation is incomplete for job switchers, as pay cuts are not necessarily seen as unfair when moving to entirely new jobs. This paper shows that loss aversion can hinder downward wage adjustments by reducing the magnitude of pay cuts for job switchers. I provide empirical support for behavioral job search by analyzing implied properties of the wage growth distribution, and I evaluate how loss aversion affects counterfactual simulations for government subsidies.\footnote{
		This theoretical insight shares similarities with models that link wage rigidity to reference-dependence \parencite[e.g. ][]{mcdonald_how_2001,ahrens_theory_2014,eliaz_reference_2014,benjamin_theory_2015,doerrenberg_asymmetric_2023,santana_theory_2024}. The contribution of this paper is corroborating these predictions with empirical facts on ``anomalies`` for job switchers. Anomalies like bunching and discontinuity have previously been used to study tax reporting \parencite{rees-jones_quantifying_2018, jones_loss_2020}, taxable income elasticity \parencite{saez_taxpayers_2010,kleven_bunching_2016}, and the extent of wage rigidity \parencite{goette_wage_2007}. In this paper, I am using similar anomalies to study how loss aversion affects wage growth for job switchers.
	}


    More broadly, this paper contributes to the growing application of behavioral economics to study the labor market. In the past, researchers used target earnings to explain why labor supply can fall with positive wage shocks.\footnote{See \colorcite{camerer_labor_1997,farber_is_2005,farber_why_2015,crawford_new_2011,thakral_daily_2021, fehr_workers_2007}} More recently, researchers began incorporating behavioral preferences into job search models to explain empirical facts that are difficult to rationalize with standard preferences\footnote{See \textcite{zimmermann_behavioral_2020} for a review.}. For example, \textcite{dellavigna_reference-dependent_2017,dellavigna_evidence_2022} use reference-dependence on prior consumption levels to explain why search effort spikes up and down around the exhaustion of unemployment benefits.\footnote{
        See \textcite{boheim_great_2011} and \colorcite{fu_social_2019} for other applications of reference dependence in job search models.
    }
    I contribute to this line of work by showing how reference-dependence on current salaries can explain anomalies in the distribution of wage growth for job switchers.

	\,


	\noindent The rest of this paper is organized into the following sections: \vspace{-6pt}
	\begin{itemize}
		\setlength{\itemsep}{-4pt}
		\item \autoref{sec:model} develops the behavioral search model and lays out testable predictions on salary growth.
		\item \autoref{sec:admin_data} describes the administrative data on job switchers in Korea.
		\item \autoref{sec:empirics} empirically tests model predictions on the distribution of salary growth.
		\item \autoref{sec:calib} uses minimum distance to fit model parameters on salary growth bins.
		\item \autoref{sec:results} provides an assessment of model fit.
		\item \autoref{sec:policy} analyzes the pass-through of subsidies to wages under loss aversion. 
		\item \autoref{sec:conclude} concludes by discussing how future research can address key limitations.
	\end{itemize}

%% file: 2_model.tex
\section{Behavioral Job Search Model} \label{sec:model}

I develop the behavioral search model in a simple static environment to highlight intuition for key predictions and their mechanisms. I do not argue in favor of any particular search model, and I show that key predictions remain unchanged in richer environments like wage bargaining and wage posting with on-the-job search. Heterogeneity implies smooth dispersion around current wages, and loss aversion disrupts this smoothness with sharp changes at the current salary due to kinked preferences.

\subsection{Labor Supply: Acceptance Decisions on Wage Offers} \label{subsec:labor_supply}

	Job seeker $i$ decides whether to accept a job offer from firm $f$ by comparing its utility with their current job. A job offer consists of wages and non-wage amenities (i.e. workplace benefits). Wages $w$ refer to logged annual salaries, and wage offers $r$ are relative to current pay $w_0$. Wage offers become realized wage growth when job seekers accept the offer to become job switchers.

	\,

	\noindent The utility of a wage offer to the job seeker is given by
	\refstepcounter{equation} \label{eq:jobseeker_utility}
	\begin{equation}
		u(r \,|\, \epsilon_{if}) = \lambda^{L(r)} \cdot r + \epsilon_{if}
		\tag*{(\theequation): Job Seeker Utility}
	\end{equation}
	\noindent where
	\begin{itemize}
		\setlength{\itemsep}{0pt}
		\item[] $r = w - w_0$ is the relative wage offer (logged pay raise above current wage $w_0$),
		\item[] $\lambda \geq 1$ is the degree of loss aversion (which nests standard preferences with $\lambda=1$),
		\item[] $L(r) = \mathbbm{1}\{r < 0\}$ is an indicator for being offered a pay cut (which triggers $\lambda^{L(r)} = \lambda$), and
		\item[] $\epsilon_{if} \sim F_\epsilon(\cdot)$ is the value of non-wage amenities relative to the current job.
	\end{itemize}

	\,

	\noindent $\lambda^{L(r)}$ is the key term that represents loss aversion, and current wage $w_0$ is the corresponding reference point for job seekers. Any offers above this level is considered a gain, while anything below this level is considered a loss. Job seekers experience steeper disutility from losses due to loss aversion, which can be seen in the marginal utility of additional pay $u'(r)$. The marginal utility for a pay cut is $\lambda$, which is larger than the marginal utility for a pay raise ($=1$). Under standard preferences $\lambda=1$, marginal utility would be constant at 1 for both pay raises and pay cuts. Job seekers are willing to take pay cuts if compensated enough with amenities, but loss aversion $\lambda > 1$ implies steeper tradeoffs between wages and non-wage amenities when offered a pay cut.

	Staying at their current job involves no change in wages or amenities ($r=0, \epsilon_{if}=0$), so the utility of staying at their current job is zero. The job seeker accepts the offer if $u(r \,|\, \epsilon_{if}) > 0$, and heterogeneity in $\epsilon_{if}$ implies that wage offer $r$ is accepted with the following probability:
	\refstepcounter{equation} \label{eq:acceptance_rate}
	\begin{align}
		p(r) = 1 - F_{\epsilon}(-\lambda^{L(r)}\cdot r)
		\tag*{(\theequation): Offer Acceptance Rate}
	\end{align}

	\noindent \autoref{fig:acp_rate} is a visual representation of \autoref{eq:acceptance_rate}. The blue line corresponds to the standard model $(\lambda = 1)$, while the orange line corresponds to the behavioral model $(\lambda > 1)$. Acceptance rates are smoothly increasing with wage offers in the standard model, but acceptance rates are kinked at the current salary in the behavioral model. Acceptance rates are identical for pay raises under both models, while they are strictly lower for pay cuts in the behavioral model. This is because job seekers are more demanding in non-wage amenities when offered a pay cut, which makes them less likely to accept. 
	
	Acceptance rate $p(r)$ increases at a steeper rate for pay cuts because $\lambda > 1$ implies that decisions are more sensitive to wages below the reference point. In the firm's perspective, this means that reducing a pay cut is more effective for avoiding offer rejections than increasing a pay raise. This ``kink'' property of the acceptance rate is important for monopsonistic firms, as I discuss below.

\subsection{Labor Demand: Profit-Maximizing Wage Offers} \label{subsec:labor_demand}

	Monopsonistic firms have labor market power, and wage offers maximize expected profits given by
	{
		\setlength{\abovedisplayskip}{7pt}
		\setlength{\belowdisplayskip}{1pt}
		\refstepcounter{equation} \label{eq:expected_profits}
		\begin{align}
			\pi(r) = p(r) \cdot (\phi_{if} - r)
			\tag*{(\theequation): Expected Profits}
		\end{align}
	}
	\noindent where
	\begin{itemize}
		\setlength{\itemsep}{0pt}
		\item[] $p(r)$ is the offer acceptance rate (given uncertainty in amenity values $\epsilon_{if}$),
		\item[] $\phi_{if}=\Psi_{f} - w_{0i}$ is labor productivity $\Psi_{f}$ relative to current wage $w_{0i}$, and
		\item[] $r = w - w_{0i}$ is the offered wage relative to the job seeker's current wage $w_{0i}$.
		\item[] (note that markup $\Psi_{f} - w$ is the same as $\phi_{if} - r$ after adding and subtracting $w_{0i}$)
	\end{itemize}
	\,

	\noindent Firms incorporate job seekers' preferences by setting wage offers that balance markup size $\phi_{if} - r$ with acceptance rate $p(r)$. Marginal profits are given by
	\refstepcounter{equation} \label{eq:marginal_profits}
	\begin{align}
		\pi'(r) = - p(r) + p'(r) \cdot (\phi_{if} - r) 
		\tag*{(\theequation): Marginal Profits}
	\end{align}

	\noindent The left panel of \autoref{fig:optimal_offers} visualizes marginal profits in \autoref{eq:marginal_profits} for the standard model (dashed line) and behavioral model (solid line) at three different levels of labor productivity $\phi_{if}$ (low, moderate, and high). Marginal profits are smooth in the standard model, while they exhibit sharp changes at $r=0$ for the behavioral model. This is because a kink in acceptance rate $p(r)$ implies a discrete jump in the slope of the acceptance rate $p'(r)$. Marginal profits are identical for pay raises in both models, while marginal profits are higher for pay cuts in the behavioral model because $\lambda > 1$ implies that additional pay is more effective at increasing acceptance rates. Highly productive workers (in green) receive identical pay raises under both models, while less productive workers (in red) receive smaller pay cuts under the behavioral model. Due to corner solutions, moderately productive workers (in yellow) who would have received a small pay cut under the standard model receive a salary match in the behavioral model.

\subsection{Salary Matching and Pay Cut Reduction} \label{subsec:salary_match_reduction}

	Setting marginal profits in \autoref{eq:marginal_profits} to zero implies the following optimality condition for wage offers
	\refstepcounter{equation} \label{eq:optimal_offer}
	\begin{align}
		\phi_{if} = r + \frac{p(r)}{p'(r)}
		\tag*{(\theequation): Optimal Wage Offers}
	\end{align}

	\noindent The right panel of \autoref{fig:optimal_offers} compares optimal wage offers for each value of productivity under standard and behavioral models. This figure highlights two key differences. First, the behavioral model predicts smaller pay cuts for each value of productivity, while pay raises remain unchanged. ``Pay cut reduction'' occurs because additional wages are particularly effective at increasing acceptance rates for pay cuts, which raises marginal profits and optimal wages. Second, the behavioral model predicts a range of productivity values that match wage offers exactly at current salaries (scatter points). ``Salary matching'' occurs because marginal profits jump down at zero in the behavioral model (in the left panel), which result in corner solutions for wage offers within a certain range of productivity values (see \hyperref[subsec:matching_wedge]{Model Appendix~\ref*{subsec:matching_wedge}} for a derivation of this ``wedge''). The intuition is that additional pay becomes much less effective at increasing acceptance rates after the reference point, so firms are not incentivized to increase offers beyond their current wage.

\subsection{Testable Predictions on the Distribution of Wage Growth} \label{subsec:model_prediction}

	\autorefapx{fig:offer_dist} shows the distribution of offered wages to job seekers (blue bars), a subset of which become realized wage growth for job switchers who accept these offers (orange bars). The model predicts three key differences in the distribution of salary growth under standard and behavioral models, which is summarized in \autoref{fig:anomaly_theory}. First, there is a mass of job switchers bunching at zero salary growth in the behavioral model, which does not exist for the standard model (``bunching''). Second, the density of salary growth is discontinuously lower for pay cuts at zero (``discontinuity'').\footnote{
        This prediction assumes that the density of wage growth is increasing around the reference point. This is true for my empirical context, as the central mass of the wage growth distribution is above zero. 
    } Third, the slope of the density exhibits sharp changes at zero (``curvature break''). The density of salary growth under the standard model does not exhibit these three properties, as the density evolves smoothly around zero. In the empirical analysis, I test for these differences using administrative data to measure salary growth for job switchers. 
 
	Note that these are predictions for loss aversion modeled as a \underline{kink} at the reference point. Another common approach is to model loss aversion as a \underline{notch} at the reference point. The notch refers to a discrete loss in utility when offers fall below the reference point, while marginal utility remains unchanged. The acceptance rate with a notch exhibits a downward jump, and the model predicts bunching along with clearly defined boundaries for missing mass below the current salary that corresponds to the size of the notch (see \cite{kleven_bunching_2016} for details). This paper focuses on loss aversion modeled as a kink since predictions implied by a notch do not align with empirical patterns observed in the administrative data.

\subsection{Wage Bargaining with Loss Aversion} \label{subsec:wage_bargaining}

	Loss aversion is a versatile framework that yields the same predictions across a variety of search models. I demonstrate this flexibility by incorporating loss aversion into two popular models: wage bargaining and wage posting with on-the-job search. I start by solving the wage bargaining model with loss aversion and derive expressions for salary matching and pay cut reduction.

	In the wage bargaining model, job seekers and firms set mutually beneficial wages through cooperative nash bargaining. Job seekers and firms are mutually aware of each others' amenity values $(\epsilon)$ and labor productivity $(\phi)$, and a successful bargain requires both parties to achieve payoffs better than their outside options. In the static search environment, the outside option for workers is to keep their current job, while the outside option for firms is to not hire the candidate.\footnote{
		In a dynamic search environment, the outside option for workers would be the future value of maintaining current employment or staying unemployed. For firms, it would be the future value of being matched with another candidate. See \textcite{cahuc_wage_2006} for a formal treatment.
	} The following bargaining regions define pairs of $(\epsilon, \phi)$ that result in successful bargains with $u(r) > 0$ for workers and $\pi(r) > 0$ for firms:
	\refstepcounter{equation} \label{eq:bargain_region}
	\begin{alignat}{2}
		& A^{+} \text{(pay raise)} \; &&= \big\{ (\epsilon, \phi): \epsilon \leq 0, \; \phi > 0, \; \phi > -\epsilon \big\} \nonumber \\ 
		& A^{-} \text{(pay cut)} \; &&= \big\{ (\epsilon, \phi): \epsilon > 0, \; \phi \leq 0, \; \phi > -\epsilon \, / \, \lambda \big\} \tag*{(\theequation): Bargaining Region} \\ 
		& A^{\pm} \text{(raise or cut)} \; &&= \big\{ (\epsilon, \phi): \epsilon > 0, \; \phi > 0, \; \phi > -\epsilon \, / \, \lambda \big\} \nonumber
	\end{alignat}

	\noindent For $(\epsilon, \phi) \in A^{+} \cup A^{-} \cup A^{\pm}$, bargained wages maximize joint payoffs weighted by bargaining power $\beta$.
	\refstepcounter{equation} \label{eq:bargain_payoff}
	\begin{equation}
		r^{*} = \arg \max_{r} \, (\lambda^{L(r)} \cdot r + \epsilon)^{\beta} \, \times \, (\phi - r)^{1-\beta}
		\tag*{(\theequation): Joint Bargaining Payoffs}
	\end{equation}

	\noindent This implies the following nash solution for bargained wages
	\refstepcounter{equation} \label{eq:bargain_solution}
	\begin{equation}
		r^{*}( \epsilon, \phi)=
		\begin{cases}
			\text{Pay Cut: } \beta \cdot \phi - \frac{1 - \beta}{ \lambda} \cdot \epsilon & \text{if } \phi < \frac{1 - \beta}{\beta \lambda} \cdot \epsilon \\
			\text{Salary Match:} \; 0 & \text{if } \phi \in \big[\frac{1 - \beta}{\beta \lambda}\cdot \epsilon, \frac{1 - \beta}{\beta} \cdot \epsilon \big] \\
			\text{Pay Raise: } \beta \cdot \phi - (1 - \beta) \cdot \epsilon & \text{if } \phi > \frac{1 - \beta}{\beta} \cdot \epsilon
		\end{cases}
		\tag*{(\theequation): Wage Bargaining Solution}
	\end{equation}

	\noindent and the following impacts of loss aversion on bargained wages:
	\begin{equation*}
		\frac{dr}{d\lambda}=
		\begin{cases}
			\frac{1 - \beta}{\lambda^2} \cdot \epsilon & \text{if } r < 0 \longmapsto \text{ Pay Cut Reduction} \\
			0 & \text{if } r \geq 0 \longmapsto \text{ Unaffected Pay Raises}
		\end{cases}
	\end{equation*}

	\noindent \autorefapx{fig:bargain_solution} visualizes the above expressions, which highlight salary matching and pay cut reduction under wage bargaining with loss aversion. Horizontal and vertical axes denote non-wage amenities $\epsilon$ and labor productivity $\phi$, and colors denote nash solutions for bargained wages (green for pay raises, red for pay cuts, white for salary match). This figure highlights three key features. First, loss aversion shrinks the region for successfully bargained pay cuts by making them less attractive to job seekers (boundary for the red region shifts up from the blue line to the orange line). Second, bargained pay cuts are smaller under loss aversion while pay raises remain unchanged (lighter red colors for behavioral model, but the same green colors). Finally, loss aversion implies a region of $(\epsilon, \phi)$ pairs that result in salary matches (between the two dashed lines). 

\subsection{Dynamic Search with Loss Aversion} \label{subsec:dynamic_search}

	Key predictions of loss aversion remain unchanged even in dynamic search environments. I incorporate loss aversion into a wage posting model with on-the-job search \parencite{burdett_wage_1998} and two additional features. First, I allow heterogeneity in labor productivity $\phi$ and non-wage amenities $\epsilon$. Second, I allow monopsonistic firms to tailor wage offers on the job seeker's reference point and labor productivity. This deviates from standard wage posting models where firms set uniform wage policies, but my setup closely mirrors realistic hiring practices: employers publish job postings, interview candidates, and tailor salary offers based on an assessment of their value and current salary.

	\subsubsection{Dynamic Search Environment}

	\noindent Firms decide how many vacancies to create in each period, and job seekers are randomly matched to vacancies. Firms make wage offers to maximize expected profits for each vacancy. The vacancy is filled if the job seeker accepts the wage offer and closed otherwise until firms reopen vacancies in the next period.

	Job seekers are either employed (E) or unemployed (U), and they are randomly matched to vacancies with offer arrival rates $\alpha_E$ and $\alpha_U$. Employed job seekers are terminated at random with job destruction rate $\delta$ (layoff risk). Workers receive flow utility from employment through wage $(w)$ and non-wage amenities $(\epsilon)$, and unemployment provides benefits $w^U$ and leisure value $\epsilon^U$. An employed job seeker accepts the wage offer if doing so yields higher expected utility than maintaining their current status, considering future opportunities and risks from layoffs.

	\subsubsection{Acceptance Rates with Dynamic Payoffs}

	\noindent As in \autoref{subsec:labor_supply}, the utility of a wage offer for the job seeker exhibits a kink at the current reference point $\tilde{w}$ due to loss aversion. Job seekers assess utility $u(\cdot | \tilde{w})$ and the expected value of future offer arrivals based on their current reference point $\tilde{w}$. \underline{Value functions for job seekers} are given by
	\begin{alignat*}{4}
		& V^{E}(w, \epsilon \,|\, \tilde{w}) 
        &&= u(w, \epsilon \,|\, \tilde{w}) 
        &&+ \beta \cdot \biggr[ 
        (1-\alpha_{E}-\delta) \cdot V^{E}(w, \epsilon \,|\, \tilde{w}) 
        &&+\alpha_{E} \cdot \bar{V}^{E}_{max}(w, \epsilon \,|\, \tilde{w}) 
        + \delta \cdot V^{U}(w^{U}, \epsilon^{U} \,|\, \tilde{w})
        \biggr] \nonumber \\
		& V^{U}(w^{U}, \epsilon^{U} \,|\, \tilde{w}) 
        &&= u(w^{U}, \epsilon^{U} \,|\, \tilde{w}) 
        &&+ \beta \cdot \biggr[ 
        (1-\alpha_{U}) \cdot V^{U}(w^{U}, \epsilon^{U} \,|\, \tilde{w}) 
        &&+ \alpha_{U} \cdot \bar{V}^{U}_{max}(w^{U}, \epsilon^{U} \,|\, \tilde{w}) \biggr] 
	\end{alignat*}
	\noindent where
	\begin{itemize}
		\setlength{\itemsep}{0pt}
        \item[] $\beta$ is the discount factor,
		\item[] $\tilde{w}$ is the reference point for job seekers ($w_0$ if employed, $w^U$ if unemployed),
		\item[] $u(w,\epsilon | \tilde{w})$ is the utility of a job offer (wages and amenities) given the reference point $\tilde{w}$,
		\item[] $V^E,V^U$ are employment and unemployment values given wages and amenities, and
		\item[] $\bar{V}^{ \{E,U\} }_{max}=E \biggr[ \max \Bigr\{ V^{\{E,U\}}(w, \epsilon \,|\, \tilde{w}), V^{E}(w', \epsilon' \,|\, \tilde{w}) \Bigr\} \biggr]$ is the expected value of future offer arrivals.
	\end{itemize}

	\noindent The wage-amenity bundle for job seekers is current wage $w_0$ and amenities $\epsilon_0$ if employed, and unemployment benefits $w^U$ and leisure value $\epsilon^U$ if unemployed. The job seeker accepts a job offer with wage $w$ and amenities $\epsilon$ if doing so gives them higher value than their current bundle. Amenities and leisure values differ across job seekers, so firms face uncertainty in whether wage offers will be accepted. The \underline{acceptance rate for wage offers} in this dynamic search environment is given by
	\begin{align*}
		p(w-\tilde{w}) = 
		\begin{cases}
			1 - P\big[ \epsilon - \epsilon_0 < -\lambda^{L(w-w_0)} \cdot (w-w_0) + \Omega^E \big] & \text{ if $\tilde{w} = w_0$ (employed)} \\
			1 - P\big[ \epsilon - \epsilon^U < -\lambda^{L(w-w^U)} \cdot (w-w^U) + \Omega^U \big] & \text{ if $\tilde{w} = w^U$ (unemployed)}
		\end{cases}
	\end{align*}

    \noindent In this dynamic search model, the offer acceptance rate is fundamentally similar to its static counterpart in \autoref{eq:acceptance_rate} with additional option value terms: $\Omega^E$ is the option value of staying at the current job, and $\Omega^U$ is the option value of staying unemployed. They consist of continuation value terms in the Bellman equation, which represent future opportunities to receive wage offers. The acceptance rate remains kinked at the reference point, which is either $w_0$ or $w^U$ depending on the employment status. The marginal profit exhibits sharp changes at the reference point due to this kink, which implies corner solutions for salary matching and pay cut reductions seen in \autoref{fig:optimal_offers}. Fundamentally, key predictions of loss aversion remain unchanged even in this dynamic search environment.

%% file: 3_data.tex
\section{Administrative Data on Job Switchers} \label{sec:admin_data}

The behavioral search model makes three predictions on the distribution of salary growth: 1) bunching, 2) discontinuity, and 3) sharp changes in curvature at zero. I empirically test these predictions by measuring salary growth for job switchers using employment registers in Korea, which I describe below. 

\subsection{Employment Registers in Korea} \label{subsec:employment_registers}

	Administrative data on 13.6 million job transitions come from annual records of matched employer-employee data maintained by Statistics Korea. Employment registers provide reliable data for workers with unemployment insurance (UI) coverage between 2015 and 2020, which notably excludes freelancers, temporary contractors, and self-employed individuals. I augment this data with worker demographics by linking employment registers with population registers, business registers, and population dynamics (for 1983-95 birth cohorts).

	Employment registers contain annual records of total earnings from each job held during the calendar year. When a worker transitions between jobs, total earnings are recorded for each job during that year. Total earnings include all forms of taxable income reported to the UI agency to calculate insurance premiums, which makes it a total compensation measure that includes base earnings, bonuses, over-time, and taxable benefits. 

\subsection{Measurement of Salary Growth} \label{subsec:measure_salary_growth}

	For each job switcher, I measure annualized salaries at previous and new jobs by dividing total earnings by the number of months worked during the transition year. Since this requires clarity on dates of hire and separation for primary jobs, the data excludes workers with multiple jobs, missing earnings, or uncertain dates (e.g. rehires, repeated contracts). The primary measure of interest is salary growth, defined as the difference in logged annualized salaries between new and previous jobs.
	\begin{equation*}
		\text{Salary Growth }(r) = \log(\text{Salary at New Job}) - \log(\text{Salary at Previous Job})
	\end{equation*}
	
	\noindent The analysis sample focuses on full-time switchers earning annualized salaries above the minimum wage and below 100 million KRW. Additionally, I focus on switchers who work for the entire month at hire and seapration to minimize measurement error. This ensures precise measurement of annual salaries so that total earnings do not reflect partial months or prorated paychecks. This is important because annualized salaries divide total earnings by the number of months worked, which would understate salary rates for switchers working partial months. This obscures the detection of bunching, which requires precise alignment between new and previous salaries. Partial months at hire understate salary growth for switchers, while partial months at separation overstate salary growth. 

\subsection{Summary Statistics} \label{subsec:summary_stats}

	\autoref{tab:conditions} shows job switchers remaining after each condition for the analysis sample. Around half of job switchers earn annualized salaries above the full-time minimum wage, and around a third of remaining switchers worked for the entire month at hire and separation. The resulting analysis sample consists of 2.1 million job switchers.

	\autoref{tab:sumstats} compares average characteristics for full-time switchers and the analysis sample. Job switchers in the analysis sample are older and more likely to be female, married, and have children. They also work longer prior to separation (employment duration), have shorter employment gaps, and are less likely to receive pay cuts. (Other differences are small but statistically significant due to large sample size). Although removing partial months is necessary to precisely measure salary growth, it comes at the expense of an analysis sample that is less representative of full-time switchers in Korea.
	
	\autoref{fig:salary_growth_dist} shows the distribution of salaries and salary growth for job switchers. Salary distributions exhibit mass points, which have also been observed in other contexts \parencite{reyes_coarse_2024,dube_monopsony_2020}. I conduct simulations to assess the impact of rounded salaries on bunching, which only partially explains the magnitude observed in the data. In the distribution of salary growth, some switchers at the tails exhibit very large pay raises and pay cuts. These large swings might come from one-time payments like sign-on bonuses or performance-based pay, and unlikely due to transitions between part-time and full-time jobs (since the analysis excludes salaries below the full-time minimum wage). Since testable predictions of loss aversion primarily pertain to the center of the distribution, I focus on the center when presenting figures to emphasize anomalies in the distribution of salary growth.

%% file: 4_empirics.tex
\section{Empirical Evidence on Anomalies} \label{sec:empirics}

The empirical segment of my analysis uses employment registers in Korea to empirically test for the presence of bunching, discontinuity, and sharp changes in curvature in the density of salary growth for switchers. I rule out unlikely explanations for anomalies based on institutional context and supporting evidence from data. I discuss several factors that can contribute to bunching, although they cannot explain the discontinuity or curvature break. However, the observed magnitude for bunching may not align with the magnitude predicted by the model due to such external factors. I corroborate my behavioral interpretation of anomalies with additional evidence from heterogeneity and similar patterns in prior studies. Loss aversion is agnostic on \underline{why} pay cuts incur greater disutility than pay raises, and I use loss aversion to frame several interpretations that can explain all three anomalies.

\subsection{Magnitudes of Anomalies} \label{subsec:anomalies}

    Measuring anomalies relies on kernel density estimates for the distribution of salary growth. \underline{Bunching} is excess mass at zero salary growth, measured as the proportion in the zero bin beyond what is expected from a smooth distribution. \underline{Discontinuity} measures the difference in density immediately above and below zero, while the \underline{curvature break} measures the change in the slope of the density at zero (i.e. difference in changes immediately above and below zero). Each salary growth bin is 0.002 log points wide, and I estimate its density by kernel-smoothing proportions in each bin with local-linear polynomials to account for boundary bias. Local-linear polynomials are weighted with the epanechnikov kernel with bandwidth 0.020, which I separately fit for pay raises and pay cuts.

    My main results focus on employed job switchers, for whom current salaries are salient as reference points. \autoref{fig:anomaly_sample} highlights three anomalies in the distribution of salary growth for switchers: 1) bunching of salary growth at zero (top panel), 2) discontinuity in its density at zero (left panel), and 3) sharp changes in its curvature at zero (right panel). The bottom panels exclude the zero bin to emphasize differences in density above and below zero. Notably, anomalies are more pronounced for disadvantaged job seekers. Bunching and discontinuity are larger for switchers with lower pay and experience (\autorefapx{fig:hetero_salary_tenure}), and they are also more pronounced for women in metro areas without college degrees (\autorefapx{fig:hetero_demogs}). This can be relevant for subsequent analyses on the pass-through of hiring subsidies, which are often intended for disadvantaged job seekers.

    \autoref{tab:anomalies} are corresponding estimates for anomalies in \autoref{fig:anomaly_sample}. Bunching measures excess mass in the zero bin. $\hat{p}_0 = 6.04\%$ of switchers are in the zero bin, while only $\hat{a}_0 = 0.64\%$ would be expected from a smooth distribution. This corresponds to an excess mass of 5.40 pp in the zero bin, which is 8.46 times larger than expected levels without loss aversion. This might even understate true bunching magnitudes if measurement error remains after excluding partial months at hire and separation. For example, some employers might disburse paychecks for December in January of the following year, which would inflate previous salaries and deflate new salaries. Since bunching requires precise alignment between new and previous salaries, bunching can be understated when there is remaining measurement error.

    Discontinuity measures the jump in density at zero salary growth. $\hat{a}_0 = 0.64$ is the percent of switchers in the zero bin based on the density of pay raises, while $\hat{b}_0 = 0.59$ is the analogous proportion based on the density of pay cuts. Density immediately above zero is 8.77\% larger than density immediately below zero, which indicates a discontinuity in the density of salary growth. The curvature break measures changes in the density’s slope at zero. Proportions for pay cuts change by $\hat{b}_0 - \hat{b}_2 = 0.0104 \; pp$ between the zero bin and the bin below it. However, proportions for pay raises only change by $\hat{a}_2 - \hat{a}_0 = 0.0016 \; pp$ in the zero bin and the bin above it. Changes in proportions immediately below zero are 6.5 times steeper than changes immediately above zero, which indicates sharp changes in curvature in the density of salary growth.

    Magnitudes for the discontinuity and curvature break can depend on the degree and kernel bandwidth for density estimation. \autorefapx{tab:anomalies_degbw} shows magnitudes for the discontinuity, curvature break, and proportions near zero using varying bandwidths and polynomial degrees for kernel estimation. The primary specification uses local-linear polynomials with bandwidth 0.020, which is close to the Rule-of-Thumb (RoT) bandwidth that minimizes weighted mean squared error (0.019 for pay cuts, 0.022 for pay raises). This is reasonable considering that each salary growth bin is 0.002 log points wide, and 1000 bins are used for estimation between -1 and +1. For local-quadratic polynomials, the RoT bandwidth is wider at 0.062 for pay cuts and 0.068 for pay raises. Although exact magnitudes vary with bandwidth and polynomial degree, all anomalies are statistically significant by large margins regardless of these choices. I maintain consistency across all figures by fixing the bandwidth to 0.020 with local-linear polynomials, which stays close to the RoT bandwidth while achieving good fit with empirical proportions in salary growth bins.

\subsection{Contributing Factors} \label{subsec:contributing_factors}

    I rule out several explanations for anomalies based on institutional context and supporting evidence from data. Anomalies cannot be explained by transitioning between minimum wage jobs, since the analyis sample only includes switchers paid above the minimum wage. Collective bargaining between unions and firms is not particularly common in Korea, and self-reporting errors are unlikely since income taxes in Korea are typically filed by professional accountants hired by employers. There are no sharp changes in average characteristics around zero salary growth (\autorefapx{fig:demogs}), and regulations on over-time pay can be ruled out since anomalies are pronounced even for smaller firms exempt from these regulations (\autorefapx{fig:anomaly_smallfirm}).\footnote{
        Small-Medium Businesses (SMBs) in Korea are often exempt from regulations on overtime pay through the Blanket Wage System.
    } I confirmed with data administrators that arbitrary reassignments of firm IDs are unlikely, and bunching magnitudes are similar for switchers with visible changes in employer attributes (\autorefapx{fig:id_swaps}).

    Several factors can contribute to bunching, although they do not account for the discontinuity or curvature break. First, some employers may believe that prior salaries are signals for the market price of labor provided by job seekers. This can explain why offers bunch at current salaries, and noisy signals can also explain why there is smooth dispersion above and below zero. However, this cannot explain why the density of pay cuts is \underline{discontinuously} lower than pay raises.
 
    Second, rounded salaries can contribute to bunching through mass points in the salary distribution. However, rounded salaries cannot explain why the density of pay cuts is discontinuously lower than pay raises. \autorefapx{fig:sim_bunch} presents simulated distributions from placebo tests to assess the impact of rounded salaries on bunching, which only offers a partial explanation for the observed magnitude. I randomly drew previous and new salaries from their empirical distributions (with replacement), and I took their difference to plot the simulated distribution of salary growth in the top panel of \autorefapx{fig:sim_bunch}. In this simulation, 0.25\% of switchers bunch at zero growth, which is small but visibly distinct. I further improved this procedure by conditioning the distribution of new salaries on each value for the previous salary. Specifically, I re-weighted the distribution of new salaries using a smooth distribution of salary growth that is shifted to be centered on each previous salary value. I took a random draw from this re-weighted distribution for new salaries and took its difference with the previous salary to calculate salary growth. The bottom panel of \autorefapx{fig:sim_bunch} plots the resulting simulation, which is a more realistic distribution of salary growth. 1.01\% of switchers bunch at zero in this simulation, which shows that rounded salaries clearly contribute to bunching but is significantly lower than the observed magnitude (6.04\%). This suggests that rounded salaries can only partially explain why job switchers bunch at zero.

    Finally, some workers may choose to reject all pay cuts because they only care about wages. This possibility can be considered in two cases: 1) all job seekers are wage-centric, and 2) there is a mixture of job seekers who only care about wages and those who also care about amenities. The first case cannot explain the co-existence of pay cuts with wage dispersion,\footnote{
        Even in job ladder models where some workers take pay cuts for higher future wages (e.g. \cite{postelvinay_distribution_2002}), creating wage dispersion requires adding heterogeneity. 
    } and the second case can explain bunching but not the discontinuity (bunching can be explained by many workers being indifferent when offered their current salaries).\footnote{
        Non-wage amenities create smooth dispersion around current salary for a subset of job seekers. The other subset of job seekers who only care about wages will reject all pay cuts, but they will also never receive pay raises since being salary matched will always make them indifferent. This would imply bunching in the distribution of salary growth, but not discontinuity.
    }

    These competing explanations do not invalidate my empirical argument for loss aversion, since they can only explain bunching and not the discontinuity or curvature break. However, the magnitude of bunching predicted by the model may not align well with the observed magnitude due to these external factors, which is further discussed in \autoref{sec:results}.

\subsection{Evidence from Heterogeneity and Prior Studies} \label{subsec:hetero_prior_studies}

    I corroborate my behavioral interpretation of anomalies with additional evidence from patterns of heterogeneity. \autoref{fig:career_gaps} shows that anomalies are more pronounced for employed job switchers without employment gaps. This is consistent with current salaries being more salient as reference points for job seekers who are continuing to receive those paychecks, while prior salaries are less salient for unemployed job seekers who no longer hold those positions. I mark job switchers as unemployed if hire and separation dates differ by at least one month between job transitions, and I mark them as employed otherwise. The top panel shows larger bunching for employed job switchers. The bottom panels show a clear discontinuity for employed switchers, but the same is not true for unemployed switchers.

    \autoref{fig:govt_private} compares bunching and discontinuity for switchers in private and public sectors. This is consistent with negotiation on precise salaries in the private-sector, while the public sector only engages in coarse negotiation based on structured pay scales set by the Korean government. Being able to precisely determine salary offers explains why bunching is observed in the private sector, but not in the public sector (top panel). However, both public and private sectors exhibit discontinuities (bottom panel), which is consistent with both employers engaging in pay cut reduction to avoid offer rejections (either by increasing salary offers or pay steps).

    Observed anomalies for job switchers are also consistent with findings from prior studies, although reference points can depend on the context. Top panels of \autorefapx{fig:prior_studies} show two examples of anomalies for job switchers from prior studies. The top left panel is from \textcite{grigsby_aggregate_2021}, which analyzes administrative payroll data from ADP. This figure shows that the distribution of wage growth for switchers exhibits the same anomalies found in this paper: bunching, discontinuity, and curvature break at zero wage growth. The top right panel is from \textcite{barbanchon_gender_2020}, which analyzes administrative data on unemployed job seekers in France. This figure shows that the distribution of re-employment wages exhibits bunching, discontinuity, and curvature break with respect to self-reported reservation wages. Considering that many wages are accepted below this value, reservation wages seem to practically function as reference points in empirical contexts.

    My behavioral search model proposes that kinked acceptance rates and corner solutions for wage offers are key mechanisms for observed anomalies. Bottom panels of \autorefapx{fig:prior_studies} show two findings from prior studies that are consistent with these mechanisms. The bottom left panel is from \textcite{roussille_bidding_2024}, which analyzes salary offers made to candidates and their decisions on an online job board. This figure shows that the probability of accepting an offer exhibits a kink at the desired salary, which serves as a reference point for job seekers. The bottom right panel is from \textcite{krueger_contribution_2016}, which analyzes survey data on unemployed job seekers in New Jersey. This figure shows that offered wages bunch at self-reported reservation wages for job seekers, which functions as another reference point that is often close to their prior wage. 

\subsection{Alternative Explanations} \label{subsec:framing}

    Loss aversion can explain all three anomalies at once, and it is an appealing interpretation because 1) it intuitively justifies why preferences would be kinked at prior salaries, and 2) it aligns with empirical evidence both in this study and in prior studies. Loss aversion does not take a stance on \underline{why} disutility for pay cuts is steeper than pay raises, which makes it an adaptable framework that embodies a wide range of explanations. I use loss aversion to frame four interpretations that can explain all three anomalies.

    First, anomalies are consistent with job seekers who care about non-wage amenities but heuristically reject all pay cuts. Since non-wage amenities allow workers to take pay cuts through compensating differentials, workers who care about amenities cannot indiscriminately reject all pay cuts unless incentives change at the current salary. This is exactly what loss aversion proposes, either as a kink or a notch. A notch implies that pay cuts incur a discrete loss in utility, while a kink implies discrete changes in marginal utility. For example, workers may heuristically refuse all pay cuts because they find them off-putting. This is reference-dependence with a notch since going below the current salary incurs a discrete loss. Alternatively, workers may unwillingly accept pay cuts but place demanding requirements on workplace benefits. This is reference-dependence with a kink since offers below the current salary incur steeper tradeoffs between wages and amenities. For job seekers in Korea, a kink better explains the data than a notch. A notch predicts clearly defined boundaries for missing mass below the reference point, which I do not see in the data (see \cite{kleven_bunching_2016} for a detailed comparison of kinks and notches).
    
    Second, anomalies are consistent with job seekers using their current salaries as benchmarks for outside options when negotiating with employers. This benchmark can be modeled as a reference point, and firms can make salary offers above or below this benchmark. Loss aversion says that salary offers below the benchmark will trigger steeper tradeoffs between wages and amenities. This still allows job seekers to accept offers below the benchmark if compensated with enough amenities.

    Third, anomalies are consistent with fairness perceptions regarding pay. Fairness can be modeled with reference-dependent preferences, as prior studies have done \parencite{kaur_nominal_2019,breza_morale_2018,mas_pay_2006,card_inequality_2012}. Pay cuts can incur displeasure due to perceived unfairness, which triggers steeper disutility than pay raises. However, what is not obvious is whether social norms on fair pay are misperceived by employers (e.g. \cite{bursztyn_misperceived_2020}). Firms may choose to salary match and reduce pay cuts because they mistakenly believe that job seekers  think pay cuts are unfair, even though this misrepresents their actual preferences. Empirical results can confirm that loss aversion shapes salary offers, but they cannot distinguish whether they are misperceived or accurately understood by employers.

    Finally, debt obligations held by job seekers is also compatible with all three anomalies. For a job seeker with monthly mortgage payments, accepting a lower salary might hinder their ability to make payments on-time due to liquidity constraints. Even without liquidity constraints, they would incur  interest payments by taking out loans. This can be modeled as a reference-point, since job seekers experience steeper disutility from financial hardship when being paid below their current salary. In this framework, job seekers would demand significant compensation in non-wage amenities to accept pay cuts since doing so requires them to draw on savings or take out loans to meet their debt obligations.

%% file: 5_calib.tex
\section{Estimation of Model Parameters} \label{sec:calib}

I use minimum distance to fit model parameters on proportions of switchers in each salary growth bin. This allows me to 1) compare model fit under standard and behavioral preferences, 2) infer plausible magnitudes for loss aversion, and 3) quantify the impact of loss aversion on the pass-through of hiring subsidies to wages. I discuss further details below.

\subsection{Minimum Distance on Salary Growth Bins} \label{subsec:mindist_procedure}

	Since job seekers selectively accept salary offers, there is censoring in the observed distribution of salary growth. \autoref{fig:datagen_process} is a diagram for the data-generating process according to the model. Firms make salary offers that maximize expected profits, which account for uncertainty in how job seekers value amenities and kinks in their acceptance rates. Job seekers decide whether to accept or decline these offers, and accepted offers become realized salary growth for job switchers. Since employment registers do not collect data on rejected offers, I do not observe the full distribution of wage offers or specific decisions made by job seekers. The observed distribution of salary growth is the censored distribution of wage offers selectively accepted by job seekers, with acceptance rate $p(r)$ being the censoring function. 

	The model has five parameters: loss aversion ($\lambda$), location and scale for labor productivity $(\mu_\phi, \sigma_\phi)$, and location and scale for non-wage amenities $(\mu_\epsilon, \sigma_\epsilon)$. I use the logistic distribution to parameterize unobserved heterogeneity, but any two-parameter distribution with an increasing inverse mills ratio is also permissible. The appendix shows that this choice is not critical, as results are similar when using the normal distribution.

	I find model parameters that minimize sum of squared distances between predicted and empirical proportions in salary growth bins. \hyperref[sec:estimation_appendix]{Estimation Appendix~\ref*{sec:estimation_appendix}} uses the model to predict proportions in each bin, which accounts for censoring through job seekers selectively accepting salary offers. Since the implications of loss aversion are most pronounced near the reference point, the estimation procedure focuses on the center of the distribution. I use salary growth bins between -0.2 and +0.2 log points, which account for three quarters of all switchers. I equally weigh bins with identity weights, since optimal weights place more weight on outskirts of the distribution farther away from the reference point.\footnote{
		This is because the optimal weight is the inverse of variance $p \cdot (1-p)$, which is smaller for bins with smaller proportions $p$.
	} I show in the appendix that these adjustments are not critical to my results, and I am able to recover exact parameters when implementing this procedure on simulated data. However, estimation results can be sensitive to including or excluding the zero bin since external factors can contribute to bunching at zero (refer to \autoref{subsec:contributing_factors}). The primary specification excludes the zero bin to avoid contamination from such factors, but the appendix reports results with the zero bin to show how this affects my results. 

\subsection{Identification and Fixed Calibrations} \label{subsec:fixed_calibrations}

	\autoref{tab:calibration_overview} summarizes which model parameters are calibrated at fixed values or estimated through minimum distance. Loss aversion $(\lambda)$ and productivity parameters ($\mu_\phi, \sigma_\phi$) are estimated with minimum distance on salary growth bins, using the procedure described above. Productivity parameters control the location and spread of the offer distribution, while loss aversion controls magnitudes of anomalies and the degree of asymmety between densities for pay raises and pay cuts.

	For identification, I calibrate location and scale parameters for amenities $(\mu_\epsilon, \sigma_\epsilon)$ at fixed values. \autorefapx{fig:offer_dist_tweaks} shows why this is necessary: the top left panel is the predicted distribution of salary growth using a baseline set of parameters, and three other panels are based on changes in parameter values. Raising $\mu_\phi$ (higher productivity), lowering $\mu_\epsilon$ (worse amenities), and lowering $\sigma_\epsilon$ (less variation in amenities) shift salary growth to higher values in observationally similar ways. Since these changes are not distinguishable from each other, minimum distance cannot separately identify $(\mu_\phi, \mu_\epsilon, \sigma_\epsilon)$. Among these parameters, I calibrate two of them at fixed values for identification. First, I fix the location parameter for amenities at $\mu_\epsilon = 0$ by assuming that non-wage amenities are equal on average at previous and new jobs. Second, I fix the scale parameter for amenities at $\sigma_\epsilon= 0.611$ by matching on the variance of non-wage amenities in \textcite{lehmann_non-wage_2025},\footnote{
		Since amenities in my model are changes between previous and new jobs, I specifically match on the variance for the difference in non-wage amenities assuming i.i.d. Their variance is in log-wage units, which aligns with my model setup.
	} which builds on \textcite{sorkin_ranking_2018} by using worker flows to decompose the total value of jobs into wage and non-wage components. This is admittedly not ideal for a model calibrated on Korean data since this variance comes from Austria, and the appendix reports sensitivity of results to other calibrations for $\sigma_\epsilon$.

%% file: 6_results.tex
\section{Assessment of Model Fit}\label{sec:results}

I provide an assessment of model fit and report on parameters obtained from minimum distance. Estimated loss aversion is $\hat{\lambda}=1.123$, which significantly improves model fit to proportions in salary growth bins compared with a standard model that restricts $\lambda=1$. The fitted model can reasonably match the discontinuity and curvature break despite not being explicit targets, but predicted bunching is four times larger than the observed magnitude. I discuss these findings in further detail below.

\subsection{Proportions in Salary Growth Bins} \label{subsec:fit_bins}

	\autoref{fig:model_fit} compares empirical and predicted proportions in non-zero salary growth bins, which shows that incorporating loss aversion significantly improves model fit. The behavioral model is better at capturing both the level and curvature of the density around zero, and it better reflects the asymmetry between densities of pay raises and pay cuts. Conversely, the standard model excessively smooths the density near zero, resulting in a poorer fit by under-predicting proportions for bins near zero while over-predicting proportions for bins farther away from zero.

	\autoref{tab:parameters_main} reports the $\chi^2$ statistic along with the $\alpha=0.05$ critical value for the quasi-likelihood ratio (QLR) test to compare minimized criterion values for nested models. The $\chi^2$ statistic for the (QLR) distance test is quite large. Incorporating loss aversion significantly lowers the squared distance between predicted and empirical proportions, which is indicative of better fit to the salary growth distribution.

\subsection{Magnitude of Loss Aversion} \label{subsec:calibrated_lossav}

	\autoref{tab:parameters_main} reports parameter estimates and their standard errors in parentheses. Estimated loss aversion is $\hat{\lambda}=1.123$, indicating that the marginal value of additional pay is 12.3\% higher for pay cuts than pay raises. This is smaller than magnitudes reported in prior studies on reference-dependence --- both in the context of job search \parencite{dellavigna_reference-dependent_2017,dellavigna_evidence_2022} and in broader empirical studies (see \colorcite{brown_meta-analysis_2024} for a meta analysis on loss aversion). To better understand these magnitudes, \autorefapx{fig:lossav_paycut_freq} compares predicted proportions in salary growth bins for $\lambda=1.123$ (this study) and $\lambda=1.955$ (average across studies in \cite{brown_meta-analysis_2024}). A larger magnitude for loss aversion implies substantially less pay cuts, which I do not observe for switchers in Korean data. However, larger magnitudes for loss aversion may better align with the absence of pay cuts observed for job stayers in US payroll data \parencite{grigsby_aggregate_2021}.

	\autorefapx{fig:criterion} shows that the minimum distance criterion exhibits a clear minimum at estimated parameters, which are not sensitive to starting values. \autorefapx{tab:parameters_adjust} also shows that parameters are not sensitive to alternative specifications. The first row are estimates from the primary specification, and subsequent rows report estimates with four adjustments: replacing kernel estimates with raw proportions in the data (row 2), parameterizing unobserved heterogeneity with the normal distribution (row 3), using optimal weights for bins (row 4), and expanding the range of salary growth from $\pm 0.2$ to $\pm 1.0$ log points. Note that productivity scale $\sigma_\phi$ are in different units for normal and logistic distributions ($\sigma^2=$ variance for normal, scaled variance for logistic). Optimal weights and expanding the range of salary growth have similar effects: they place more weight on the outskirts of the distribution. Higher $\sigma_\phi$ increases the spread of the productivity distribution to better fit the tails at the expense of under-fitting proportions near the zero bin.\footnote{
		Note that the optimal weighting matrix is the inverse covariance matrix of proportions in salary growth bins. Bins $b$ farther away from zero receive larger weights because they have smaller proportions $p_b$ with variance $p_b*(1-p_b)$.
	} As discussed in \autoref{subsec:summary_stats}, very large magnitudes for pay raises or pay cuts can come from one-time payments like sign-on bonuses or performance-based pay. Since my model does not seek to explain tail values, the primary specification focuses on achieving better fit for moderate values of salary growth.

	\autorefapx{fig:calib_params} shows the sensitivity of estimates to fixed calibrations for the amenity scale parameter $\sigma_\epsilon$, with the scatter point indicating the primary specification that matches the variance of non-wage amenities in \textcite{lehmann_non-wage_2025}. As expected, productivity location $\hat{\mu}_\phi$ is increasing with fixed calibrations for amenity scale $\sigma_\epsilon$. This is because both parameters control the location of the salary growth distribution and are not separately identified (refer to \autoref{subsec:fixed_calibrations}). Higher variance in non-wage amenities imply less sensitivity to wages, which lowers salary offers. This is offset by higher productivity among job seekers, which increases salary offers.

	\autorefapx{fig:calib_params} also shows that loss aversion $\hat{\lambda}$ is decreasing with fixed calibrations for amenity scale $\sigma_\epsilon$. To understand these patterns, note that the acceptance rate flattens out with increasing amenity scale $\sigma_\epsilon$. Optimality conditions for wage offers in \autoref{eq:optimal_offer} imply $\phi = r + \frac{1}{\lambda^{L(r)} } \cdot \frac{1 - F_\epsilon(-\lambda^{L(r)} \cdot r )}{f_\epsilon(-\lambda^{L(r)} \cdot r )}$. The mills ratio term $\frac{1-F_\epsilon(\cdot)}{f_\epsilon(\cdot)}$ is increasing in $\sigma_\epsilon$ due to lower density of amenities at zero, which implies that the same magnitude of loss aversion $\lambda$ has a larger impact on salary offer $r$. These changes are offset with a lower magnitude for loss aversion, so that the overall salary impact of loss aversion remains the same.

\subsection{Anomaly Magnitudes} \label{subsec:fit_anomalies}

	\autoref{fig:anomaly_comparison} compares anomalies predicted by the model with their observed magnitudes. The fitted model can reasonably match the discontinuity and curvature break despite not being explicit targets, but the predicted magnitude of bunching is four times larger than observed bunching at zero. Earlier discussions in \autoref{sec:empirics} noted that observed bunching can either be understated due to measurement error or overstated due to external contributing factors. There is suggestive evidence that bunching is likely understated, as loss aversion implied by bunching is significantly lower than what is implied by all other segments of the distribution. When including the zero bin in the minimum distance procedure, the model closely matches bunching by substantially lowering loss aversion (\autorefapx{tab:parameters_sub}), which comes at the expense of under-predicting the discontinuity and curvature break (\autorefapx{fig:anomaly_comparison_zero}). Furthermore, including the zero bin significantly worsens model fit by making it unable to match the density or its curvature (\autorefapx{fig:model_fit_zero}). Since observed bunching at zero appears to lack consistency with other segments of the distribution, the primary specification excludes the zero bin to avoid worsening fit in all other bins. However, the distance test in \autorefapx{tab:parameters_sub} shows that loss aversion is still a significant improvement over the standard model even when including the zero bin, which is the central message of this paper.

%% file: 7_policy.tex
\section{Implications for Hiring Subsidies} \label{sec:policy}

I use a wide range of values for loss aversion to discuss its implications on wages, welfare, and the pass-through of hiring subsidies. Larger values of loss aversion imply smaller and less frequent pay cuts, and employers are more likely to match offers at current salaries. Loss aversion lowers the pass-through of hiring subsidies to wages for two reasons: 1) higher elasticity for pay cuts lowers tax incidence on workers, and 2) salary matches constrain subsidizied offers made by employers. These effects remain unchanged even with salary history bans as long as employers can imperfectly observe current salaries with noise. I cannot quantify welfare implications since behavioral utility is not transferrable between workers and firms, but I make several qualitative comments on individual components of welfare. I discuss these aspects in further detail below.  

\subsection{Implications on Salary Offers} \label{subsec:salary_implications}

	The left panel of \autorefapx{fig:wage_implications} shows how loss aversion affects the prevalence of pay raises, pay cuts, and salary matches, and the right panel is an analogous figure for average salary offers. These figures highlight three specific values of loss aversion: $\lambda = 1$ (no loss aversion), $\lambda = 1.123$ (this study), and $\lambda = 1.955$ (average across studies in a meta analysis by \cite{brown_meta-analysis_2024}). Green lines for pay raises are flat in both panels, indicating that loss aversion does not affect pay raises in terms of prevalence or magnitude. However, red lines indicate that pay cuts are less frequent and smaller in magnitude for larger values of loss aversion. The yellow line in the left panel shows that salary matching is increasingly common for larger values of loss aversion, which is an important channel for limiting the pass-through of hiring subsidies to wages. Altogether, higher values of loss aversion imply more frequent salary matches and less frequent pay cuts, which increase the overall average of salary offers received by job seekers (purple line).

	\autorefapx{fig:salary_impact_adjust} shows that these implications are not sensitive to alternative specifications, and \autorefapx{fig:salary_impact_calib} shows they are also not sensitive to fixed calibrations for the amenity scale parameter $\sigma_\epsilon$. This may seem counterintuitive since loss aversion does depend on $\sigma_\epsilon$, but this is consistent with the earlier discussion in \autoref{subsec:calibrated_lossav}. Higher variance for non-wage amenities flattens the acceptance rate, which amplifies the impact of loss aversion. This is offset by smaller calibrated values for loss aversion, which leaves overall wage impacts of loss aversion unchanged.

\subsection{Pass-Through of Hiring Subsidies to Wages} \label{subsec:pass_through}

	It is common for governments to subsidize employers for hiring certain workers, including recent subsidies introduced by the Korean government. There are diverse motives for hiring subsidies, which range from offsetting labor costs for critical industries to promoting career opportunities for disadvantaged job seekers. An important question in Public Finance is the extent to which hiring subsidies benefit workers or firms. Workers fully benefit from the subsidy if the entire amount passes on to higher wages, and firms fully benefit if wages remain unchanged. Many policy decisions rely on understanding the pass-through of hiring subsidies to wages.

	\,

	\noindent Consider a hiring subsidy of size $\Delta$ given to employers. Subsidized profits are given by
	\refstepcounter{equation} \label{eq:subsidized_profits}
	\begin{equation}
		\pi(r) = p(r) \cdot (\phi_{if} + \Delta - r)
		\tag*{(\theequation): Subsidized Profits}
	\end{equation}

	\noindent Since the subsidy raises the value of labor to firms, this is equivalent to shifting the productivity distribution so that $\phi_{if} + \Delta \sim F_\phi( \mu_\phi+\Delta, \sigma_\phi)$. It is otherwise equivalent to the model developed in \autoref{sec:model}, and the left panel of \autoref{fig:policy_implications} shows the distribution of salary growth for a hypothetical subsidy $\Delta$ that raises the value of labor to employers. Unsurprisingly, hiring subsidies shift the distribution of salary growth to higher values. 
	
	\autoref{fig:passthru_lossav} shows that neglecting loss aversion would overstate the pass-through of hiring subsidies to wages, which is perhaps less obvious. The left panel plots the overall pass-through of hiring subsidies, measured as the salary increase divided by the size of the subsidy. The right panel plots pass-through separately for marginal and inframarginal job seekers. ``Inframarginal'' refers to job seekers who would accept salary offers regardless of the subsidy (salary offer without the subsidy is already high enough to be accepted). ``Marginal'' refers to job seekers who would only accept the salary offer if augmented with the subsidy (salary offer is only high enough with the subsidy). The right panel shows that loss aversion can especially lower pass-through for inframarginal job seekers who may lack better outside options. This can be especially relevant for disadvantaged job seekers, who are often intended beneficiaries of hiring subsidies and also exhibit more pronounced anomalies (\autorefapx{fig:hetero_salary_tenure} and \autorefapx{fig:hetero_demogs}). These implications are not sensitive to alternative specifications or calibrations for the amenity scale parameter, as can be seen in \autorefapx{fig:impact_sensitivity}.

	\autoref{fig:passthru_constrained} shows the mechanisms through which loss aversion lowers the pass-through of hiring subsidies. The right panel plots salary offers for each value of labor productivity $\phi$, and the left panel plots the implied pass-through as the salary increase divided by the size of the subsidy. Differences between solid and dashed lines represent the impact of loss aversion on the pass-through of hiring subsidies, which is largely divided into three segments of job seekers. 

    \subsubsection*{Case 1: Slightly lower pass-through for pay cuts}
    
		Job seekers with low productivity receive pay cuts, whose acceptance decisions are more sensitive to pay due to loss aversion. Higher elasticity of labor supply lowers tax incidence on workers, which implies lower pass-through to wages. This can be seen in the left panel of \autoref{fig:passthru_constrained}, where the solid line for the behavioral model is below the dashed line for the standard model for low values of productivity $\phi_{if}$.
	
	\subsubsection*{Case 2: No impact on pass-through for pay raises}

        Job seekers with high productivity receive pay raises, which are not affected by loss aversion. As a result, loss aversion does not affect the pass-through of hiring subsidies for these workers. This can be seen in the left panel of \autoref{fig:passthru_constrained}, where solid and dashed lines overlap for high values of productivity $\phi_{if}$.
	
	\subsubsection*{Case 3: Significantly lower pass-through for salary matches}

        For a large subset of job seekers with moderate productivity, offers are salary matched either with the subsidy or without the subsidy. This can be seen in the right panel of \autoref{fig:passthru_constrained}, where solid lines for the behavioral model exhibit flat regions for moderate values of productivity. Pass-through for the behavioral model is the difference between solid lines, which is constrained when subsidized offers (in green) or unsubsidized offers (in red) are salary matched. The intuition is that job seekers with moderately low productivity receive pay cuts without the subsidy, but positive impacts of the subsidy are constrained due to salary matching. Job seekers with moderately high productivity receive pay raises with the subsidy, but even unsubsidized offers are bounded below due to salary matching. In both cases, loss aversion lowers the pass-through of hiring subsidies through salary matching at subsidized or unsubsidized offers.

\subsection{Persistent Effects Under Salary History Bans} \label{subsec:salary_history_bans}

	The optimal wage offer in \autoref{eq:optimal_offer} assumes that the employer knows the job seeker's current wage, which allows them to incorporate reference-dependent preferences. Given these assumptions, it is natural to ask whether wage implications of loss aversion depend on precise knowledge of current salaries. This is likely true in Korea, where it is common for employers to require documentation of current salaries before officially releasing job offers. However, this may not be true in Europe and the United States, where salary history bans prohibit employers from directly asking candidates about their current salaries. 
	
	Somewhat surprisingly, implications of loss aversion remain unchanged even with salary history bans as long as employers have other ways of imperfectly observing current salaries with noise. This is likely true in real-world settings, where it is common practice to ask candidates for ``salary expectations'' or ``fair compensation'' at the application and interview stages of the hiring process. 
	
	Consider an extension where firms imperfectly observe current salaries so that \underline{perceived} salaries are $\tilde{w}_{0i} = w_{0i} + \eta$ with noise $\eta \sim F_\eta(0, \sigma_\eta)$. Since firms cannot directly optimize on relative wage offer $r$, they indirectly optimize on noisy profits based on perceived offers $\tilde{r} = w - \tilde{w}_{0i}$ and perceived productivity $\tilde{\phi}_{if}=\Psi_f - \tilde{w}_{0i}$.
	\refstepcounter{equation} \label{eq:perceived_profits}
	\begin{equation}
		\pi(\tilde{r}) = p(\tilde{r}) \cdot (\tilde{\phi}_{if} - \tilde{r})
		\tag*{(\theequation): Noisy Profits}
	\end{equation} 

	\noindent This extension is mostly similar to the model developed in \autoref{sec:model} except for two key differences. First, the actual wage offer received by job seekers is $r=\tilde{r} + \eta$ while firms believe they are offering $\tilde{r}$. Second, firms cannot salary match exactly at current wage $w_{0i}$ due to imperfect information, so they approximately match at perceived salaries $\tilde{w}_{0i}$.

	The right panel of \autoref{fig:policy_implications} shows the distribution of salary growth under a salary history ban, where precise bunching at $r=0$ is replaced with excess mass near zero due to imperfect salary matching. Otherwise, implications of loss remain unchanged, and productivity is simply replaced with ``perceived'' productivity in \autoref{fig:optimal_offers}. Employers continue to salary match at perceived salaries $\tilde{r}=0$ and reduce perceived pay cuts $\tilde{r} < 0$. Salary offers in the right panel of \autoref{fig:passthru_constrained} are replaced with ``perceived'' salary offers. Subsidy pass-through in the left panel is the same since perception noise $\eta$ is differenced out. The common intuition behind these results is that imperfect observation of current salaries limits employers' ability to precise control salary offers, but their underlying motives remain unchanged. 

\subsection{Qualitative Discussion on Welfare} \label{subsec:welfare}
	
	I cannot quantify welfare implications since behavioral utility is not transferrable between workers and firms. However, I can make several qualitative statements about individual components of welfare. In terms of social welfare, mutually beneficial matches are less likely to form under loss aversion. This is not apparent from the static model discussed throughout the paper, but it can be seen through extensions with wage bargaining (\autoref{subsec:wage_bargaining}) and dynamic search (\autoref{subsec:dynamic_search}). \autoref{eq:bargain_region} defines the region of $(\epsilon, \phi)$ for successful bargains, which is the colored region in \autoref{fig:bargain_solution}. In the wage bargaining model, successfully bargained pay cuts are less likely to occur under loss aversion. \hyperref[subsec:steady_state]{Model Appendix~\ref*{subsec:steady_state}} solves for optimal vacancies in the dynamic search model, with the main result being that it is less profitable for firms to open vacancies because they face higher wage costs and more rejections under loss aversion. Across both extensions, the common welfare implication is that more frequent rejections of pay cuts prevent the formation of mutually beneficial matches between firms and job seekers.

	Firms are worse off under loss aversion because hiring workers is more costly due to salary matching and pay cut reduction. Job seekers benefit from higher wages, but less job opportunities are available to them due to lower vacancies. Implications on equilibrium employment require in-depth analyses that I do not pursue in this paper. On one hand, employment increases since job seekers receive higher salary offers on average (right panel of \autoref{fig:wage_implications}), which they are more likely to accept. On the other hand, employment decreases since firms open less vacancies. Matching models like \textcite{pissarides_equilibrium_2000} are likely important to this discussion, which is a promising application for loss aversion.

%% file: 8_conclude.tex
\section{Directions for Future Research}\label{sec:conclude}

	This paper examined wage implications of loss aversion in three steps. In the first step, I developed a behavioral search model with predictions on the distribution of salary growth. In the second step, I tested these predictions with administrative data in Korea, which confirmed the existence of anomalies in the distribution of salary growth for job switchers. In the third step, I calibrated model parameters with minimum distance and showed that loss aversion significantly improves the quality of predicted proportions in salary growth bins. The calibrated model makes it to quantify implications of loss aversion on the pass-through of hiring subsidies to wages. 

	I conclude with a discussion on key limitations of this paper and how they can be addressed in future research. First, the static model cannot say much about implications of loss aversion on equilibrium employment. This is likely better suited for matching models of labor markets that incorporate loss aversion, which is a promising direction for future research. Second, measurement error can hinder the detection of bunching, despite best attempts to limit them by excluding partial months at hire and separation. This limitation can be addressed with better data on salary rates rather than total earnings, which would likely come from unemployment insurance agencies that use this information to calculate monthly premiums. Finally, the magnitude of loss aversion depends on a fixed calibration for the scale parameter of non-wage amenities, which can be addressed in several ways. One approach is to replicate \textcite{lehmann_non-wage_2025} on Korean data to estimate the variance of non-wage amenities. Another approach is to collect data on accepted and rejected offers, possibly through choice experiments administered on surveys. Since loss aversion is a relatively nascent development in the job search literature, these improvements would provide a valuable foundation for future work building on this intersection.

%% file: 0_figures.tex
\section*{Main Figures}
\addcontentsline{toc}{subsection}{Main Figures}

\vspace*{\fill}
\begin{figure}[H]
    \makebox[\textwidth][c]{\includegraphics[width=0.60\textwidth]{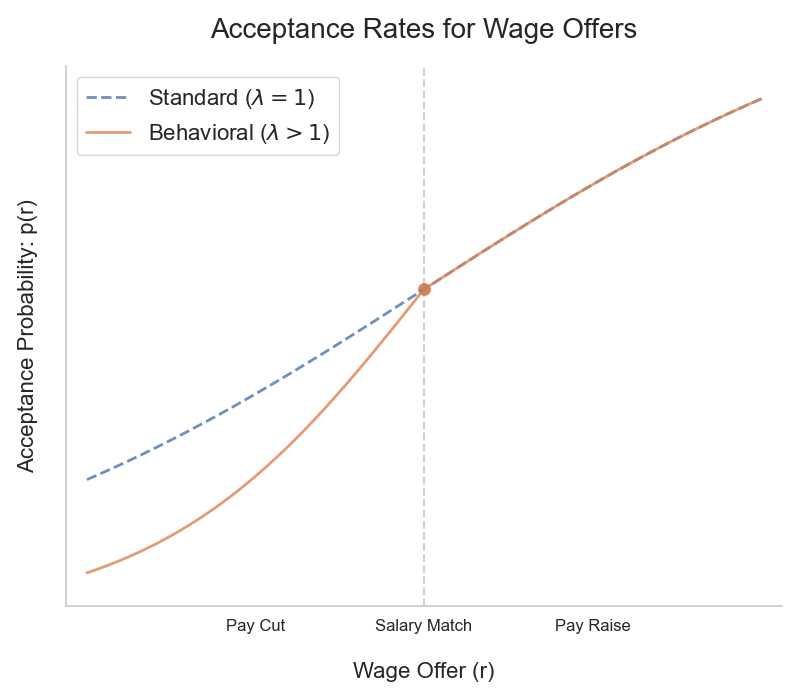}}

    \caption{Acceptance Rates on Wage Offers (Labor Supply)} \label{fig:acp_rate}
    \caption*{\normalsize
        This figure compares the probability of accepting wage offers in \autoref{eq:acceptance_rate} under the standard model $(\lambda=1)$ and behavioral model $(\lambda>1)$. The horizontal axis is offered wage relative to the current wage $(r)$, defined as the difference in (logged) offered wage $w$ and the job seeker's current wage $w_0$. Salary match $(r=0)$ means that the offered wage is identical to the job seeker's current wage. The vertical axis is the probability of accepting the wage offer $p(r)$, given uncertainty in how job seekers value non-wage amenities $\epsilon_{if}$. For the behavioral model, the scatter point denotes a kink in the acceptance rate, with a steeper slope for pay cuts than pay raises.
    }
\end{figure}
\vspace*{\fill}
\clearpage

\newgeometry{top=0.1in, bottom=0.8in, left=0.8in, right=0.8in}

\vspace*{\fill}
\begin{figure}[H]

    \makebox[\textwidth][c]{%
        \includegraphics[width=0.47\paperwidth]{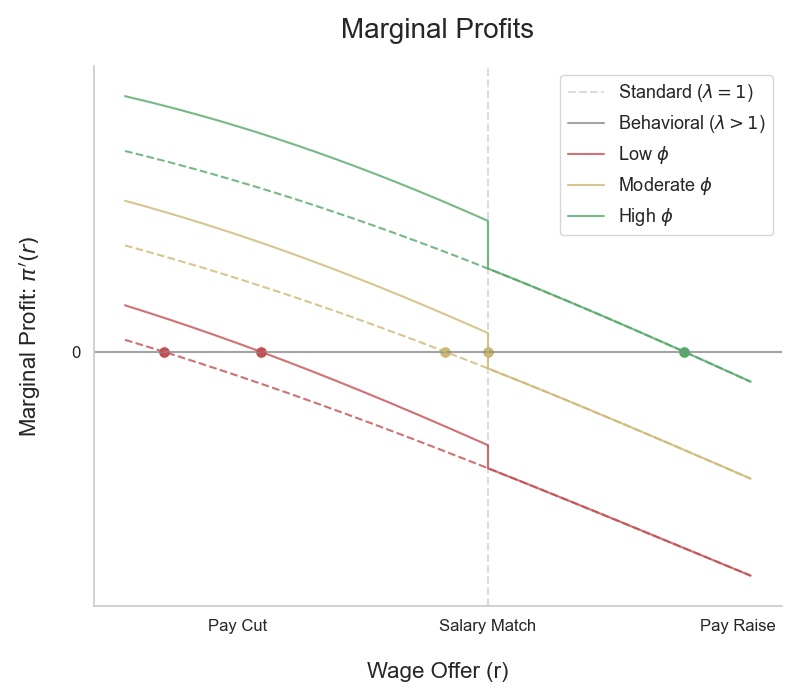}%
        \includegraphics[width=0.47\paperwidth]{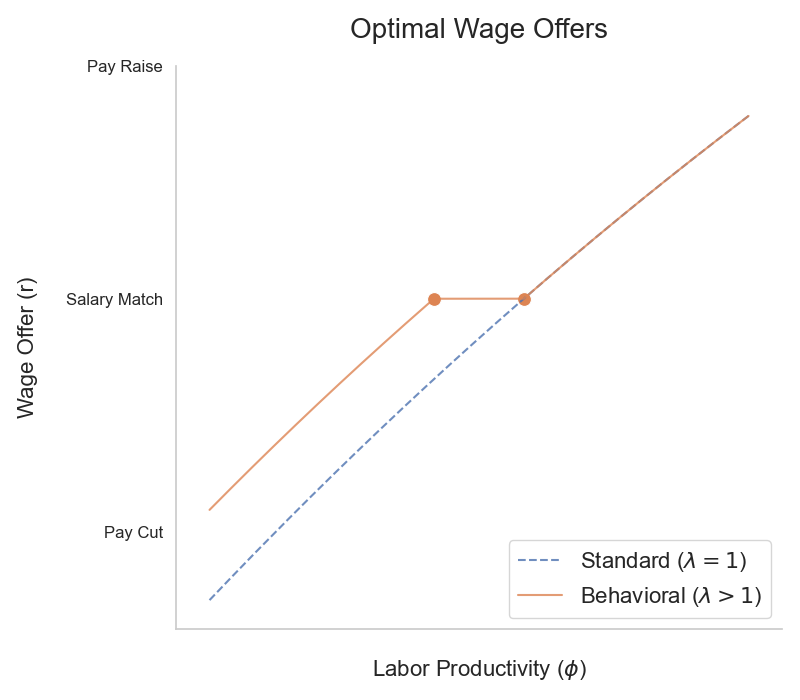}
    }

    \caption{Marginal Profits and Optimal Wage Offers (Labor Demand)} \label{fig:optimal_offers}
    \caption*{\normalsize 

        The left panel compares marginal profits in \autoref{eq:marginal_profits} under the standard model $(\lambda=1)$ and behavioral model $(\lambda>1)$ at three levels of labor productivity $\phi$ (low, moderate, high). The horizontal axis is offered wage relative to the current wage $(r)$, defined as the difference in (logged) offered wage $w$ and the job seeker's current wage $w_0$. Salary match $(r=0)$ means that the offered wage is identical to the job seeker's current wage. The vertical axis is marginal profit $\pi'(r)$ at the offered wage $r$, and scatter points denote optimal wage offers that set marginal profits to zero. Optimal wage offers are pay raises $r>0$ for high productivity (in green) and pay cuts $r<0$ for low productivity (in red). For moderate productivity (in yellow), optimal offers are pay cuts in the standard model (dashed line) and salary matches in the behavioral model (solid line). 

        \, 

        \noindent The right panel compares optimal wage offers in \autoref{eq:optimal_offer} under the standard model $(\lambda=1)$ and behavioral model $(\lambda>1)$. The horizontal axis is labor productivity relative to the current wage $(\phi)$, defined as the difference in (logged) labor productivity $\Psi$ and the job seeker's current wage $w_{0i}$. The vertical axis is offered wage relative to the current wage $(r)$, defined as the difference in (logged) offered wage $w$ and the job seeker's current wage $w_0$. Salary match $(r=0)$ means that the offered wage is identical to the job seeker's current wage. Scatter points denote upper and lower boundaries for productivity values $\phi \in [\underaccent{\bar}{\phi}, \bar{\phi}]$ that result in salary matches. 
    }
\end{figure}
\vspace*{\fill}
\clearpage

\begin{figure}[ht]
    \makebox[\textwidth][c]{\includegraphics[height=0.35\textheight]{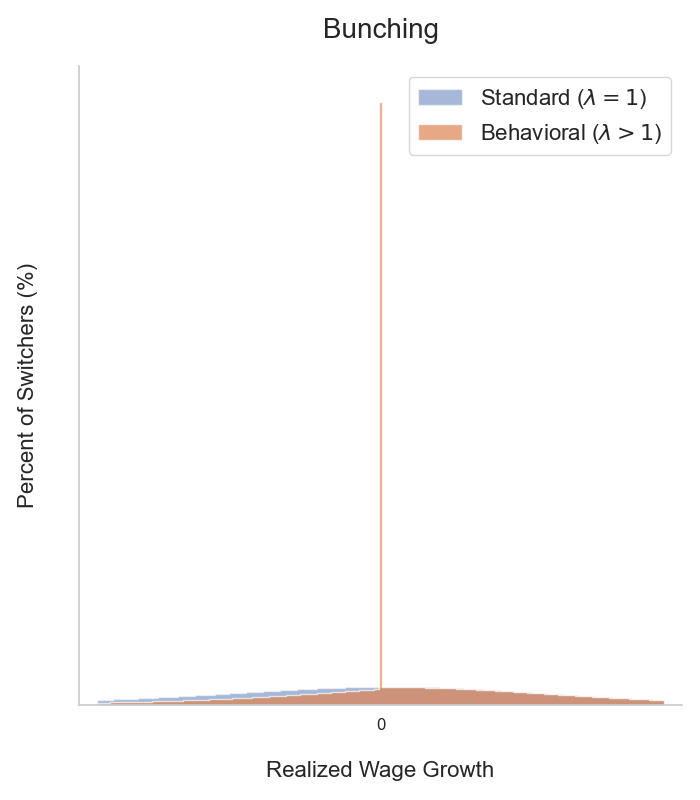}}
    \vspace{1em}

    \makebox[\textwidth][c]{%
        \includegraphics[height=0.35\textheight]{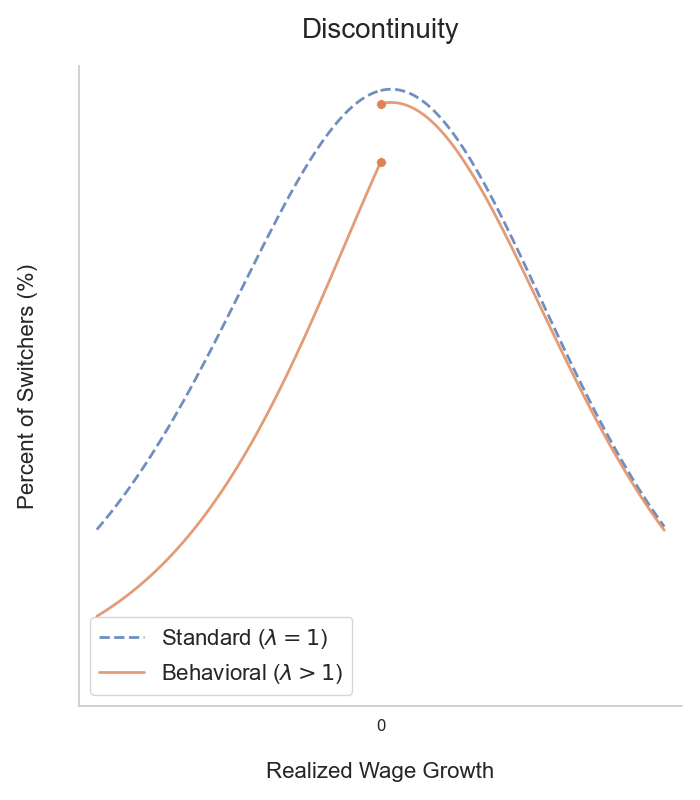}%
        \includegraphics[height=0.35\textheight]{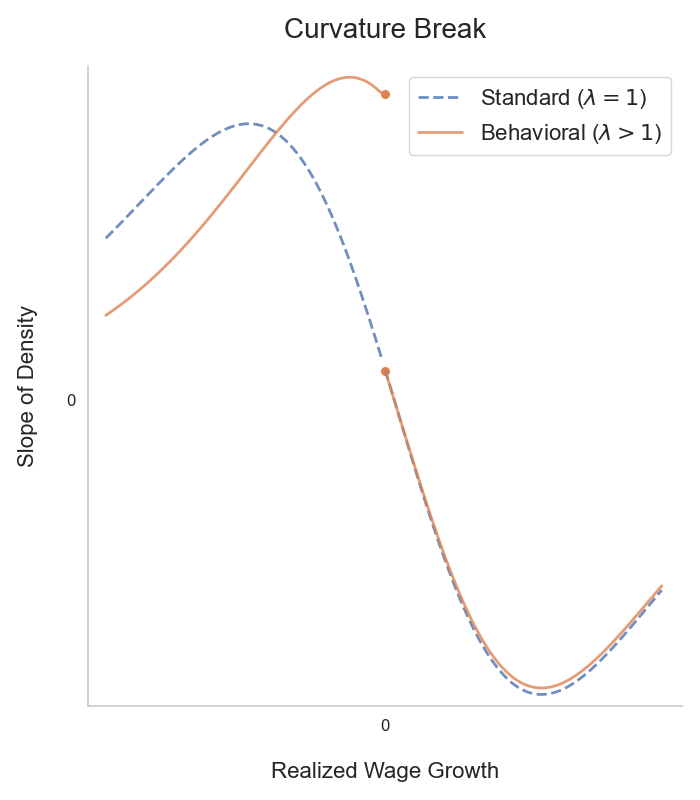}
    }

    \caption{Testable Predictions on the Distribution of Wage Growth} \label{fig:anomaly_theory}
    \caption*{\normalsize 
        This figure emphasizes key differences in the distribution of salary growth for job switchers (orange bars in \autorefapx{fig:offer_dist}) under the standard model $(\lambda=1)$ and behavioral model $(\lambda>1)$. \autoref{subsec:model_prediction} discusses three key predictions from the behavioral model, which are 1) bunching at zero wage growth (top panel), 2) discontinuous density at zero (left panel), and 3) sharp changes in its curvature at zero (right panel). The horizontal axis is realized wage growth for job switchers, defined as the difference in (logged) accepted wage $w$ and their prior wage $w_0$. Salary match $(r=0)$ means that the accepted wage is identical to their prior wage. The vertical axes of top and left panels denote the percent of job switchers in the conditional density of accepted wage offers, and the vertical axis of the right panel denotes the slope of this density. Left and right panels exclude the zero bin to emphasize differences in density above and below zero. 
    }
\end{figure}
\clearpage

\vspace*{\fill}
\begin{figure}[H]

    \makebox[\textwidth][c]{\includegraphics[height=0.35\textheight]{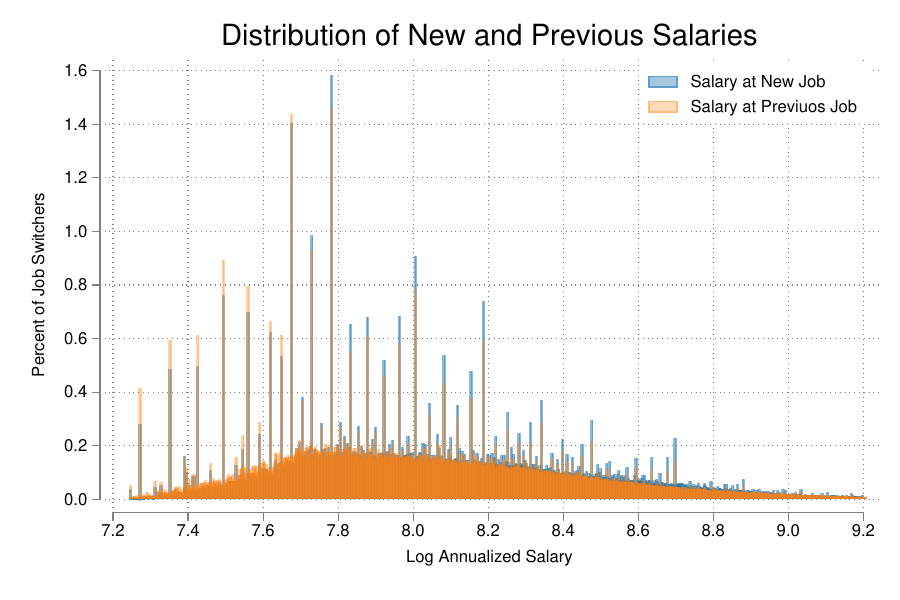}}

    \makebox[\textwidth][c]{\includegraphics[height=0.35\textheight]{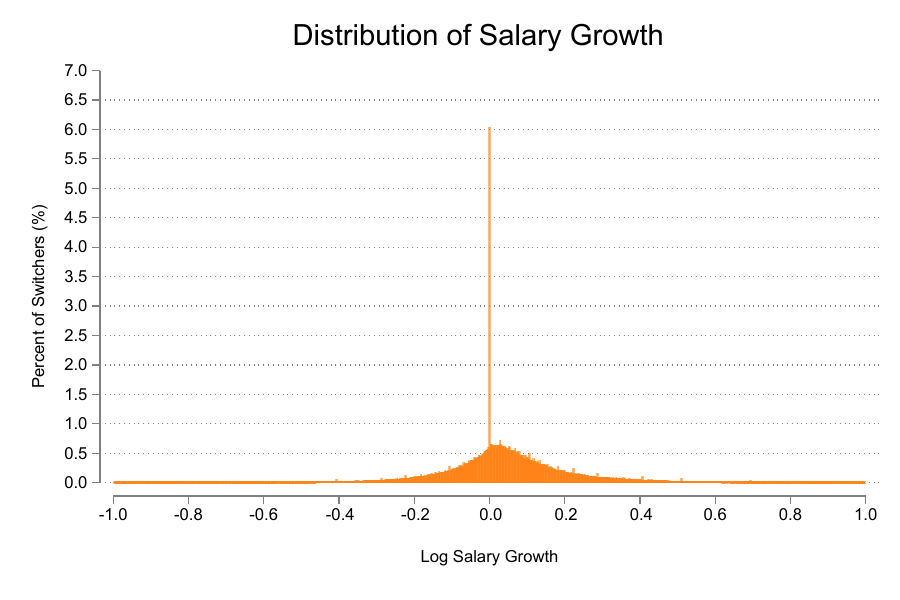}}

    \caption{Distribution of Salaries and Salary Growth for Job Switchers in Korea} \label{fig:salary_growth_dist}
    \caption*{\normalsize
        This figure shows the distribution of previous salaries, new salaries, and salary growth for job switchers in the analysis sample (discussed in \autoref{subsec:summary_stats}). Annualized salaries divide total earnings by the number of months worked during the transitioning year. The analysis sample consists of job switchers working entire months at hire and separation and earning below 100 million KRW and above the full-time minimum wage (non-inclusive). 
    }
\end{figure}
\vspace*{\fill}
\clearpage

\begin{figure}[ht]
    \makebox[\textwidth][c]{\includegraphics[height=0.37\textheight]{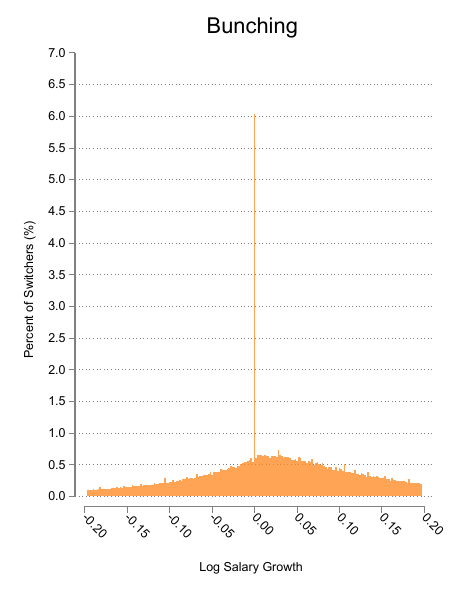}}

    \makebox[\textwidth][c]{%
        \includegraphics[height=0.37\textheight]{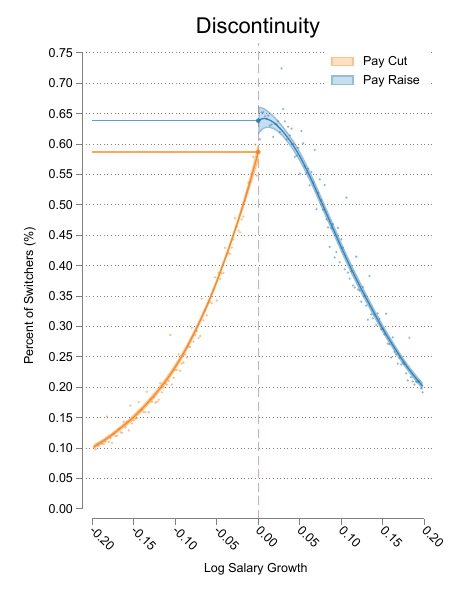}
        \includegraphics[height=0.37\textheight]{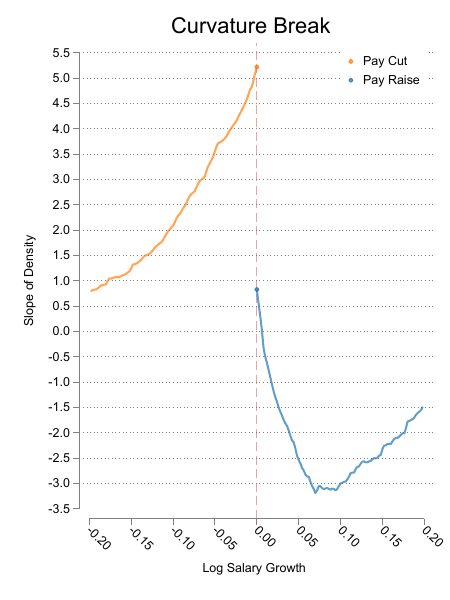}
    }
    
    \caption{Anomalies in the Distribution of Salary Growth for Job Switchers in Korea} \label{fig:anomaly_sample}
    \caption*{\normalsize
        This figure emphasizes three anomalies in the distribution of salary growth for job switchers without employment gaps in the analysis sample (discussed in \autoref{subsec:anomalies}). Corresponding estimates for anomalies can be found in \autoref{tab:anomalies}. The horizontal axis is log salary growth, defined as the difference in log annualized salaries at new and previous jobs. Bars in the top panel and scatter points in the left panel correspond to percentages of switchers in salary growth bins, each of which are 0.002 log points wide. Lines in the left panel are kernel-density estimates (with shaded 95\% confidence intervals) based on proportions smoothed with local-linear polynomials that account for boundary bias. Local-linear polynomials are weighted with the epanechnikov kernel with bandwidth 0.020, which is separately fit for pay cuts (in orange) and pay raises (in blue). The right panel is the slope of the density curve, approximated as changes in kernel-density estimates in adjacent bins. The bottom panels exclude the zero bin to emphasize differences above and below zero.
    }
\end{figure}
\clearpage

\vspace*{\fill}
\begin{figure}[H]
    \makebox[\textwidth][c]{\includegraphics[height=0.38\textheight]{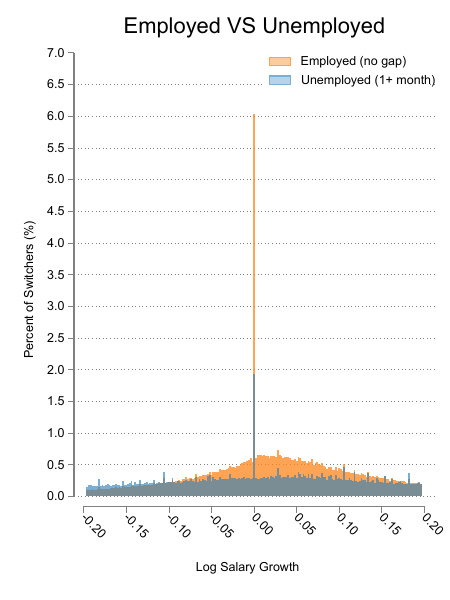}}

    \makebox[\textwidth][c]{\includegraphics[height=0.38\textheight]{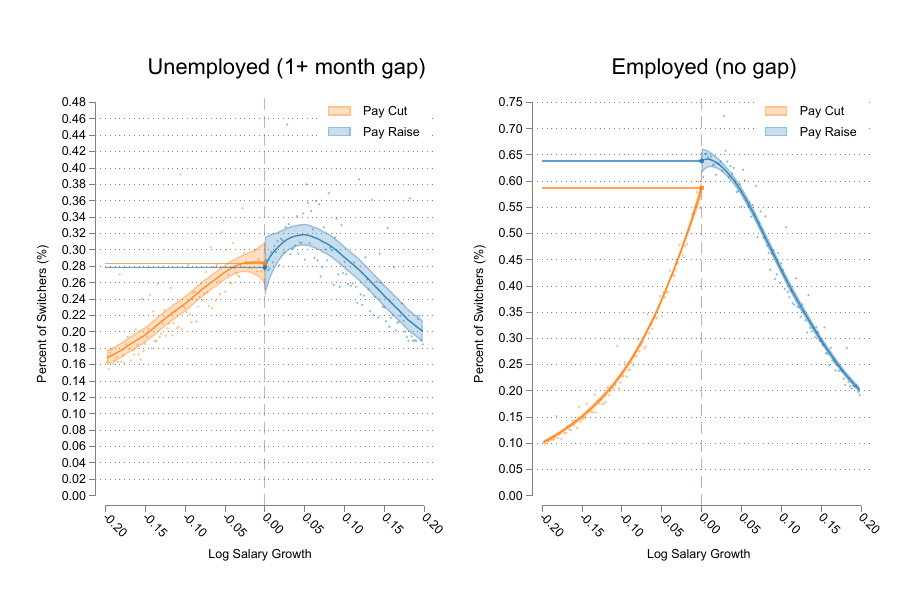}}
    
    \caption{Comparison of Employed and Unemployed Job Switchers} \label{fig:career_gaps}
    \caption*{\normalsize 
        This is an implementation of \autoref{fig:anomaly_sample} that compares bunching and discontinuity for employed and unemployed job seekers (discussed in \autoref{subsec:hetero_prior_studies}). I mark job switchers as unemployed if hire and separation dates differ by at least one month between jobs, and I mark them as employed otherwise (i.e. no employment gaps).  
    }
\end{figure}
\vspace*{\fill}
\clearpage

\vspace*{\fill}
\begin{figure}[H]

    \makebox[\textwidth][c]{\includegraphics[height=0.38\textheight]{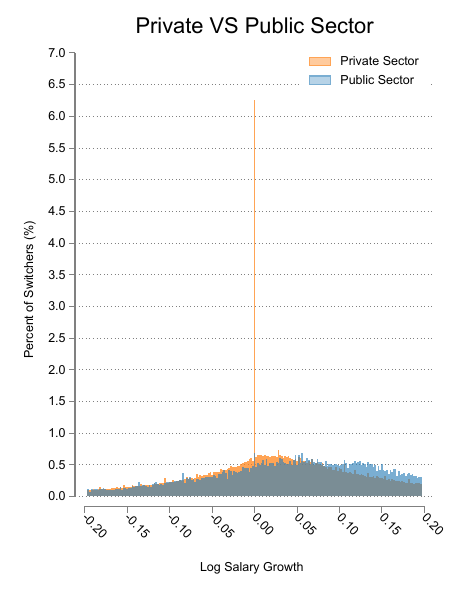}}

    \makebox[\textwidth][c]{\includegraphics[height=0.38\textheight]{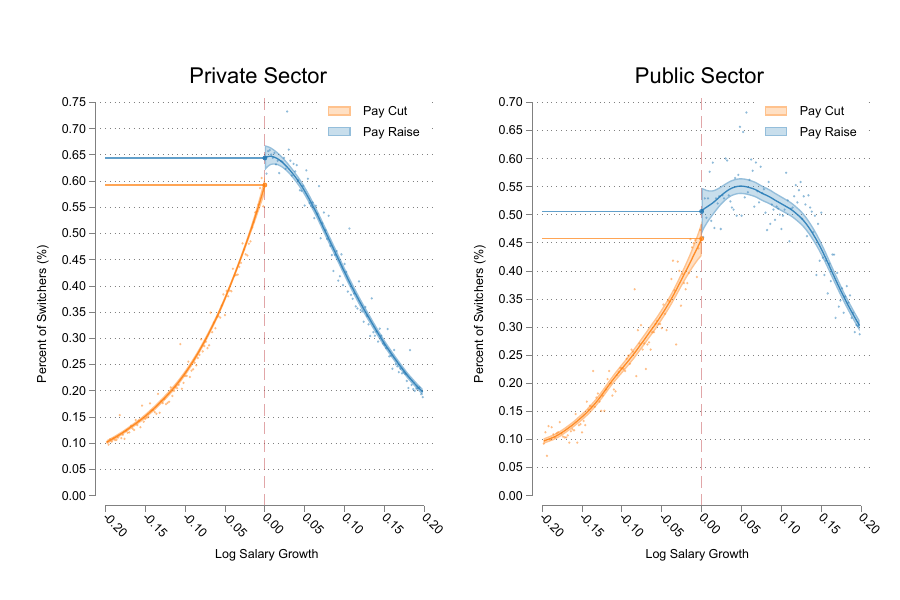}}

    \caption{Comparison of Job Switchers in Private and Public Sectors} \label{fig:govt_private}
    \caption*{\normalsize 
        This is an implementation of \autoref{fig:anomaly_sample} that compares bunching and discontinuity for job switchers in public and private sectors (discussed in \autoref{subsec:hetero_prior_studies}).
    }
\end{figure}
\vspace*{\fill}
\clearpage

\vspace*{\fill}
\begin{figure}[H]
    \makebox[\textwidth][c]{\includegraphics[width=0.90\paperwidth]{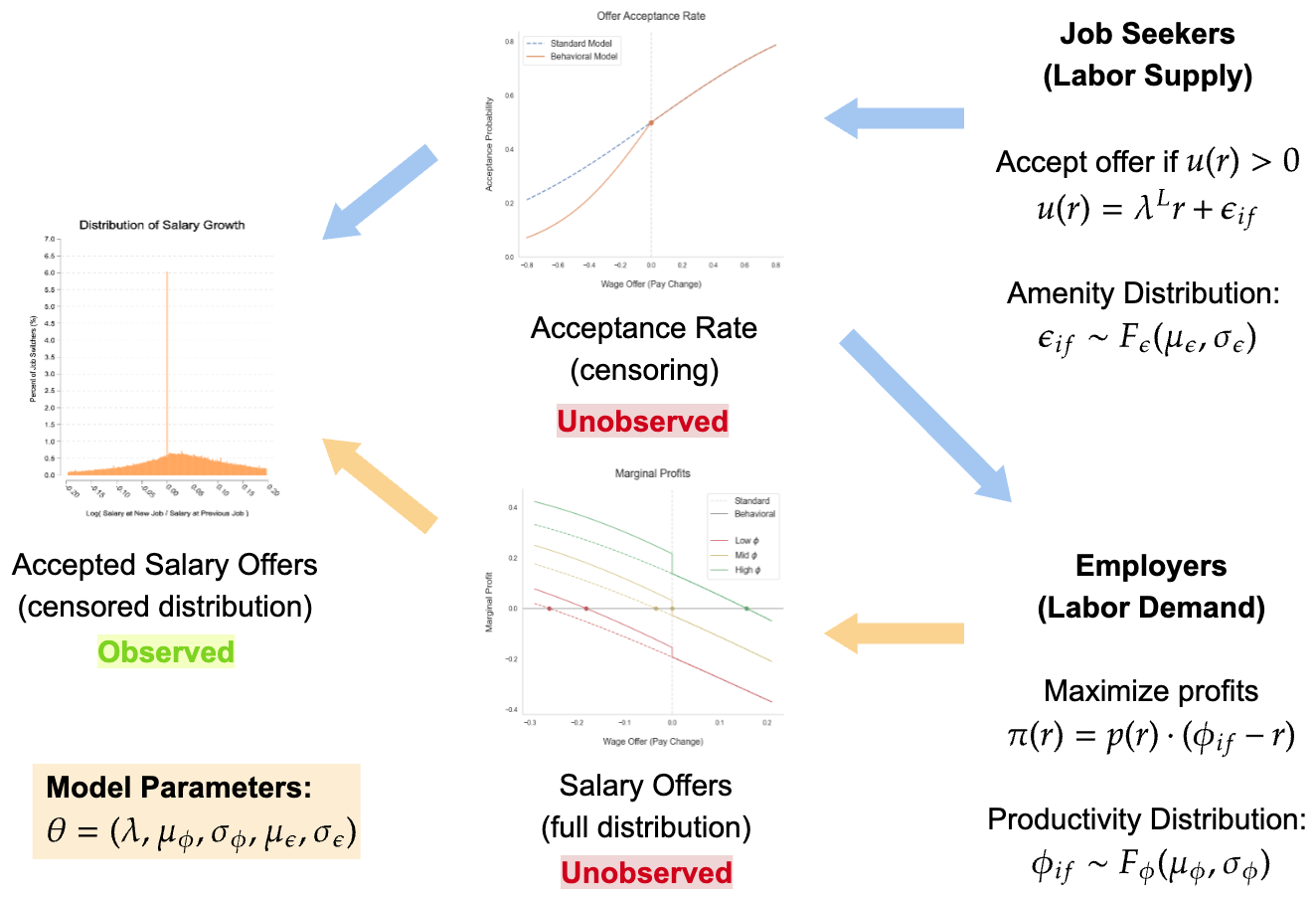}}

    \caption{Data Generating Process for the Distribution of Salary Growth} \label{fig:datagen_process}
    \caption*{\normalsize 
        This figure summarizes the data-generating process for salary growth according to the model (discussed in \autoref{subsec:mindist_procedure}). The model has five parameters: loss aversion ($\lambda$), location and scale for labor productivity $(\mu_\phi, \sigma_\phi)$, and location and scale for non-wage amenities $(\mu_\epsilon, \sigma_\epsilon)$. \hyperref[sec:estimation_appendix]{Estimation Appendix~\ref*{sec:estimation_appendix}} derives predicted proportions in each bin that takes into account censoring due to selective acceptances of salary offers. 
    }
\end{figure}
\vspace*{\fill}
\clearpage

\vspace*{\fill}
\begin{figure}[H]

    \makebox[\textwidth][c]{%
        \includegraphics[width=0.47\paperwidth]{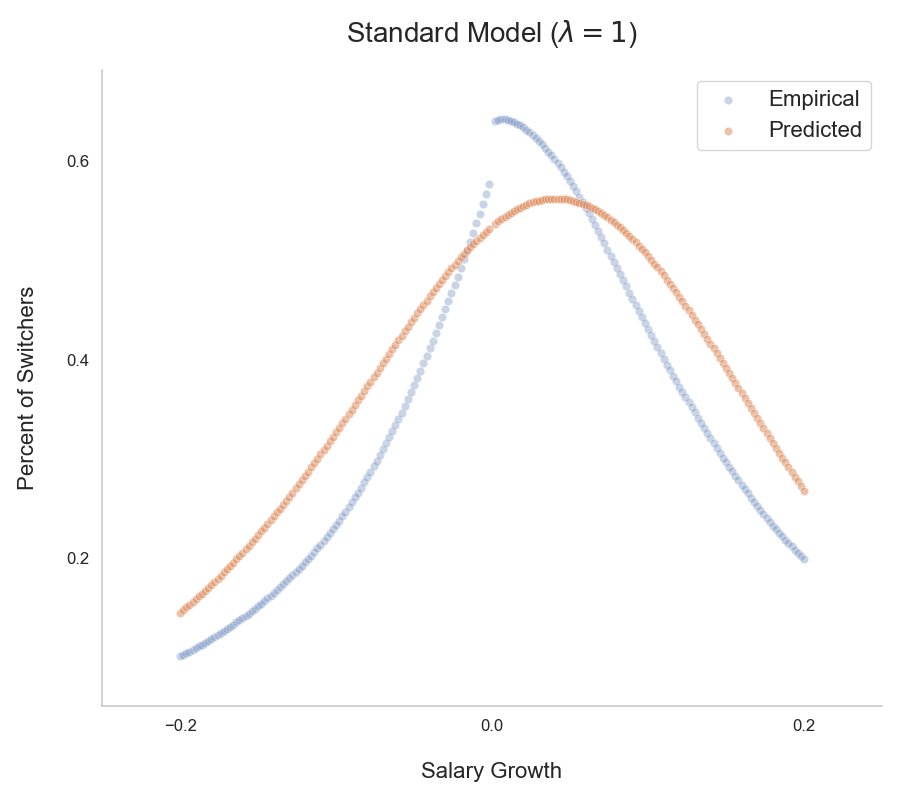}%
        \includegraphics[width=0.47\paperwidth]{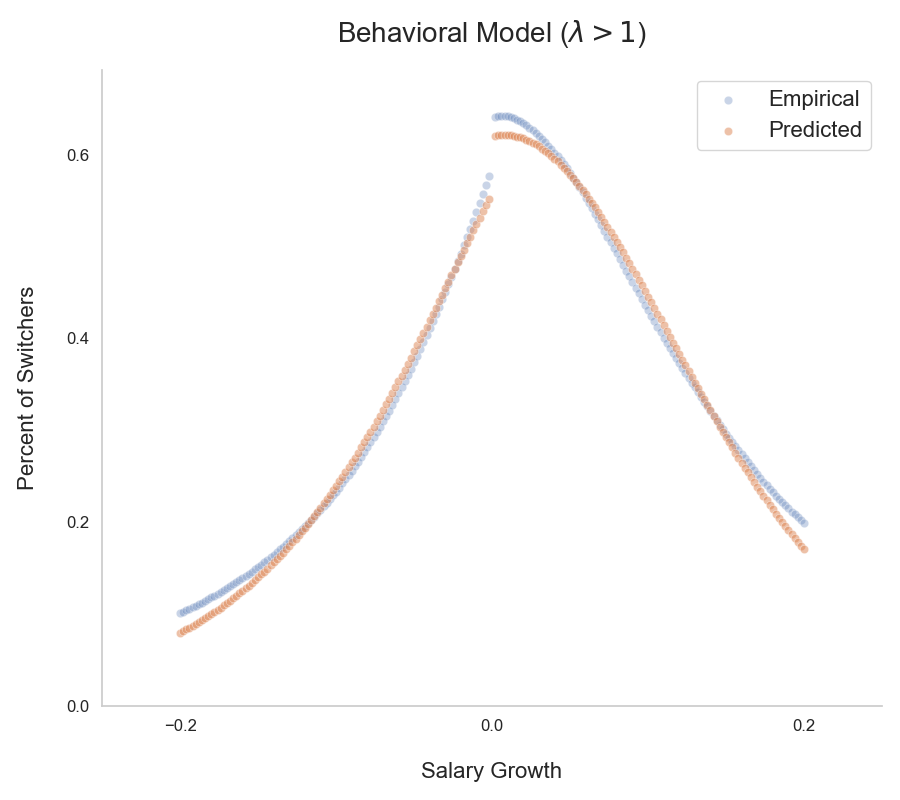}
    }

    \caption{Model Fit to Proportions in Salary Growth Bins} \label{fig:model_fit}
    \caption*{\normalsize 
        This figure compares empirical and predicted proportions in non-zero salary growth bins (discussed in \autoref{subsec:fit_bins}). Corresponding parameters are in \autoref{tab:parameters_main}, which minimize squared distances between predicted proportions (in orange) and empirical proportions (in blue). See \hyperref[sec:estimation_appendix]{Estimation Appendix~\ref*{sec:estimation_appendix}} for a derivation of predicted proportions in each bin. The horizontal axis is log salary growth, and scatter points denote percentages of switchers in each bin (in increments of 0.002 log points). Empirical proportions are kernel density estimates for job switchers without employment gaps in the analysis sample.
    }
\end{figure}
\vspace*{\fill}
\clearpage

\vspace*{\fill}
\begin{figure}[H]

    \makebox[\textwidth][c]{\includegraphics[width=0.95\textwidth]{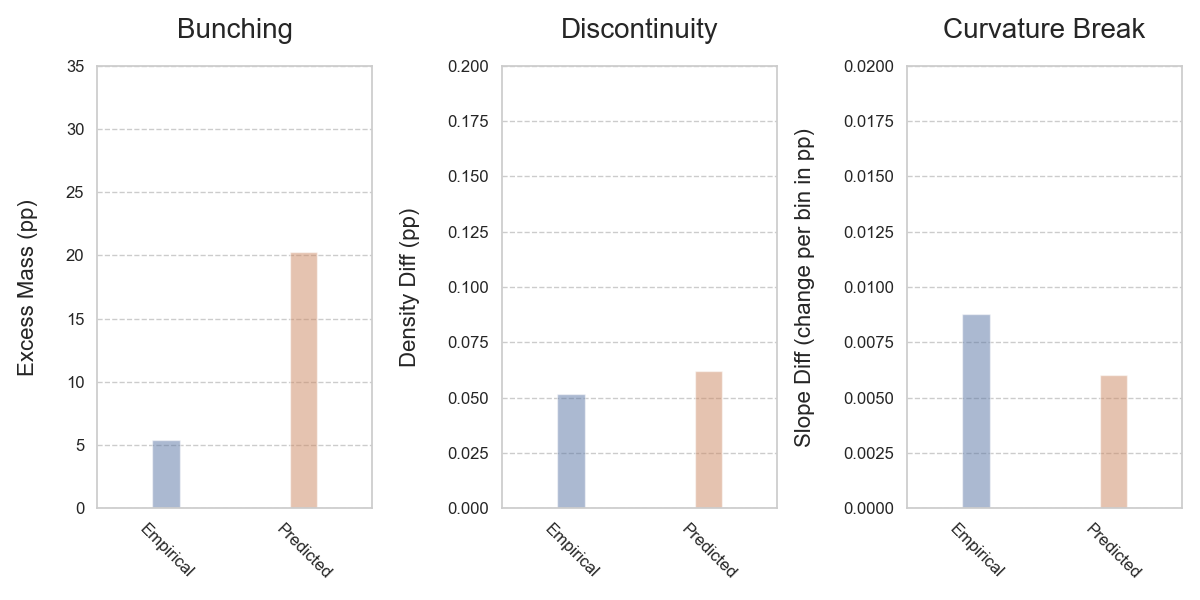}}

    \caption{Empirical and Predicted Magnitudes of Anomalies} \label{fig:anomaly_comparison}
    \caption*{\normalsize 
        This figure compares anomalies predicted by the fitted model with their empirical magnitudes (discussed in \autoref{subsec:fit_anomalies}). Empirical anomalies correspond to \autoref{tab:anomalies}, and predicted anomalies are based on parameters in \autoref{tab:parameters_main}.
    }
\end{figure}
\vspace*{\fill}
\clearpage

\vspace*{\fill}
\begin{figure}[H]

    \makebox[\textwidth][c]{%
        \includegraphics[width=0.47\paperwidth]{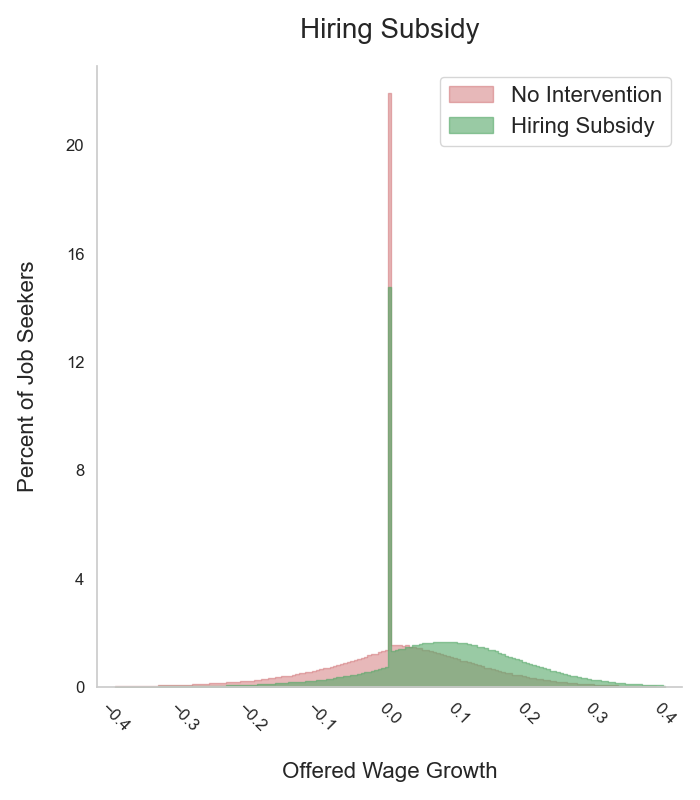}%
        \includegraphics[width=0.47\paperwidth]{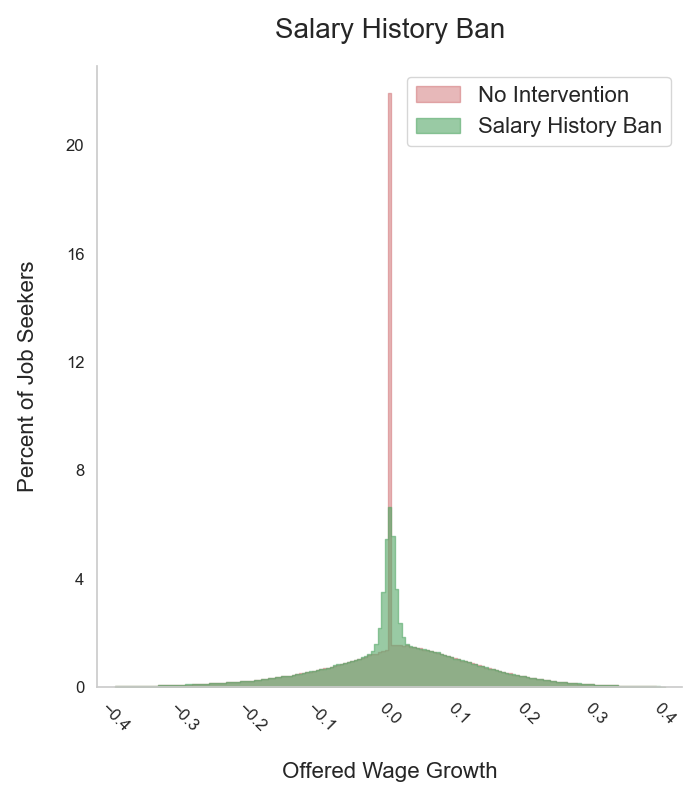}
    }

    \caption{Distribution of Wage Offers Under Policy Interventions} \label{fig:policy_implications}
    \caption*{\normalsize
        This figure uses the fitted model to predict the distribution of salary offers under two policy interventions. The horizontal axis is the wage offer relative to the job seeker's current wage (offered wage growth). The left panel is for a hypothetical hiring subsidy that raises the value of labor to employers $\phi_{if}$, which is modeled in \autoref{subsec:pass_through}. The right panel is for a salary history ban that only allows employers to indirectly observe current salaries, which is modeled in \autoref{subsec:salary_history_bans}. Predicted distributions are based on parameters in \autoref{tab:parameters_main}.
    }
\end{figure}
\vspace*{\fill}
\clearpage

\vspace*{\fill}
\begin{figure}[H]

    \makebox[\textwidth][c]{%
        \includegraphics[width=0.47\paperwidth]{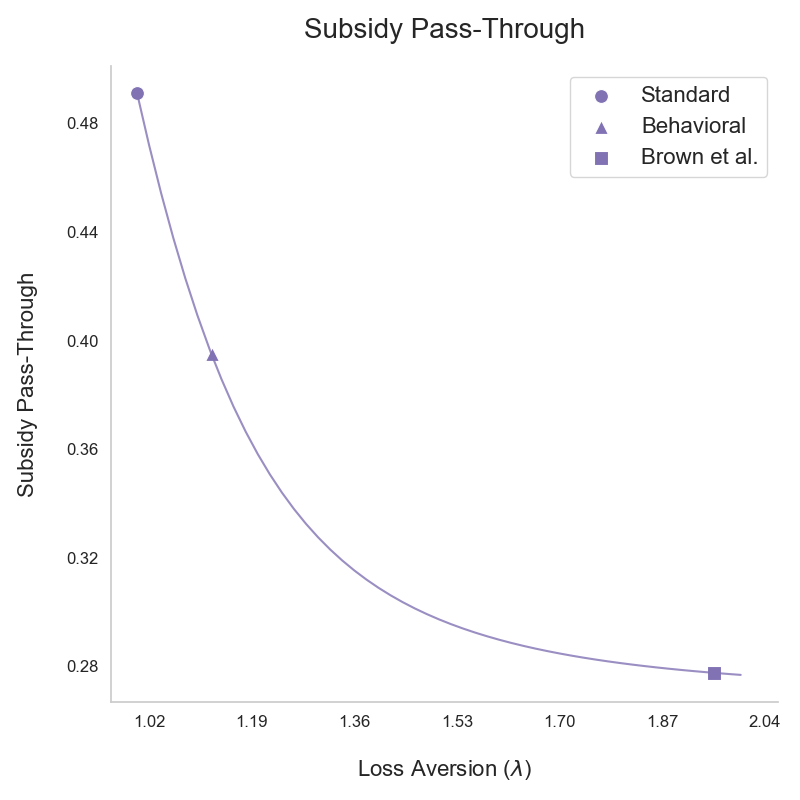}%
        \includegraphics[width=0.47\paperwidth]{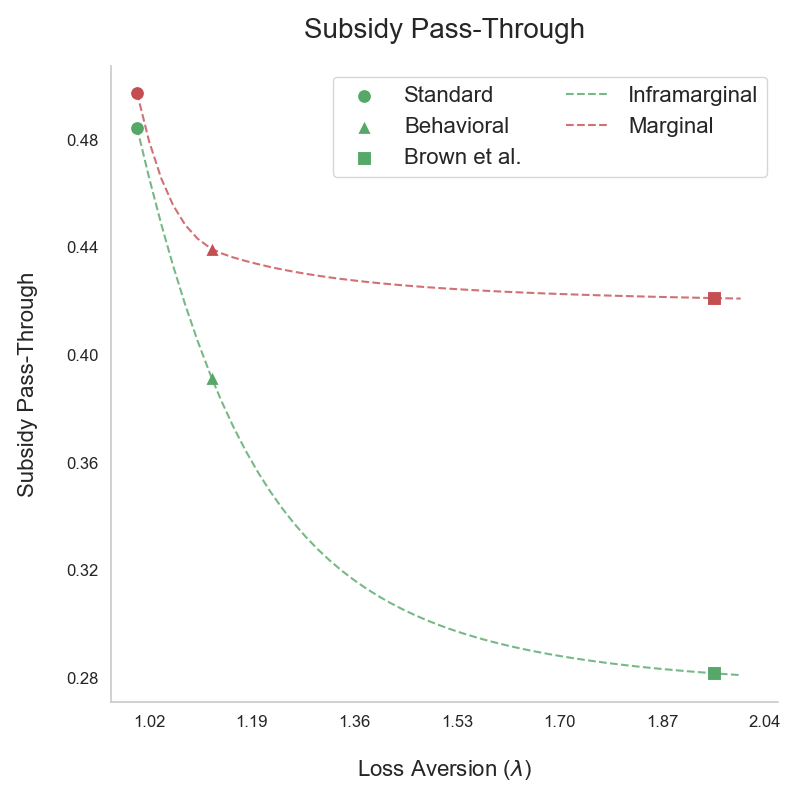}
    }

    \caption{Implications of Loss Aversion on Subsidy Pass-Through} \label{fig:passthru_lossav}
    \caption*{\normalsize 
        This figure uses the fitted model to predict the pass-through of hiring subsidies to salaries received by job seekers (discussed in \autoref{subsec:pass_through}). Pass-through is defined as the increase in salary offer divided by the size of the hiring subsidy. The left panel plots predicted pass-through for each value of loss aversion, holding other parameters fixed at \autoref{tab:parameters_main}. The right panel plots pass-through separately for inframarginal and marginal job seekers. ``Inframarginal'' refers to job seekers who would accept salary offers regardless of the subsidy (offer is already high enough to be accepted without the subsidy). ``Marginal'' refers to job seekers who would only accept their salary offer if augmented with the subsidy (offer is only high enough with the subsidy).
    }
\end{figure}
\vspace*{\fill}
\clearpage

\vspace*{\fill}
\begin{figure}[H]

    \makebox[\textwidth][c]{%
        \includegraphics[width=0.47\paperwidth]{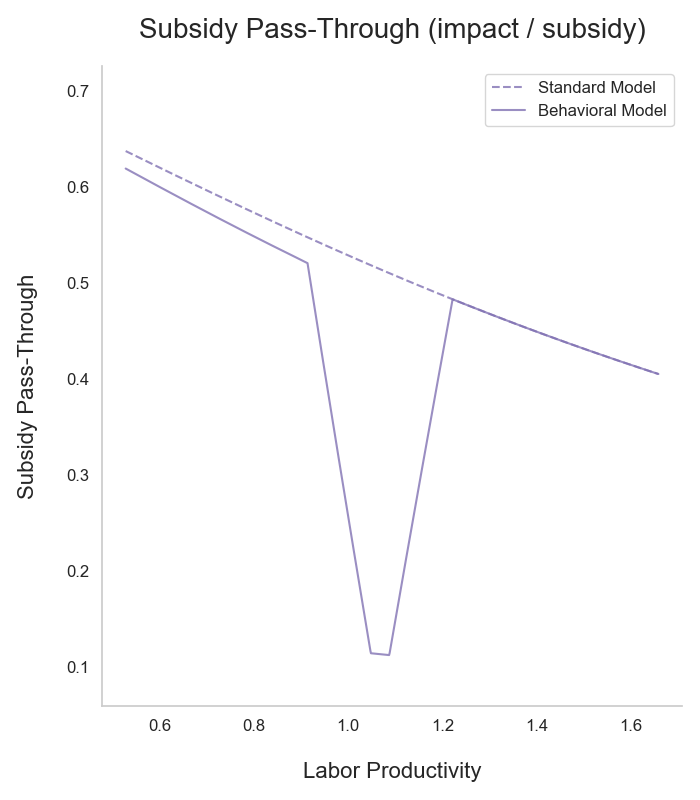}%
        \includegraphics[width=0.47\paperwidth]{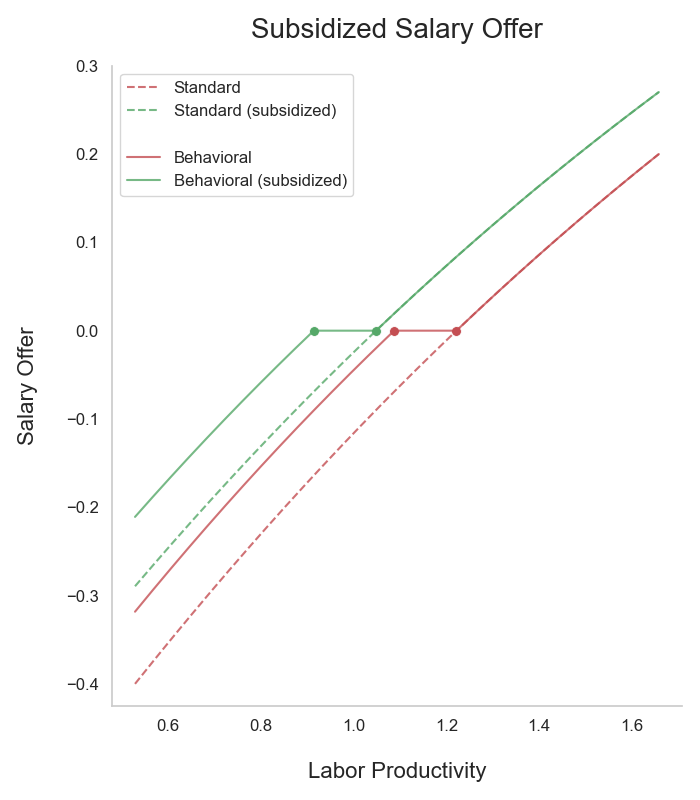}
    }

    \caption{Mechanisms for Lower Pass-Through Under Loss Aversion} \label{fig:passthru_constrained}
    \caption*{\normalsize 
        This figure shows mechanisms behind how loss aversion lowers the pass-through of hiring subsidies to salaries (discussed in \autoref{subsec:pass_through}). The right panel plots salary offers for each value of labor productivity $\phi$, and the left panel plots the implied pass-through as the difference with and without the subsidy, divided by the size of the subsidy. Dashed lines correspond to the standard model $(\lambda=1)$, and solid lines correspond to the behavioral model $(\lambda>1)$
    }
\end{figure}
\vspace*{\fill}
\clearpage

\restoregeometry

%% file: 0_tables.tex
\section*{Main Tables}
\addcontentsline{toc}{subsection}{Main Tables}

\vspace*{\fill}
\begin{table}[H]
    \caption{Analysis Sample Conditions} \label{tab:conditions}
    \begingroup
    \renewcommand{\arraystretch}{2.5}
    \makebox[\textwidth][c]{\input{figtable/table/conditions.tex}}
    \endgroup

    \vspace{24pt}
    \caption*{\normalsize 
        This table shows the proportion of job switchers remaining after each analysis sample condition in \autoref{subsec:measure_salary_growth}. 
        Full-time switchers earn annualized salaries below 100 million KRW and above the full-time minimum wage (non-inclusive). Annualized salaries divide total earnings by the number of months worked during the transitioning year. The full-time minimum wage is calculated as hourly minimum wage $\times$ 209 hours/month $\times$ 12 months, where monthly hours are full-time standards set by the Ministry of Employment and Labor. The last condition refers to switchers who work for the entire month at hire and separation.
    }
\end{table}
\vspace*{\fill}
\clearpage

\begin{table}[H]
    \caption{Summary Statistics} \label{tab:sumstats}
    \begingroup
    \renewcommand{\arraystretch}{1.4}
    \makebox[\textwidth][c]{\input{figtable/table/sumstats.tex}}
    \endgroup

    \vspace{24pt}
    \caption*{\normalsize 
        This table of summary statistics compares the analysis sample with full-time job switchers earning above the minimum wage (discussed in \autoref{subsec:summary_stats}). These switchers correspond to the second and last rows in \autoref{tab:conditions}.
        All averages are at the person level. Prior salary is annualized salary at the previous job, and pay change divides the new salary by the previous salary (not logged). Employment gap is the number of months between separation from the previous job and hire at the new job. Employment duration is the length of employment at the previous job prior to separation. Spousal employment status is conditional on being married. Education is observed for married switchers in 1983-95 birth cohorts.
    }
\end{table}
\clearpage

\newgeometry{top=0.1in, bottom=0.8in, left=0.8in, right=0.8in}

\vspace*{\fill}
\begin{table}[H]
    \caption{Estimated Anomalies and Bootstrapped Standard Errors} \label{tab:anomalies}
    \begingroup
    \renewcommand{\arraystretch}{1.5}
    \makebox[\textwidth][c]{\input{figtable/table/anomalies.tex}}
    \endgroup

    \vspace{24pt}
    \caption*{\normalsize 
        This table reports estimates for anomalies observed in \autoref{fig:anomaly_sample} and discussed in \autoref{subsec:anomalies}.
        Estimates for bin proportions are in percentages, and estimates for anomalies are in percentage point differences. Standard errors are bootstrapped with 10,000 iterations. These measurements correspond to switchers without employment gaps in the analysis sample. The zero bin corresponds to log salary growth values between -0.001 and +0.001, with analogous definitions for the -0.002 bin and +0.002 bin. $\hat{p}$ are raw proportions in the data, while $\hat{a}$ (pay raises) and $\hat{b}$ (pay cuts) are kernel-density estimates based on proportions smoothed with local-linear polynomials that account for boundary bias.  Local-linear polynomials are weighted with the epanechnikov kernel with bandwidth 0.020, which is separately fit for pay cuts and pay raises. \\ 
        \, \\ 
        \noindent Bunching is excess mass at zero salary growth, which is measured as the proportion in the zero bin $(\hat{p}_0)$ beyond expected levels from a smooth density $\hat{a}_0$. Discontinuity is the jump in density at zero, measured as the difference in density immediately above zero $(\hat{a}_0)$ and below zero $(\hat{b}_0)$. Curvature break measures the change in the slope of the density at zero, measured as the difference in density changes immediately below zero $(\hat{b}_0 - \hat{b}_2)$ and above zero $(\hat{a}_2 - \hat{a}_0)$. 
    }
\end{table}
\vspace*{\fill}
\clearpage

\vspace*{\fill}
\begin{table}[H]
    \caption{Calibrations for Model Parameters} \label{tab:calibration_overview}
    \begingroup
    \renewcommand{\arraystretch}{2}
    \makebox[\textwidth][c]{\input{figtable/table/calibrations.tex}}
    \endgroup

    \vspace{24pt}
    \caption*{\normalsize 
        This table summarizes how values are set for each parameter in the model (discussed in \autoref{subsec:fixed_calibrations}). $\lambda \geq 1$ is the degree of loss aversion, $(\mu_\phi,\sigma_\phi)$ are location and scale parameters for the distribution of labor productivity, and $(\mu_\epsilon,\sigma_\epsilon)$ are location and scale parameters for the distribution of non-wage amenities. Unobserved heterogeneity is parameterized with the logistic distribution.
    }
\end{table}
\vspace*{\fill}
\clearpage

\vspace*{\fill}
\begin{table}[H]
    \caption{Parameter Estimates and Standard Errors} \label{tab:parameters_main}
    \begingroup
    \renewcommand{\arraystretch}{2.5}
    \makebox[\textwidth][c]{\input{figtable/search/table/parameters_main.tex}}
    \endgroup

    \vspace{24pt}
    \caption*{\normalsize 
        Parameters and standard errors correspond to \autoref{fig:model_fit}, which are discussed in \autoref{subsec:calibrated_lossav}. $\lambda \geq 1$ is the degree of loss aversion, while $(\mu_\phi,\sigma_\phi)$ are location and scale parameters for labor productivity (logistic distribution). Parameter estimates and standard errors are obtained through minimum distance on salary growth bins, which is described in \autoref{subsec:mindist_procedure}. Salary growth bins are equally weighted proportions in non-zero bins between -0.2 and +0.2 log points in increments of 0.002. Standard errors (in parentheses) are based on the sandwich formula with a bootstrapped covariance matrix for empirical proportions. Location and scale parameters for non-wage amenities $(\mu_\epsilon, \sigma_\epsilon)$ are calibrated at fixed values in \autoref{tab:calibration_overview}. The Quasi-Likelihood Ratio (QLR) test compares the minimum distance criterion for two nested models and rejects the null of equal performance when the $\chi^2$ statistic exceeds the critical value for significance level $\alpha=0.05$ (in parentheses).
    }
\end{table}
\vspace*{\fill}
\clearpage

\restoregeometry

%% file: figtable/table/conditions.tex
\centering
\begin{tabular}{lcc}
\toprule[2pt]
\textbf{Sample Condition} & \textbf{N} & \textbf{\%} \\ \hline
All Job Switchers & 13,582,267 & 100\% \\ 
+ Full-Time Switchers & 7,260,601 & 53.5\% \\ 
........ above full-time min wage & 7,462,836 & 54.9\% \\ 
........ below 100m KRW & 13,363,665 & 98.4\% \\ 
+ Full Months at Hire/Separation & 2,089,484 & 15.4\% \\
\bottomrule[2pt]
\end{tabular}

%% file: figtable/table/sumstats.tex
\centering
\begin{tabular}{l|cc|cc|cc}
\toprule[2pt]
\multicolumn{1}{c}{} & \multicolumn{2}{c}{\textbf{\makecell[c]{Full-Time\\(Above Min Wage)}}} & \multicolumn{2}{c}{\textbf{Analysis Sample}} & \multicolumn{2}{c}{} \\
\cmidrule(r){2-3} \cmidrule(r){4-5}
\multicolumn{1}{c}{} & Mean & \multicolumn{1}{c}{Std Dev} & Mean & \multicolumn{1}{c}{Std Dev} & Diff & Std Err \\
\midrule
Prior Salary (10k KRW)          & 3077.9 & 1387.0 & 3200.7 & 1462.1 & -122.80   & (1.13) \\
Avg Pay Change (new / old)      & 1.069 & 0.365 & 1.063 & 0.320     & 0.0065    & (0.0003) \\
Pay Cut (binary)                & 0.439 & 0.496 & 0.394 & 0.489     & 0.0457    & (0.0004) \\
Pay Raise (binary)              & 0.541 & 0.498 & 0.560 & 0.496     & -0.0194   & (0.0004) \\
Pay Match (binary)              & 0.020 & 0.139 & 0.046 & 0.209     & -0.0262   & (0.0002) \\
Employment Gap (months)         & 5.06  & 8.39  & 4.35  & 7.89      & 0.705     & (0.006) \\
Employment Duration (months)    & 27.2  & 37.5  & 35.0  & 44.4      & -7.86     & (0.03) \\
Age (years)                     & 38.8  & 12.0  & 41.9  & 12.6      & -3.08     & (0.01) \\
Female (binary)                 & 0.372 & 0.483 & 0.400 & 0.490     & -0.0275   & (0.0004) \\
Married (binary)                & 0.524 & 0.499 & 0.614 & 0.487     & -0.0899   & (0.0004) \\
Parent (binary)                 & 0.519 & 0.500 & 0.626 & 0.484     & -0.1072   & (0.0005) \\
Spousal Employment | Married (binary) & 0.760 & 0.427 & 0.752 & 0.432     & 0.0072    & (0.0005) \\
HS Grad (binary)                & 0.233 & 0.423 & 0.238 & 0.426     & -0.0054   & (0.0013) \\
College Grad (binary)           & 0.756 & 0.429 & 0.751 & 0.432     & 0.0051    & (0.0013) \\
Area: Seoul (binary)            & 0.215 & 0.411 & 0.198 & 0.399     & 0.0172    & (0.0003) \\
Area: Seoul - Adjacent (binary) & 0.351 & 0.477 & 0.340 & 0.474     & 0.0110    & (0.0004) \\
Area: Other Metro (binary)      & 0.184 & 0.388 & 0.200 & 0.400     & -0.0156   & (0.0003) \\
Area: Rural (binary)            & 0.249 & 0.432 & 0.261 & 0.439     & -0.0126   & (0.0004) \\
Government (binary)             & 0.034 & 0.181 & 0.041 & 0.198     & -0.0069   & (0.0002) \\
Registered Corporation (binary) & 0.741 & 0.438 & 0.736 & 0.441     & 0.0051    & (0.0004) \\
Employment Size (persons)       & 1062.4 & 7505.2 & 1103.0 & 8348.8 & -40.5 & (6.5) \\ \hline
Number of Switchers & \multicolumn{2}{c|}{N = 7,260,601} & \multicolumn{2}{c|}{N = 2,089,484} & & \\
\bottomrule[2pt]
\end{tabular}

%% file: figtable/table/anomalies.tex
\centering
\begin{tabular}{lcc}
\toprule[2pt]
\textbf{} & \textbf{Estimate} & \textbf{Std Err} \\
\midrule
Bunching: ($\hat{p}_0 - \hat{a}_0$) & 5.3997 & (0.0209) \\
Discontinuity: ($\hat{a}_0 - \hat{b}_0$) & 0.0515 & (0.0044) \\
Curvature Break: ($\hat{b}_0 - \hat{b}_2$) - ($\hat{a}_2 - \hat{a}_0$) & 0.0088 & (0.0006) \\
\midrule
$\hat{b}_2$: Percent in -0.002 Bin & 0.5767 & (0.0027) \\
$\hat{b}_0$: Percent in Zero Bin (kernel estimate: pay cuts)  & 0.5871 & (0.0030) \\
$\hat{p}_0$: Percent in Zero Bin & 6.0382 & (0.0205) \\
$\hat{a}_0$: Percent in Zero Bin (kernel estimate: pay raises) & 0.6386 & (0.0032) \\
$\hat{a}_2$: Percent in +0.002 Bin & 0.6402 & (0.0029) \\
\bottomrule[2pt]
\end{tabular}

%% file: figtable/table/calibrations.tex
\centering
\begin{tabular}{lc}
    \toprule[2pt]
    \multicolumn{1}{c}{\textbf{Parameter}} & \multicolumn{1}{c}{\textbf{Calibration}} \\
    \midrule
     $\lambda$: Loss Aversion & \multirow{3}{*}{\makecell[c]{Estimated with minimum\\ \, \\distance on salary growth bins}} \\
    $\mu_\phi$: Productivity Location &  \\
    $\sigma_\phi$: Productivity Scale & \\
    \midrule
    $\mu_\epsilon$: Amenity Location & \makecell[c]{\, \\ Assume equal amenities on average \\ \, \\ at previous and new jobs $(\mu_\epsilon=0)$ \\ \,} \\
    \midrule
    $\sigma_\epsilon$: Amenity Scale & \makecell[c]{\, \\ Match the variance of non-wage\\ \, \\amenities in \colorcite{lehmann_non-wage_2025} $(\sigma_\epsilon=0.611)$ \\ \,} \\
    \bottomrule[2pt]
\end{tabular}

%% file: figtable/search/table/parameters_main.tex
\begin{tabular}{lcc}
\toprule[2pt]
\textbf{Parameter} & \textbf{Standard} & \textbf{Behavioral} \\
\midrule
$\lambda$: Loss Aversion & $\cdot$ & \makecell[c]{1.1235 \\ (0.0004)} \\
$\mu_\phi$: Productivity Location & \makecell[c]{1.2503 \\ (0.0003)} & \makecell[c]{1.1953 \\ (0.0003)} \\
$\sigma_\phi$: Productivity Scale & \makecell[c]{0.1855 \\ (0.0002)} & \makecell[c]{0.1594 \\ (0.0003)} \\
QLR Test & \multicolumn{2}{c}{\makecell[c]{274.35 \\(CV: 3.84)}}  \\
\bottomrule[2pt]
\end{tabular}

%% file: a1_model.tex
\section{Model Appendix} \label{sec:model_appendix}

This model appendix provides derivations for 1) the wedge of labor productivity for offer matching at current salaries, and 2) employment in steady-state equilibrium for the dynamic search model introduced in \autoref{subsec:dynamic_search}.

\subsection{Productivity Wedge for Salary Matching} \label{subsec:matching_wedge}

	Let $f_\epsilon(\cdot)$ and $F_\epsilon(\cdot)$ denote the density and cumulative distribution functions for non-wage amenities, respectively. $f_\phi(\cdot)$ and $F_\phi(\cdot)$ are corresponding functions for labor productivity. $p(r)$ is the offer acceptance rate, with $p_L(r)= 1 - F_{\epsilon}(-\lambda r)$ denoting acceptance rates for pay cuts and $p_G(r)=1 - F_{\epsilon}(-r)$ denoting acceptance rates for pay raises.

	\,

	\noindent Optimality conditions for wage offers in \autoref{eq:optimal_offer} allow me to infer labor productivity from accepted wage offers. Let $\phi_L(r)$ and $\phi_G(r)$ denote labor productivity implied by pay cuts and pay raises, given by
	\begin{alignat*}{5}
		& \phi_L(r) &&= \; r \; &&+ \; \frac{1}{\lambda} \cdot &&\frac{1-F_\epsilon(-\lambda r)}{f_\epsilon(-\lambda r)} \; && \text{ for } r < 0 \; \text{(pay cut)} \\ 
		& && && && && \\ 
		& \phi_G(r) &&= \; r \; &&+ &&\frac{1-F_\epsilon(-r)}{f_\epsilon(-r)} \; && \text{ for } r > 0 \; \text{(pay raise)}
	\end{alignat*}

	\noindent Both expressions are monotonically increasing in wage offer $r$ as long as $F_\epsilon(\cdot)$ is well-behaved with an increasing inverse mills ratio (e.g. normal, logistic). Implied productivity $\phi$ is always $\phi_G(r)$ in a standard model, but $\lambda > 1$ under loss aversion implies $\phi_L(r) < \phi_G(r)$ for any fixed pay cut $r < 0$ (i.e. smaller pay cuts given the same level of productivity).

	\,

	\noindent Corner solutions at $r = 0$ for a range of $\phi$ imply the following productivity wedge for salary matching:
	\begin{align*}
		\phi \in \big[ \, \phi_L(0), \; \phi_G(0) \, \big] \; \text{ for } r = 0 \; \text{(salary matching)}
	\end{align*}

	\clearpage

\subsection{Steady-State Employment in Dynamic Search} \label{subsec:steady_state}

	This sub-appendix solves for employment in steady-state equilibrium for the dynamic search model introduced in \autoref{subsec:dynamic_search}. In each period, firms decide how many vacancies to create given convex costs and uncertainty in current salaries they will encounter among job seekers.\footnote{	
		Convex vacancy costs deviate from the standard setup in \textcite{burdett_wage_1998}. Convex costs pin down a unique optimum for vacancies, which prevents firms from hiring infinitely many workers that generate infinitesimally small profits.
	} Once vacacies are randomly matched to job seekers, firms observe their current salaries and set wage offers to maximize expected profits given uncertainty in their valuation of non-wage amenities. The vacancy is filled if the job seeker accepts the wage offer, and it is otherwise closed until firms reopen vacancies in the next period.

	\subsubsection*{Expected Profits from Each Vacancy}

		\noindent In steady-state equilibrium, vacancies and expected profits remain stable across time. Firms maximize expected profits for each vacancy by tailoring wage offers to the job seeker's current salary and labor productivity. For a given match between firm $f$ and job seeker $i$, expected profits are given by
		{
			\setlength{\abovedisplayskip}{5pt}
			\setlength{\belowdisplayskip}{5pt}
			\begin{equation*}
				\pi(w \, | \, \Psi_{f}, \tilde{w}_i) = p(w - \tilde{w}_i) \cdot (\Psi_{f} - w)
			\end{equation*} 
		}
		where 
		\begin{itemize}
			\setlength{\itemsep}{0pt}
			\item[] $\Psi_{f}$ is the firm's labor productivity,
			\item[] $\tilde{w}$ is the current wage ($w_{0i}$ if employed, $w^U$ if unemployed), and
			\item[] $p(w - \tilde{w})$ is the offer acceptance rate.
		\end{itemize}

		\noindent Since the acceptance rate is kinked at $w = \tilde{w}$ due to loss aversion, optimal wage offers implied by marginal profits can differ across equally productive workers based on their current wage. For a firm with productivity $\Psi_{f}$, expected profits generated by a vacancy are averaged over the distribution of current wages.
		{
			\setlength{\abovedisplayskip}{5pt}
			\setlength{\belowdisplayskip}{5pt}
			\begin{equation*}
				\bar{\pi}(\Psi_f) 
				= \int_{\tilde{w}} p\Big(w^*(\Psi_f, \tilde{w}_i) - \tilde{w}_i \Big) 
				\cdot \Big(\Psi_{f} - w^*(\Psi_f, \tilde{w}_i)\Big) \, dG(\tilde{w})
			\end{equation*}
		}
		where
		\begin{itemize}
			\setlength{\itemsep}{0pt}
			\item[] $G(\tilde{w})$ is the distribution of current wages among job seekers, and
			\item[] $w^*(\Psi_f, \tilde{w}_i)$ is the optimal wage offer with current wage $\tilde{w}_i$.
		\end{itemize}

	\subsubsection*{Optimal Vacancy Creation}

		\noindent In each period, firms decide how many vacancies to create given convex costs, which I assume to be quadratic for the sake of discussion. Firms set the number of vacancies $(J)$ to maximize total profits net of vacancy costs, which is given by
		\begin{equation*}
			J \cdot \bar{\pi}(\Psi_f) - c \cdot J^2
		\end{equation*}
		\noindent where $c$ is the quadratic cost coefficient for vacancy creation. The number of vacancies that maximize total profits are given by
		\begin{equation*}
			J^*(\Psi_f) = \frac{\bar{\pi}(\Psi_f)}{2c}
		\end{equation*}

		\noindent Loss aversion $(\lambda)$ lowers expected profits for vacancy creation $\bar{\pi}(\Psi_f)$ by lowering acceptance rates and raising wage costs when offering pay cuts. As a result, firms open fewer vacancies in steady-state equilibrium but face higher acceptance rates since pay cuts are smaller and less frequent under loss aversion. Fewer vacancies lower steady-state employment, while higher acceptance rates raise employment. 

		\,

		\noindent A key limitation of this dynamic setup is that offer arrival rates $\alpha_E, \alpha_U$ do not depend on the number of vacancies, which can be formalized through matching models (see \cite{pissarides_equilibrium_2000} for an example). I did not pursue this path because this level of depth is beyond the scope of my paper, but it is nonetheless a promising direction that I hope to pursue for future research.

%% file: a2_estimation.tex
\section{Estimation Appendix} \label{sec:estimation_appendix}

This estimation appendix lays out expressions for the predicted proportion of job switchers in each salary growth bin according to the model. Predicted proportions are compared against their empirical counterparts by minimizing their squared distance through minimum-distance estimation. 

\,

\subsubsection*{Total Density of Accepted Wage Offers}

    \noindent \hyperref[subsec:matching_wedge]{Model Appendix~\ref*{subsec:matching_wedge}} solves for productivity values implied by pay cuts $\phi_L(r)$, pay raises $\phi_G(r)$, and salary matching $\phi \in [\phi_L(0), \phi_G(0)]$. Implied productivity values $\phi_L(r)$ and $\phi_G(r)$ are monotonically increasing in $r$, so the uncensored density of wage offers $f_r(r)$ can be expressed in terms of productivity density $f_\phi(\cdot)$ evaluated at implied productivity values:
    \begin{alignat*}{3}
        & f_r(r) &&= f_\phi \big(\phi_L(r)\big) \quad && \text{for} \quad r < 0 \; \text{(pay cut density)} \\
        & f_r(r) &&= f_\phi \big(\phi_G(r)\big) \quad && \text{for} \quad r > 0 \; \text{(pay raise density)}
    \end{alignat*}

    \noindent This implies the following integral for the total density of \underline{accepted} wage offers (given parameters $\theta$):

    \begin{align*}
    A(\theta) = 
    {\color{black}  \underbrace{
        \int_{r < 0} 
        f_\phi \big(\phi_L(r)\big) 
        \cdot \phi_L'(r)
        \cdot p_{L}(r) \, dr}_{\text{Pay Cuts}}
    }
    {\color{black} \; + \; \underbrace{
        \int_{r > 0} 
        f_\phi \big(\phi_G(r)\big) 
        \cdot \phi_G'(r)
        \cdot p_{G}(r) \, dr}_{\text{Pay Raises}}
    }
    {\color{black} \; + \; \underbrace{
        p(0) \cdot 
        \int_{\phi_{L}(0)}^{\phi_{G}(0)} 
        f_{\phi}(\phi) d \phi
    }_{\text{Salary Matches}}
    }
    \end{align*}

    \,

    \noindent This integral is not analytically tractable, but it can be numerically approximated by discretizing the salary growth distribution into small bins. This scaling factor $A(\theta)$ is used to infer predicted proportions in the conditional distribution of accepted offers.

    \,

    \clearpage

\subsubsection*{Predicted Proportions in Salary Growth Bins}

    \noindent The proportion of job switchers in a salary growth bin is the probability that a wage offer $r$ falls within a specified range $[\underaccent{\bar}{r}, \bar{r}]$. This can be expressed as the probability that productivity falls within its implied range $[\phi(\underaccent{\bar}{r}), \phi(\bar{r})]$. For small enough bins, predicted proportions in the censored distribution of \underline{accepted} wage offers can be approximated by
    \begin{align*}
        P \big( r \in [ \, \underaccent{\bar}{r}, \; \bar{r} \, ] \, \big) = 
        \begin{cases}
            p_L(r) \cdot \Big[ 
                F_\phi \big( \phi_L(\bar{r}) \big) - 
                F_\phi \big( \phi_L(\underaccent{\bar}{r}) \big) 
            \Big] / A(\theta) & \text{if } r < 0 \quad \text{(prop. for pay cuts)} \\
            p_G(r) \cdot \Big[ 
                F_\phi \big( \phi_G(\bar{r}) \big) - 
                F_\phi \big( \phi_G(\underaccent{\bar}{r}) \big) 
            \Big] / A(\theta) & \text{if } r > 0 \quad \text{(prop. for pay raises)} \\
            p(0) \; \; \cdot \Big[ 
                F_\phi \big( \phi_G(0) \big) - 
                F_\phi \big( \phi_L(0) \big) 
            \Big] / A(\theta) & \text{if } r = 0 \quad \text{(prop. for salary matches)} \\
        \end{cases}
    \end{align*}

    \noindent Note that wage offers are relative to current salaries, so that accepted wage offers are realized salary growth for job switchers. I use the above expression to numerically approximate the predicted proportion of job switchers in each salary growth bin given parameters $\theta=(\lambda, \mu_\phi, \sigma_\phi, \mu_\epsilon, \sigma_\epsilon)$.

%% file: 0_figures_apx.tex
\section*{Appendix Figures}
\addcontentsline{toc}{subsection}{Appendix Figures}

\vspace*{\fill}
\begin{figure}[ht]
    \makebox[\textwidth][c]{%
        \includegraphics[width=0.47\paperwidth]{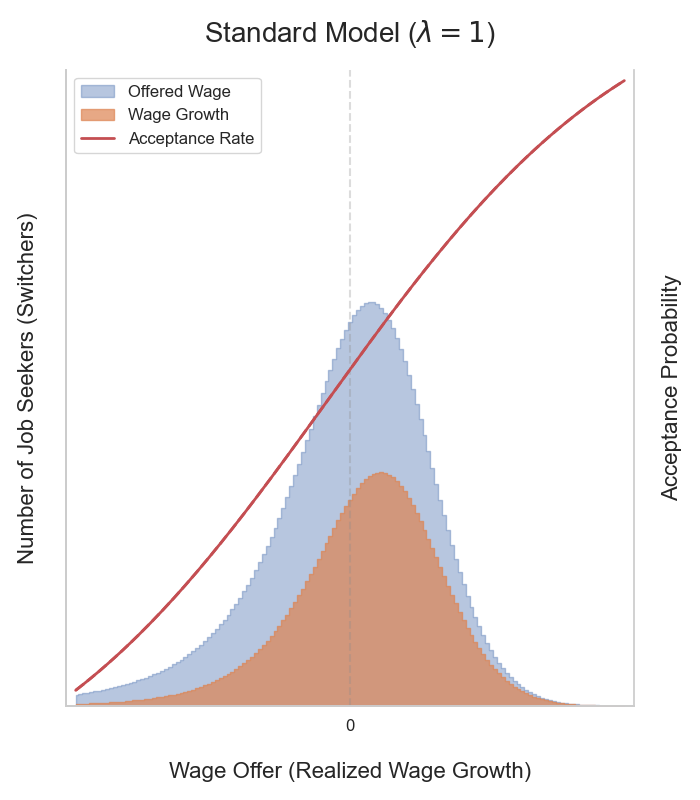}%
        \includegraphics[width=0.47\paperwidth]{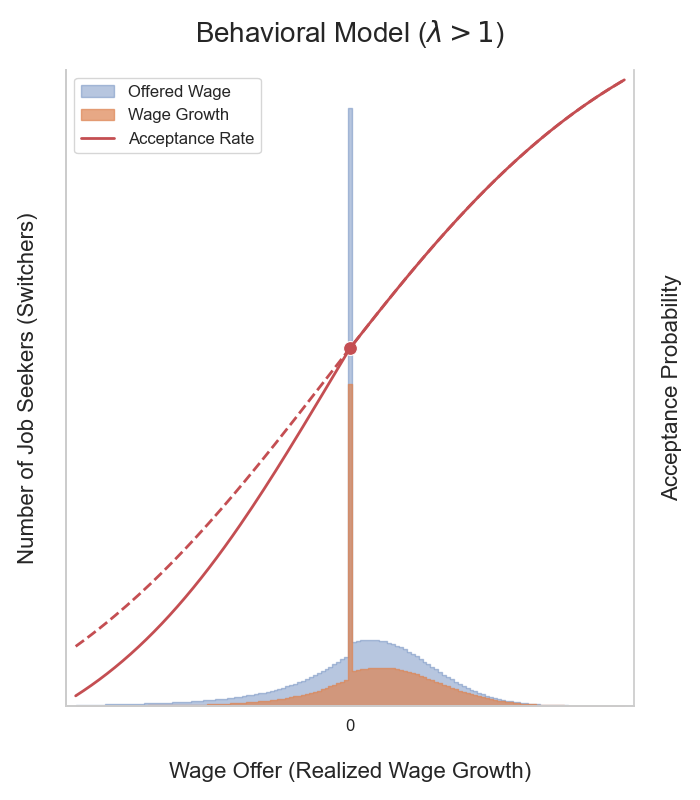}
    }

    \caption{Distribution of Wage Offers and Salary Growth} \label{fig:offer_dist}
    \caption*{\normalsize 
        This figure compares the distribution of wage offers and realized wage growth given unobserved heterogeneity in labor productivity and non-wage amenities, which is discussed in \autoref{subsec:salary_match_reduction}. A subset of wage offers (in blue) are accepted by job seekers and become realized wage growth for job seekers (in orange). The horizontal axis is offered wage relative to the current wage $(r)$, defined as the difference in (logged) offered wage $w$ and the job seeker's current wage $w_0$. Salary match $(r=0)$ means that the offered wage is identical to the job seeker's current wage. The left vertical axis denotes the number of job seekers in each wage offer bin, and the right vertical axis denotes the number of job switchers in each realized wage growth bin. The red line is the acceptance rate for wage offers, with the corresponding right vertical axis denoting the acceptance probability given uncertainty in how job seekers value non-wage amenities $\epsilon_{if}$. 
    }
\end{figure}
\vspace*{\fill}
\clearpage

\newgeometry{top=0.1in, bottom=0.8in, left=0.8in, right=0.8in}

\vspace*{\fill}
\begin{figure}[ht]

    \makebox[\textwidth][c]{%
        \includegraphics[width=0.47\paperwidth]{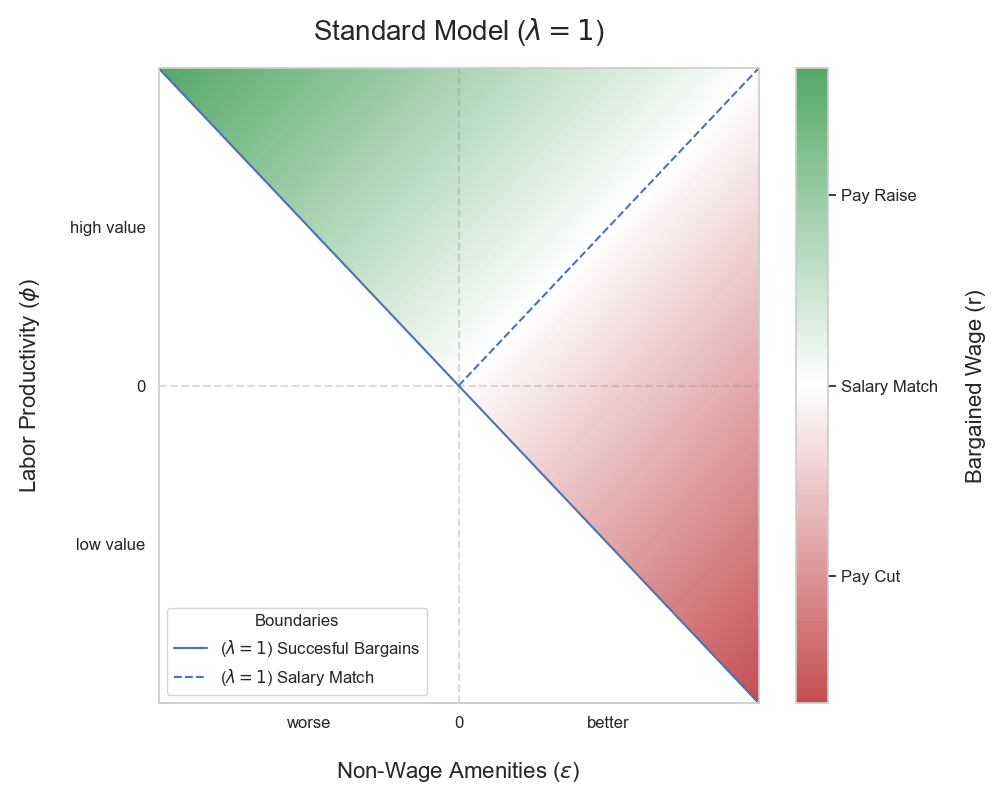}%
        \includegraphics[width=0.47\paperwidth]{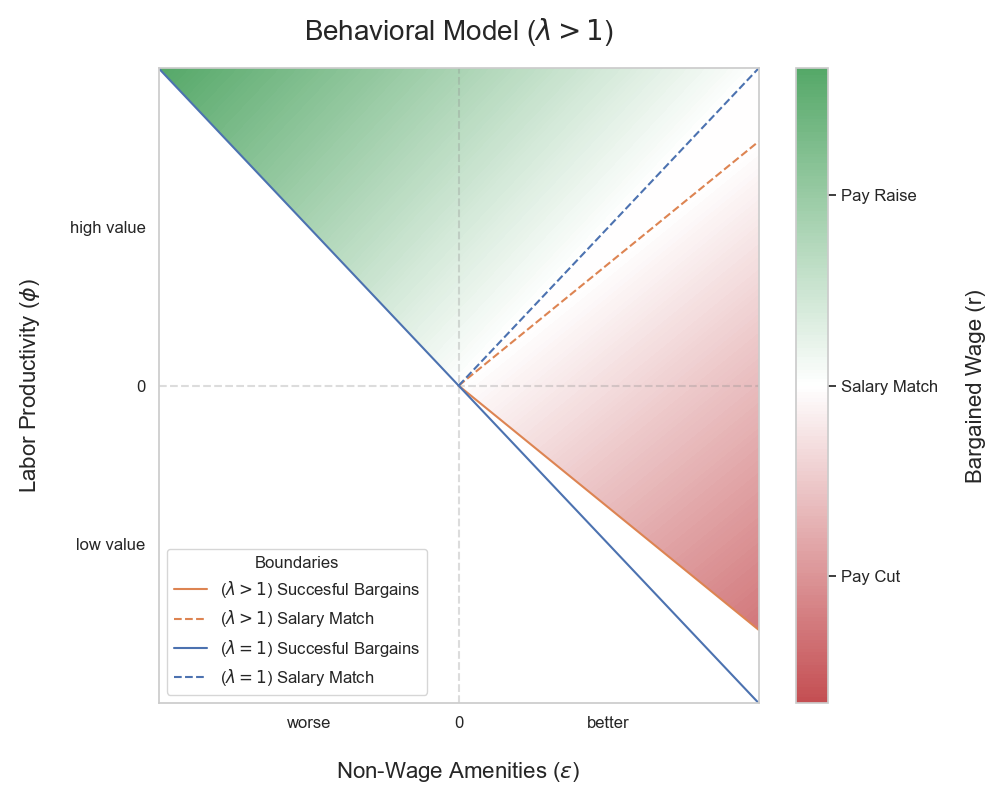}
    }

    \caption{Wage Bargaining With Loss Aversion} \label{fig:bargain_solution}
    \caption*{\normalsize 
        This figure compares successfully bargained wages under the standard model $(\lambda=1)$ and behavioral model $(\lambda>1)$. Colored regions correspond to successful bargains in \autoref{eq:bargain_region}, and colors denote bargained wage values in \autoref{eq:bargain_solution} (red for pay cuts, green for pay raises). Lighter colors denote smaller magnitudes, and white denotes a salary match $(r=0)$. The horizontal axis denotes the value of non-wage amenities $\epsilon_{if}$, and the vertical axis denotes the value of labor productivity $\phi_{if}$. The solid blue line denotes the region for successful bargains in the standard model, and the solid orange line is the updated region for successfully bargained pay cuts in the behavioral model. In the standard model, the dashed blue line defines $(\epsilon_{if},\phi_{if})$ values for salary matching $(r=0)$. In the behavioral model, the white area between the two dashed lines define the corresponding region for salary matching. 
    } 
\end{figure}
\vspace*{\fill}
\clearpage

\begin{figure}[H]
    \makebox[\textwidth][c]{\includegraphics[height=0.37\textheight]{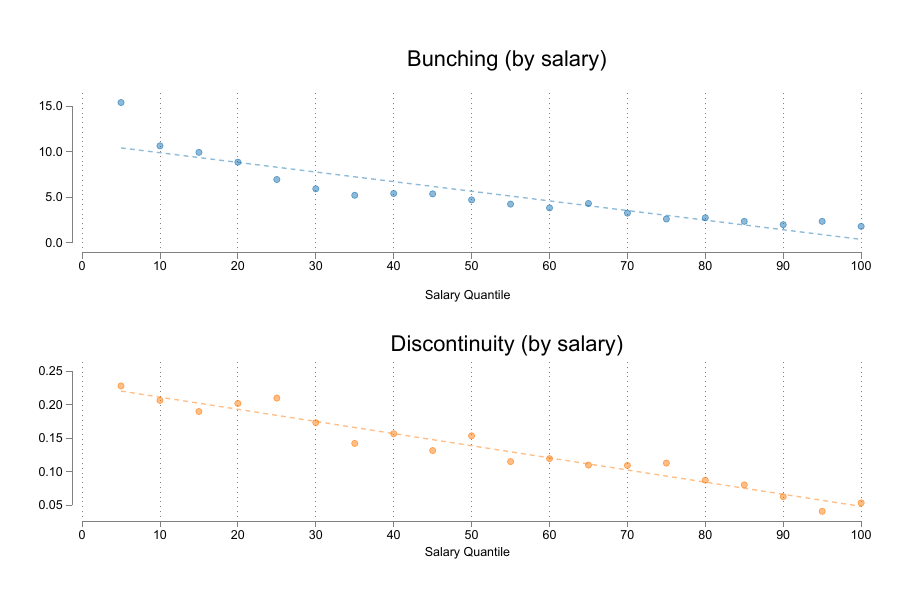}}

    \makebox[\textwidth][c]{\includegraphics[height=0.37\textheight]{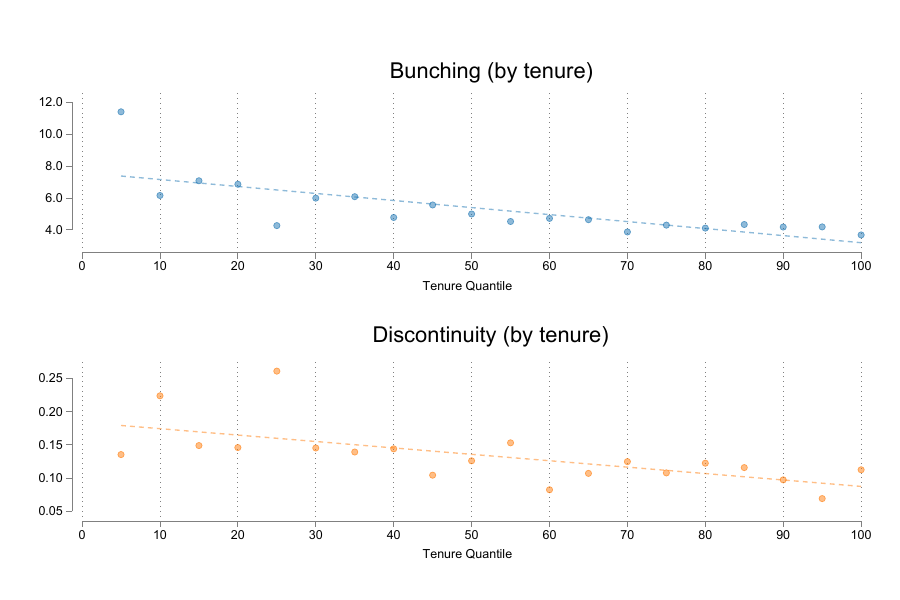}}
    
    \caption{Bunching and Discontinuity by Quantile Bins for Salary and Tenure} \label{fig:hetero_salary_tenure}
    \caption*{\normalsize 
        This figure plots bunching and discontinuity for job switchers in each of the 20 quantile bins for prior salary and tenure (work experience in months since 2015), which is discussed in \autoref{subsec:anomalies}. Bunching and discontinuity is measured in the same way as \autoref{tab:anomalies}. Dashed lines denote best linear fit across bin-level scatter points. 
    } 
\end{figure}
\clearpage

\vspace*{\fill}
\begin{figure}[H]
    \makebox[\textwidth][c]{\includegraphics[height=0.38\textheight]{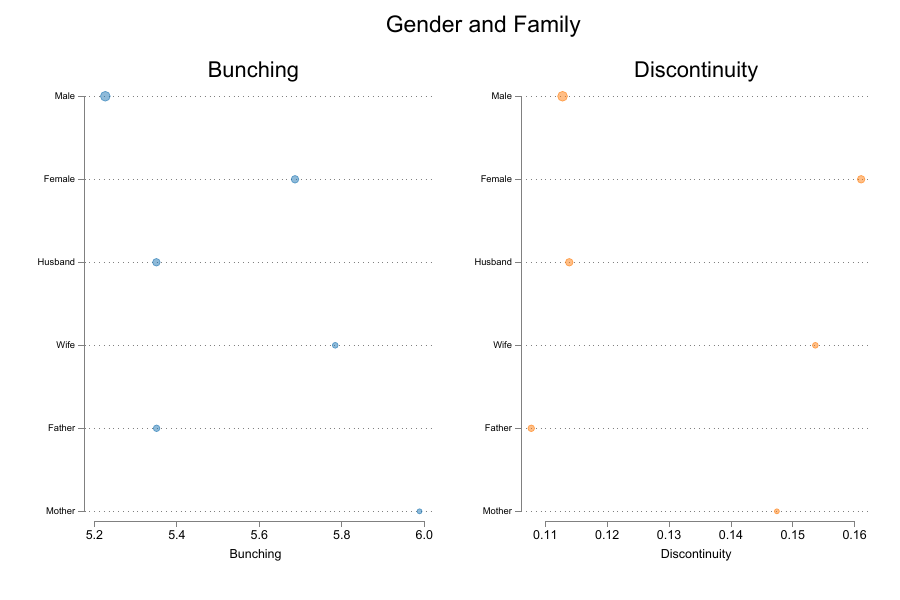}}
    \vspace{1em} 

    \makebox[\textwidth][c]{\includegraphics[height=0.38\textheight]{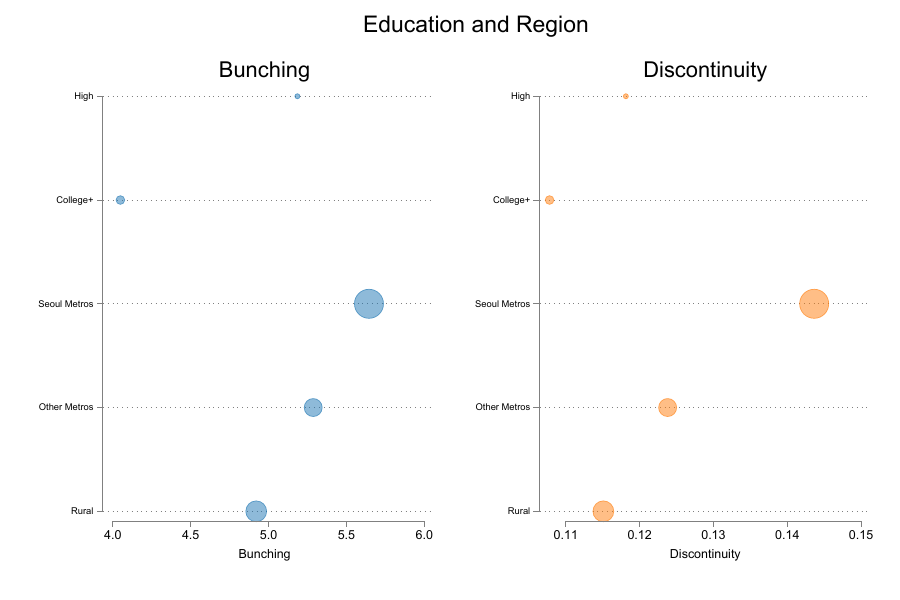}}
    
    \caption{Bunching and Discontinuity by Job Seeker Characteristics} \label{fig:hetero_demogs}
    \caption*{\normalsize 
        This figure plots bunching and discontinuity for job switchers with specified characteristics (discussed in \autoref{subsec:anomalies}). Bunching and discontinuity is measured in the same way as \autoref{tab:anomalies}. The size of scatter points indicate the number of job switchers corresponding to that characteristic.
    } 
\end{figure}
\vspace*{\fill}
\clearpage

\begin{figure}[H]

    \makebox[\textwidth][c]{%
        \includegraphics[width=0.37\paperwidth]{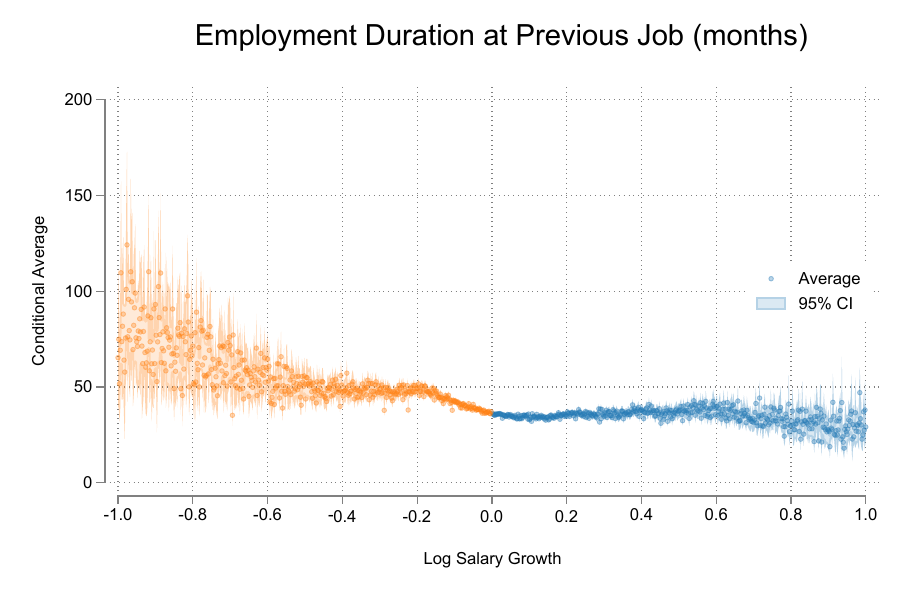}
        \includegraphics[width=0.37\paperwidth]{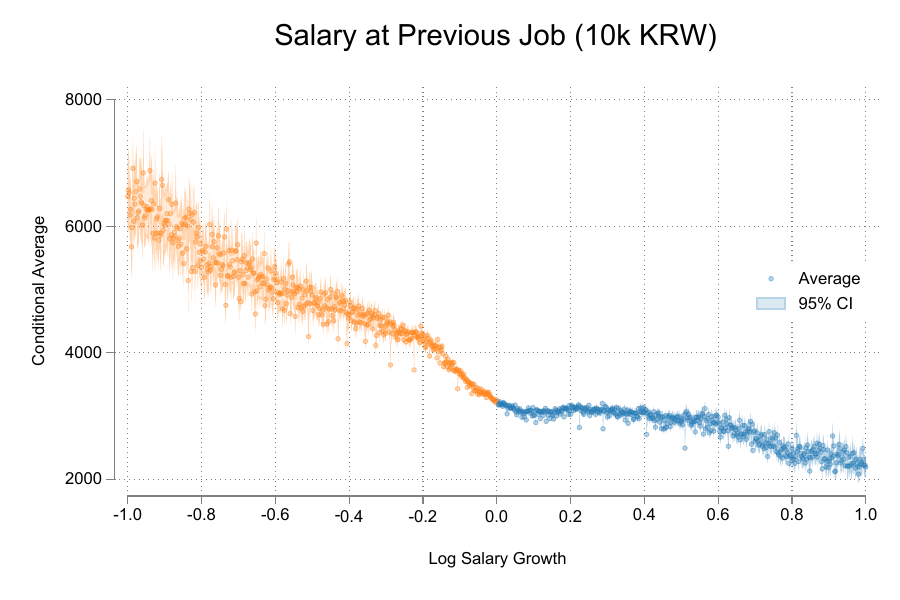}
    }

    \makebox[\textwidth][c]{%
        \includegraphics[width=0.37\paperwidth]{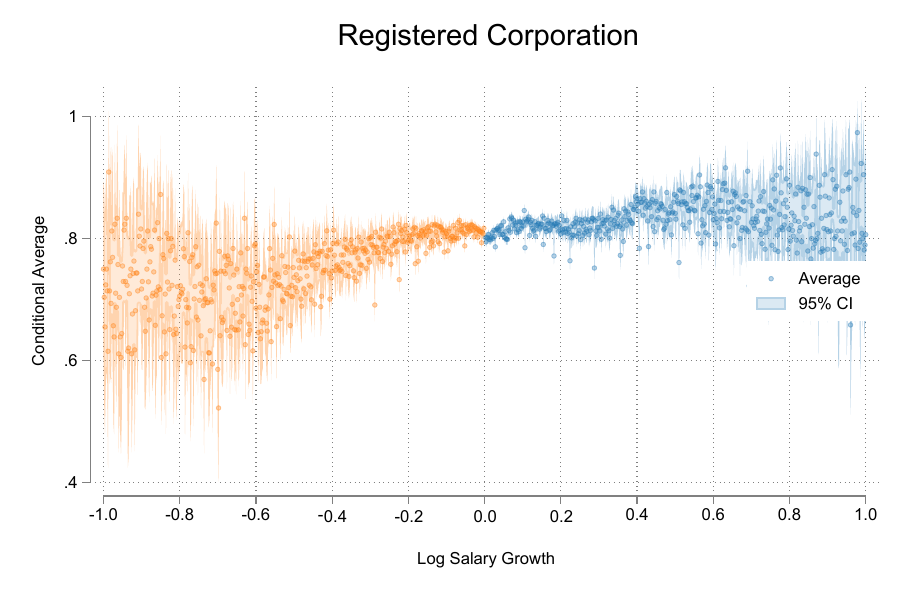}    
        \includegraphics[width=0.37\paperwidth]{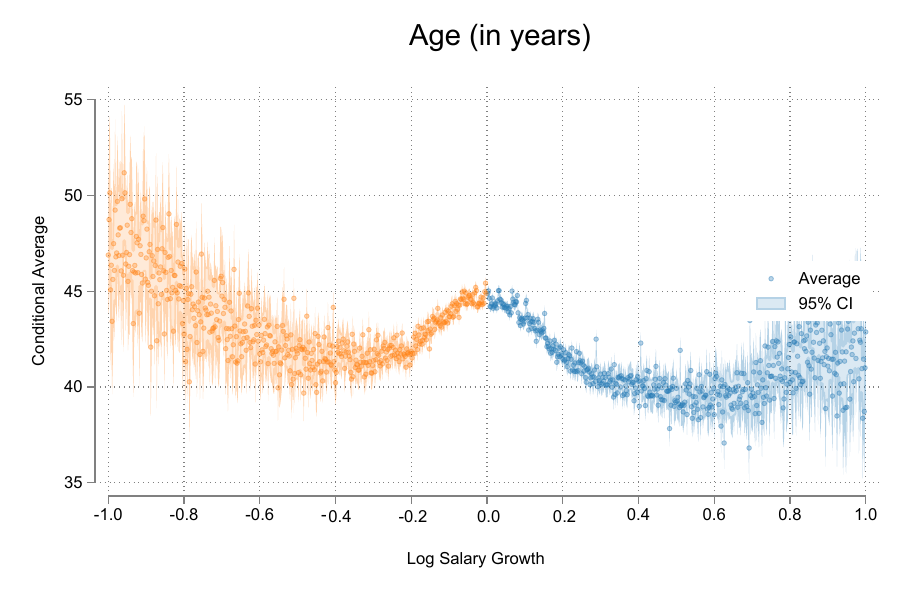}
    }

    \makebox[\textwidth][c]{%
        \includegraphics[width=0.37\paperwidth]{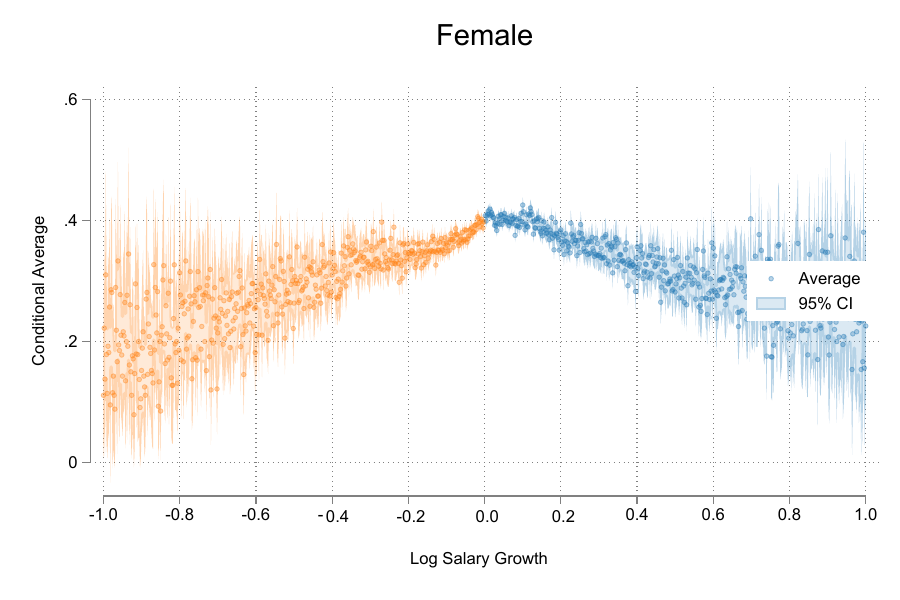}    
        \includegraphics[width=0.37\paperwidth]{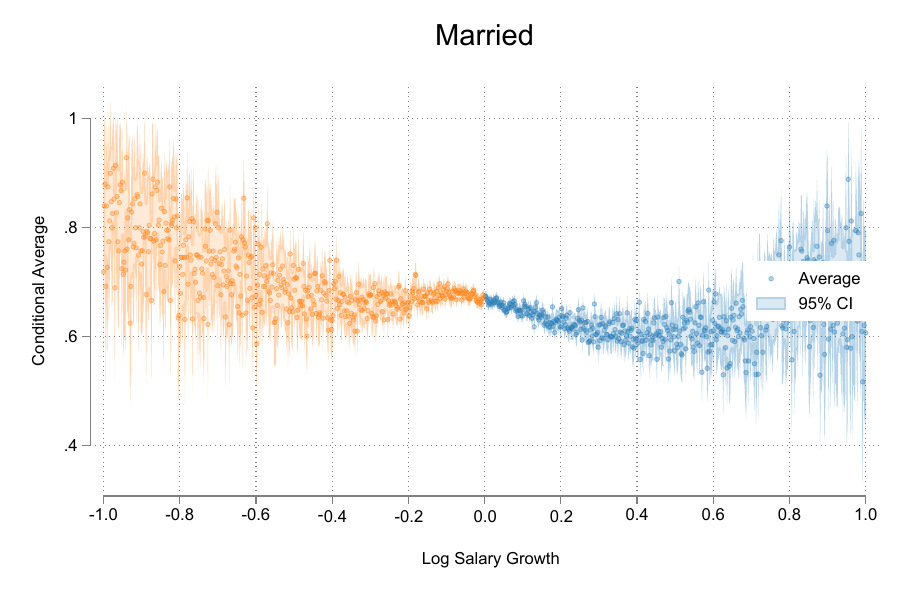}
    }

    \makebox[\textwidth][c]{%
        \includegraphics[width=0.37\paperwidth]{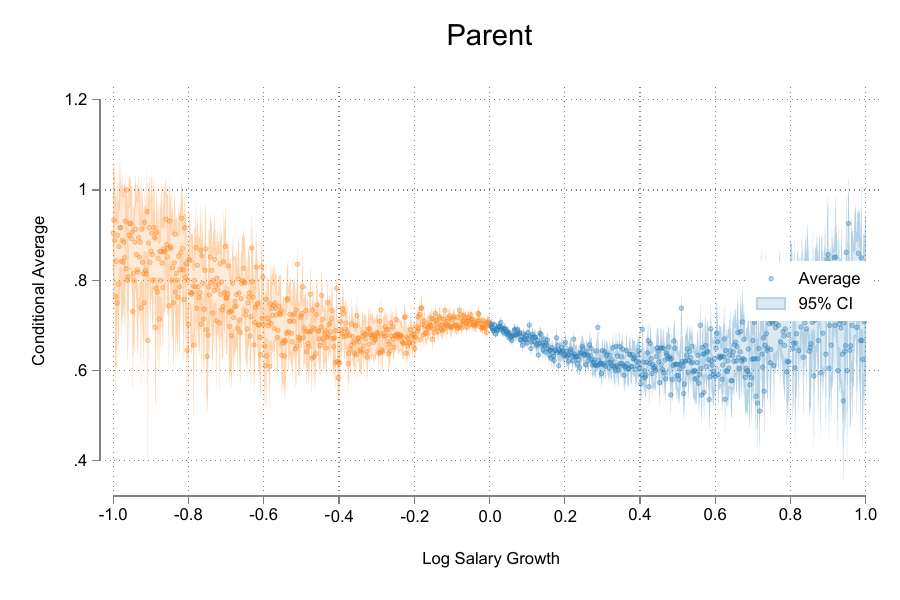}    
        \includegraphics[width=0.37\paperwidth]{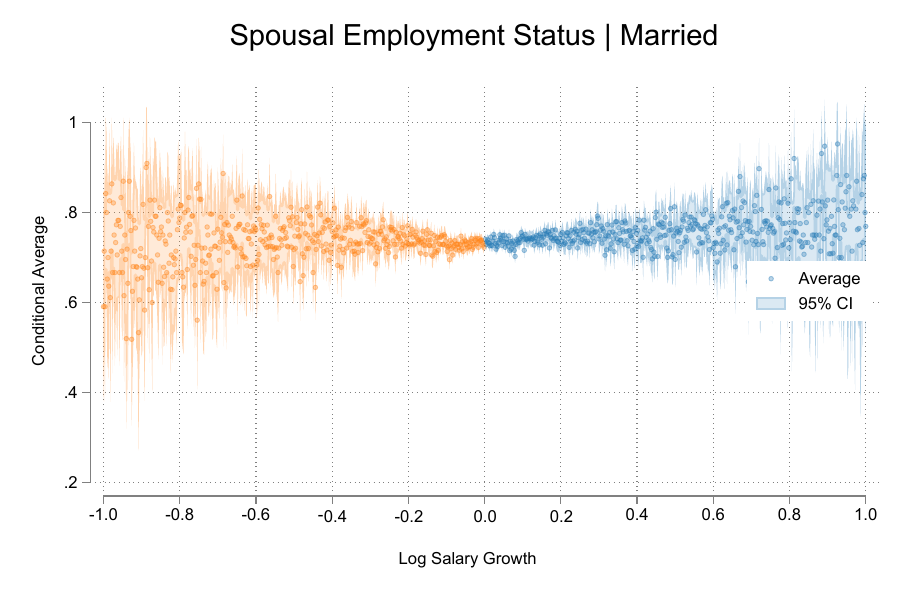}
    }

    \caption{Average Characteristics in Each Salary Growth Bin} \label{fig:demogs}
    \caption*{\normalsize 
        This figure plots average characteristics in each salary growth bin (discussed in \autoref{subsec:contributing_factors}). Scatter points are bin-level average values for each characteristic, and shaded areas are 95\% confidence intervals. 
    }
\end{figure}
\clearpage

\begin{figure}[ht]
    \makebox[\textwidth][c]{\includegraphics[height=0.38\textheight]{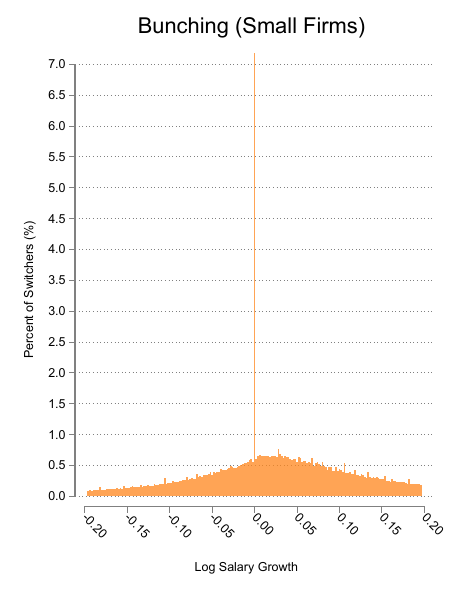}}
    \vspace{1em}

    \makebox[\textwidth][c]{%
        \includegraphics[height=0.38\textheight]{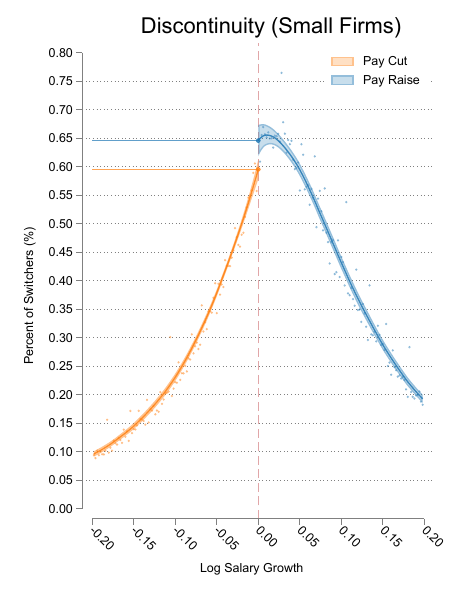}
        \includegraphics[height=0.38\textheight]{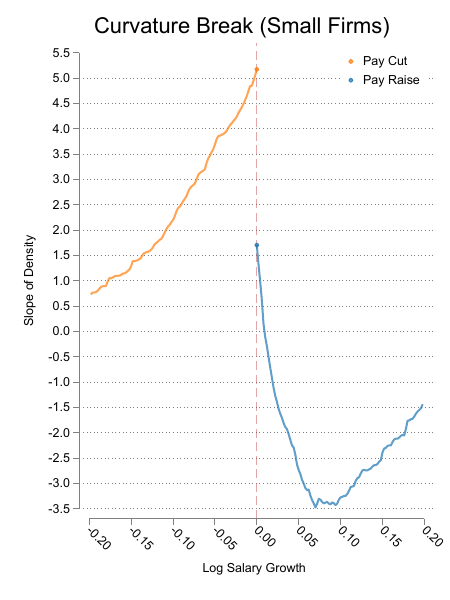}
    }

    \caption{Anomalies for Job Switchers in Small-Medium Businesses (SMBs)} \label{fig:anomaly_smallfirm}
    \caption*{\normalsize 
        This is an implementation of \autoref{fig:anomaly_sample} for job switchers in Small-Medium Businesses (SMBs), who are often exempt from overtime pay through the Blanket Wage System (discussed in \autoref{subsec:contributing_factors}).
    } 
\end{figure}
\clearpage

\vspace*{\fill}
\begin{figure}[H]
    \makebox[\textwidth][c]{\includegraphics[height=0.43\textheight]{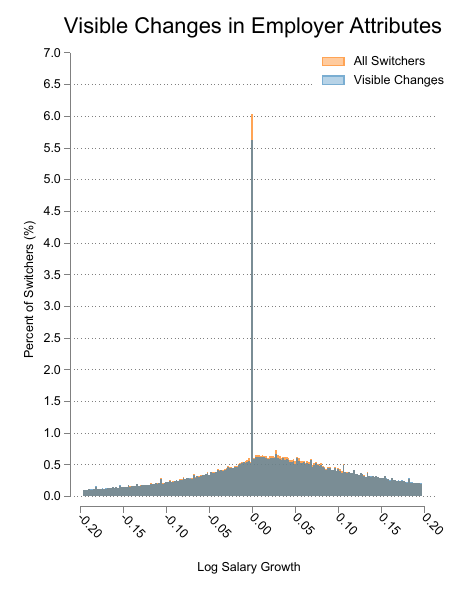}}
    
    \caption{Bunching for Job Switchers with Visible Changes in Firm Attributes} \label{fig:id_swaps}
    \caption*{\normalsize 
        This is an implementation of \autoref{fig:anomaly_sample} for job switchers with visible changes in employer attributes, which rules out arbitrary reassignment of firm IDs (discussed in \autoref{subsec:contributing_factors}). Considered attributes are corporate registration status, organization category, designated corporate scale, and public/private sectors. 
        }
\end{figure}
\vspace*{\fill}
\clearpage

\vspace*{\fill}
\begin{figure}[H]

    \makebox[\textwidth][c]{\includegraphics[height=0.35\textheight]{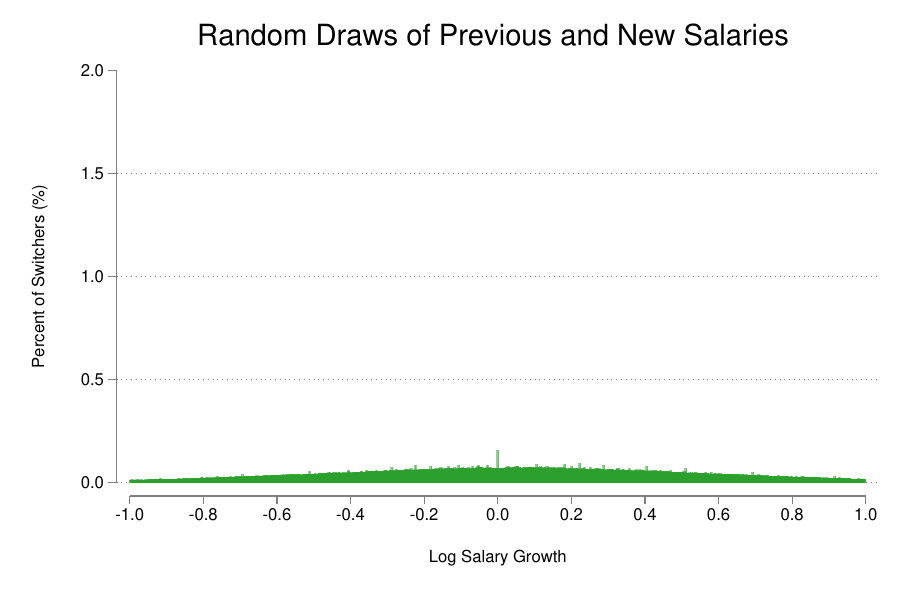}}

    \makebox[\textwidth][c]{\includegraphics[height=0.35\textheight]{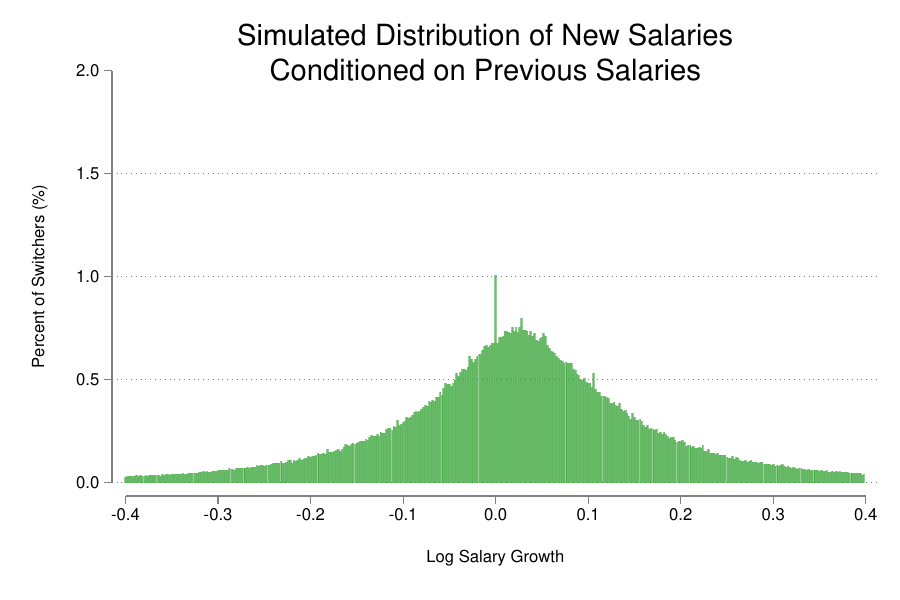}}

    \caption{Placebo Tests to Assess the Impact of Rounded Salaries on Bunching} \label{fig:sim_bunch}
    \caption*{\normalsize 
        This figure shows simulated distributions from placebo tests that assess the impact of rounded salaries on bunching magnitudes (discussed in \autoref{subsec:contributing_factors}). The top panel takes random draws of new and previous salaries (with replacement) from their empirical distributions in \autoref{fig:salary_growth_dist}, and their differences are plotted as salary growth. The bottom panel improves this procedure by conditioning the distribution of new salaries on each previous salary. Specifically, I re-weighted the distribution of new salaries using a smooth distribution of salary growth that was shifted to be centered on each previous salary value. I took a random draw for the new salary from this re-weighted distribution and took its difference with the previous salary to calculate salary growth. 
    }
\end{figure}
\vspace*{\fill}
\clearpage

\vspace*{\fill}
\begin{figure}[H]

    \makebox[\textwidth][c]{%
        \includegraphics[width=0.47\paperwidth]{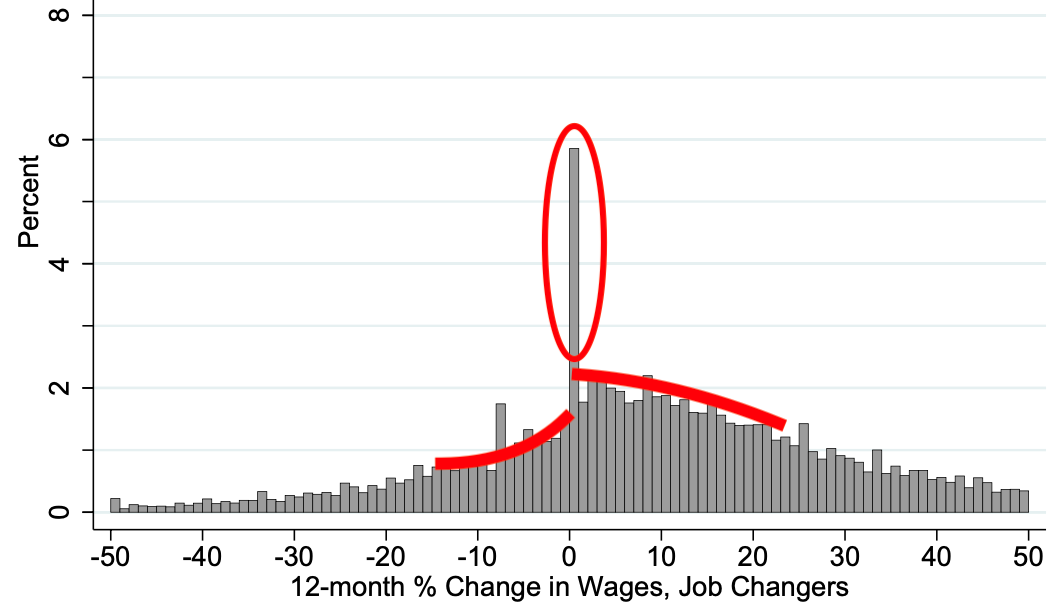}%
        \includegraphics[width=0.47\paperwidth]{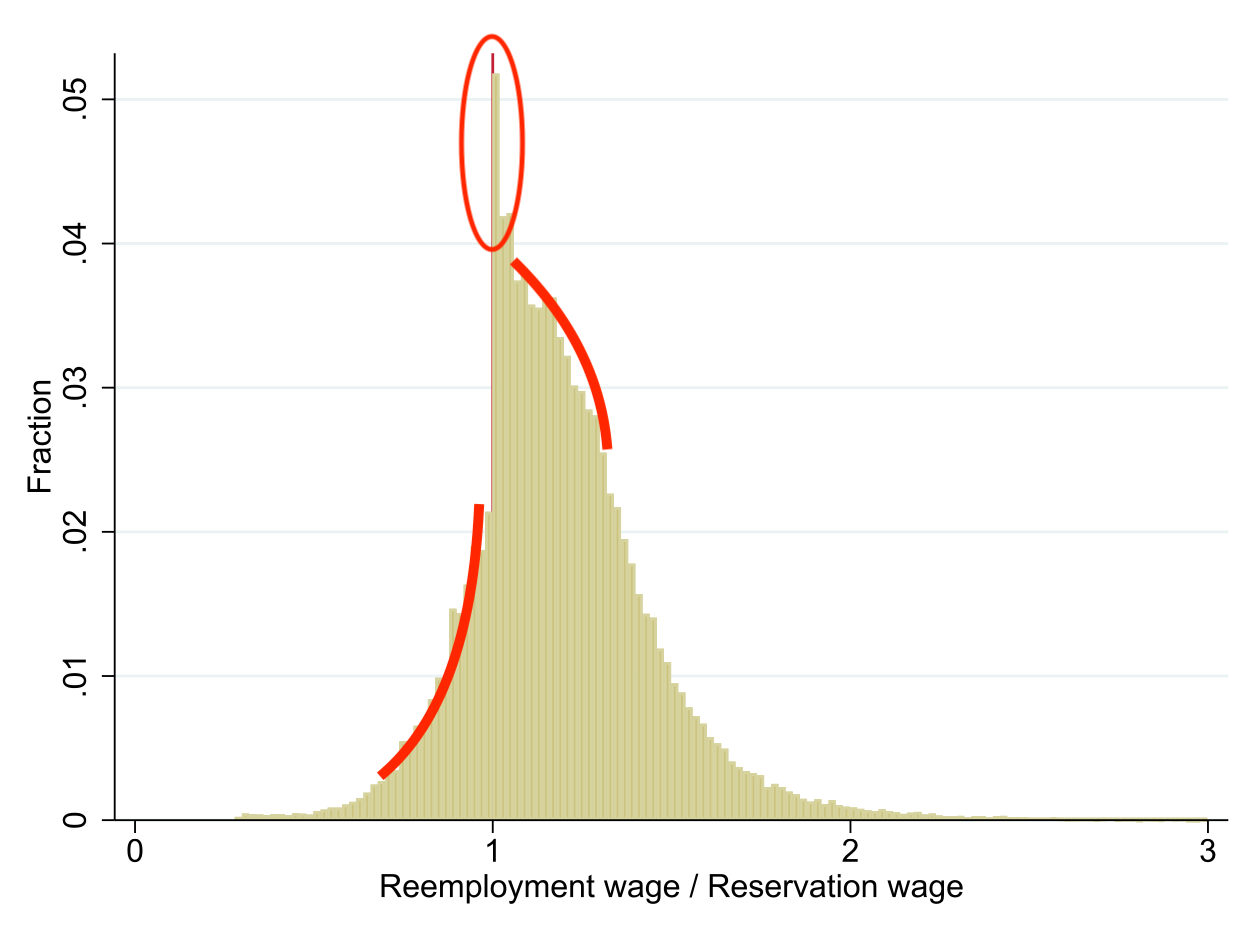}
    }

    \makebox[\textwidth][c]{%
        \includegraphics[width=0.47\paperwidth]{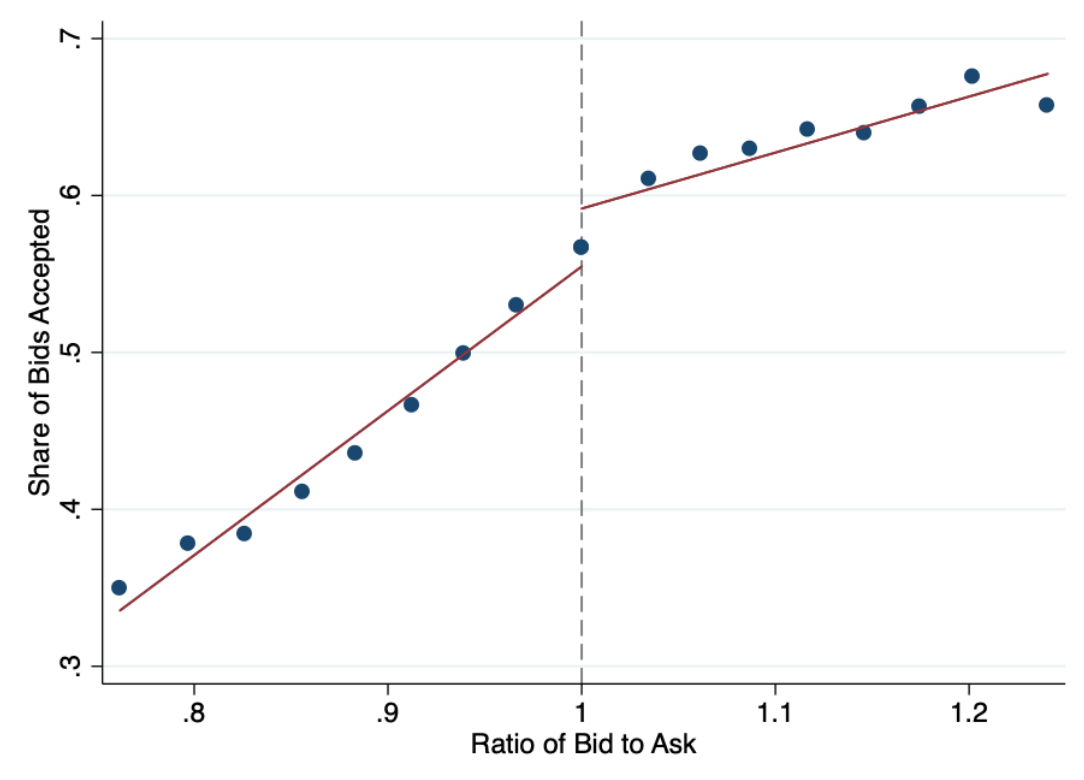}%
        \includegraphics[width=0.47\paperwidth]{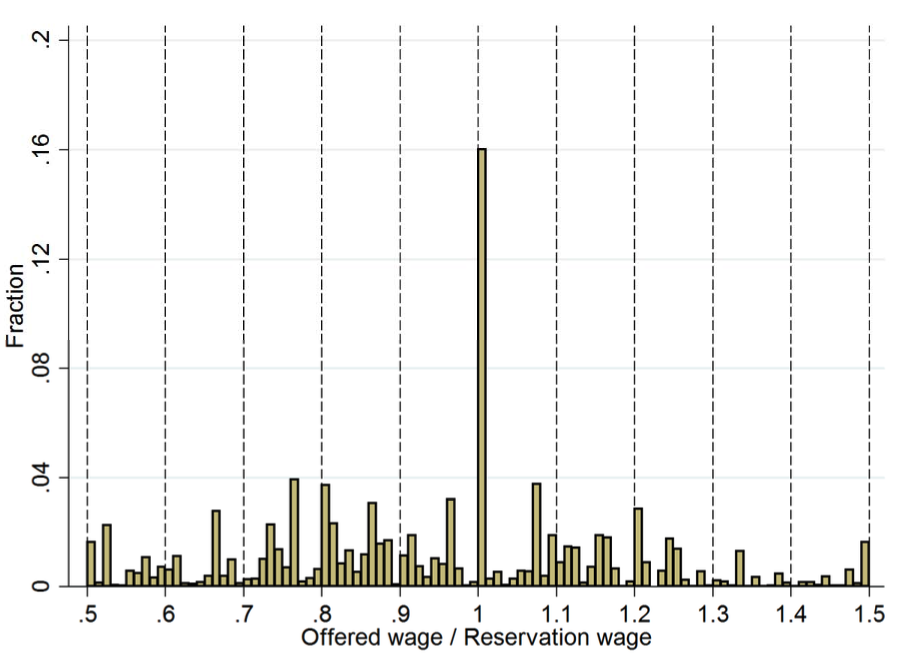}
    }
    
    \caption{Supporting Evidence for Anomalies and Key Mechanisms from Prior Studies} \label{fig:prior_studies}
    \caption*{\normalsize 
    
        Top panels show anomalies for job switchers from two prior studies (discussed in \autoref{subsec:hetero_prior_studies}). The top left panel is Figure 8b from \textcite{grigsby_aggregate_2021}, which analyzes administrative payroll data from ADP. The top right panel is Figure 2b from \textcite{barbanchon_gender_2020}, which analyzes administrative data on unemployed job seekers in France. Red lines have been annotated by us, not the original authors. 

        \,

        Bottom panels show two findings from prior studies that are consistent with kinked acceptance rates and corner solutions for wage offers. The bottom left panel is Figure 2B from \textcite{roussille_bidding_2024}, which analyzes salary offers made to candidates and their decisions on an online job board. The bottom right panel is Figure 8 from \textcite{krueger_contribution_2016}, which analyzes survey data on unemployed job seekers in New Jersey. 
    }
\end{figure}
\vspace*{\fill}
\clearpage

\begin{figure}[ht]
    \makebox[\textwidth][c]{%
        \includegraphics[height=0.38\textheight]{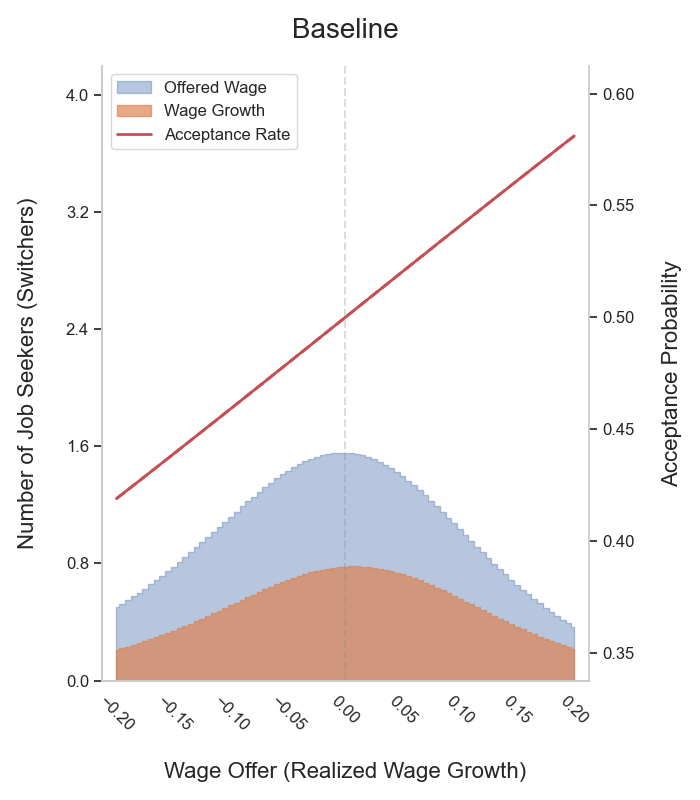}%
        \includegraphics[height=0.38\textheight]{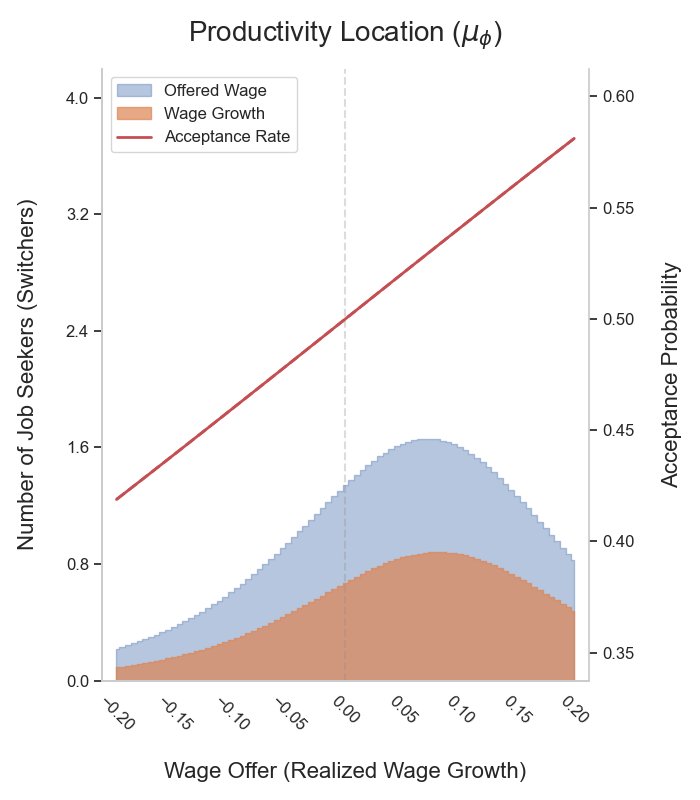}
    }

    \makebox[\textwidth][c]{%
        \includegraphics[height=0.38\textheight]{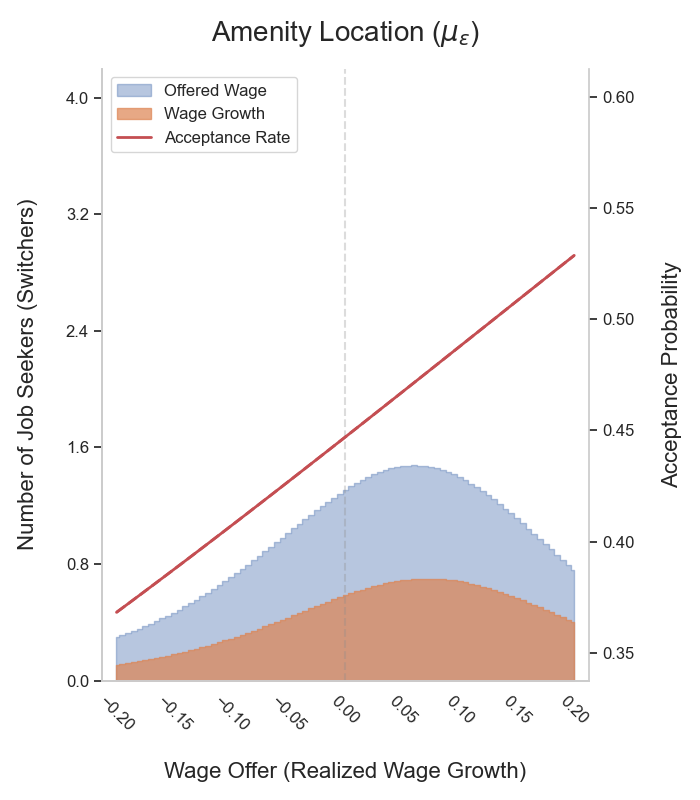}%
        \includegraphics[height=0.38\textheight]{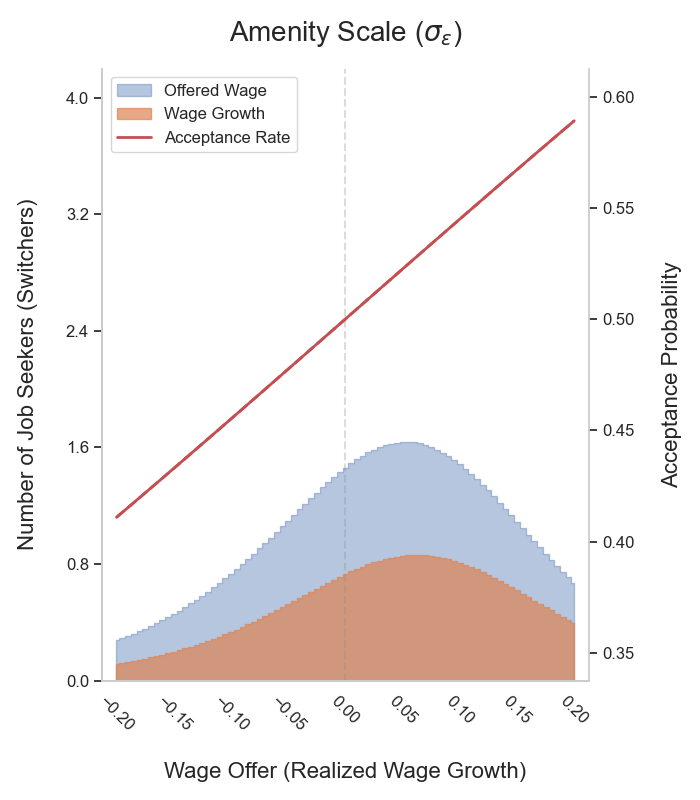}
    }

    \caption{Parameter Changes with Observationally Similar Increases in Salary Growth} \label{fig:offer_dist_tweaks}
    \caption*{\normalsize 
        This is an implementation of \autoref{fig:offer_dist} with various parameter values for a standard model without loss aversion (discussed in \autoref{subsec:fixed_calibrations}). The top left panel is based on a baseline set of parameters, and three other panels represent parameter changes that result in observationally similar increases in salary growth. The top right panel raises productivity location $\mu_\phi$, the bottom left panel lowers amenity location $\mu_\epsilon$, and the bottom right panel lowers amenity scale $\sigma_\epsilon$.
    }
\end{figure}
\clearpage

\vspace*{\fill}
\begin{figure}[ht]

    \makebox[\textwidth][c]{%
        \includegraphics[width=0.47\paperwidth]{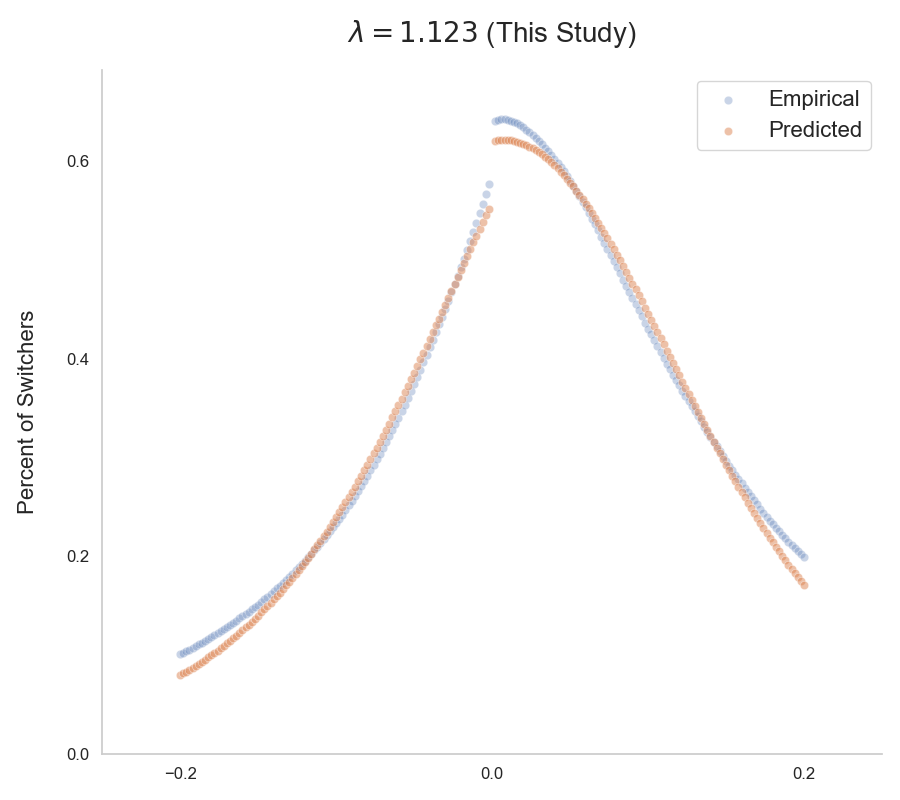}%
        \includegraphics[width=0.47\paperwidth]{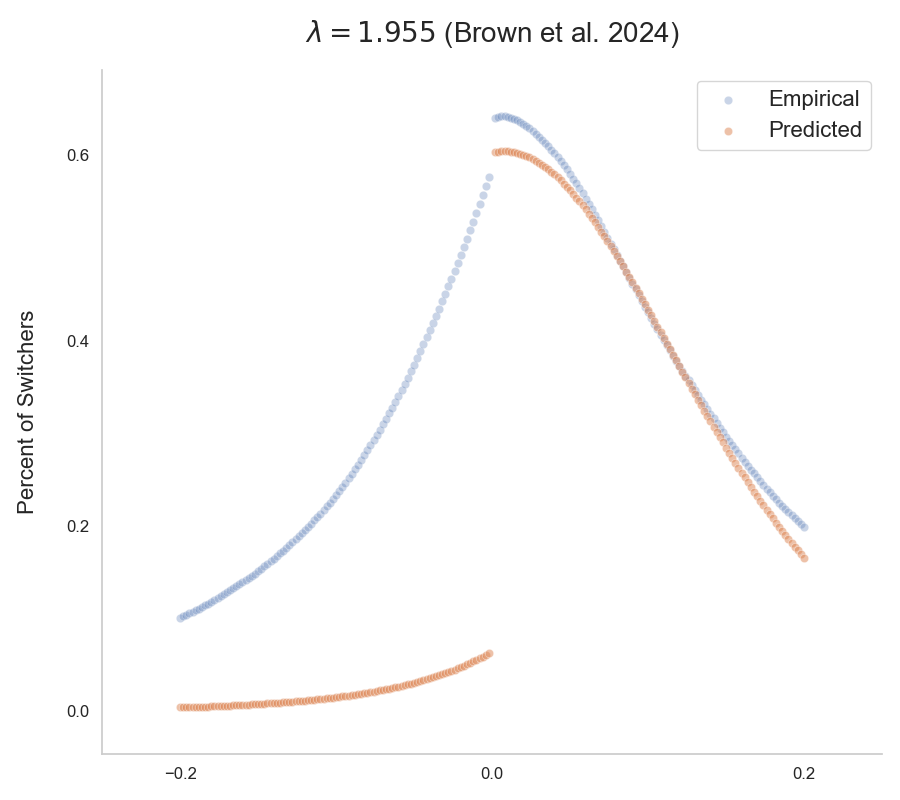}
    }

    \caption{Comparison of Model Fit at Two Values of Loss Aversion} \label{fig:lossav_paycut_freq}
    \caption*{\normalsize
        This is an implementation of \autoref{fig:model_fit} with two different values for loss aversion (discussed in \autoref{subsec:calibrated_lossav}). Predicted proportions in the left panel are based on loss aversion in this study ($\lambda=1.123$), and the right panel is based on average loss aversion across studies in a meta analysis by \cite{brown_meta-analysis_2024} ($\lambda=1.955$).
    }
\end{figure}
\vspace*{\fill}
\clearpage

\vspace*{\fill}
\begin{figure}[ht]

    \makebox[\textwidth][c]{%
        \includegraphics[width=0.47\paperwidth]{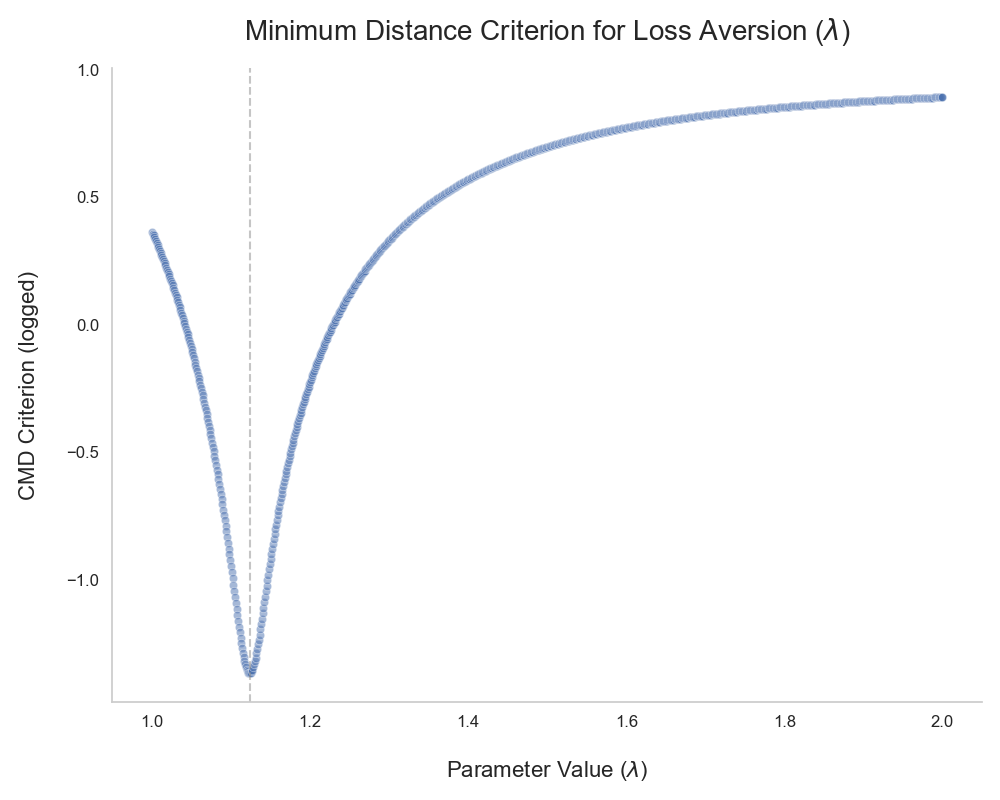}%
        \includegraphics[width=0.47\paperwidth]{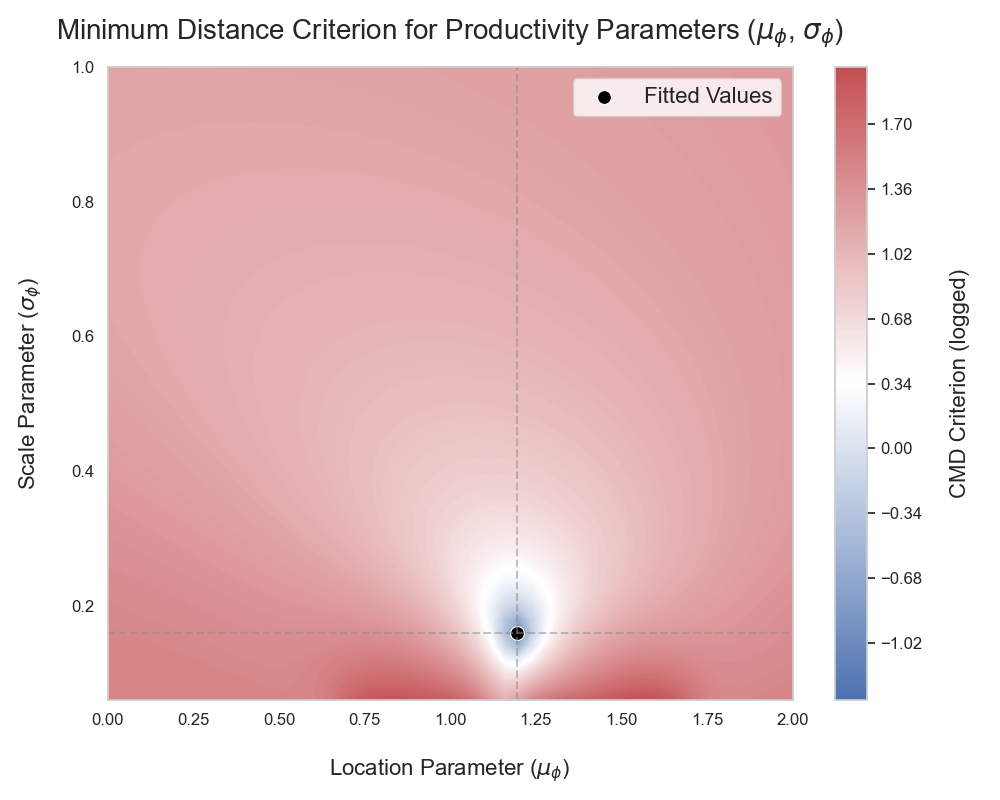}
    }

    \caption{Minimum Distance Criterion Given Parameter Values} \label{fig:criterion}
    \caption*{\normalsize 
        This figure plots the minimum distance criterion for each parameter value (discussed in \autoref{subsec:calibrated_lossav}). The minimum distance criterion is the sum of squared distances between empirical and predicted proportions in each salary growth bin. The left panel plots logged criterion values given loss aversion $\lambda$. The right panel plots logged criterion values given location and scale parameters $(\mu_\phi, \sigma_\phi)$, with red and blue denoting high and low values (respectively). Parameters that minimize the criterion correspond to \autoref{tab:parameters_main}.
    } 
\end{figure}
\vspace*{\fill}
\clearpage

\vspace*{\fill}
\begin{figure}[ht]

    \makebox[\textwidth][c]{\includegraphics[width=0.90\paperwidth]{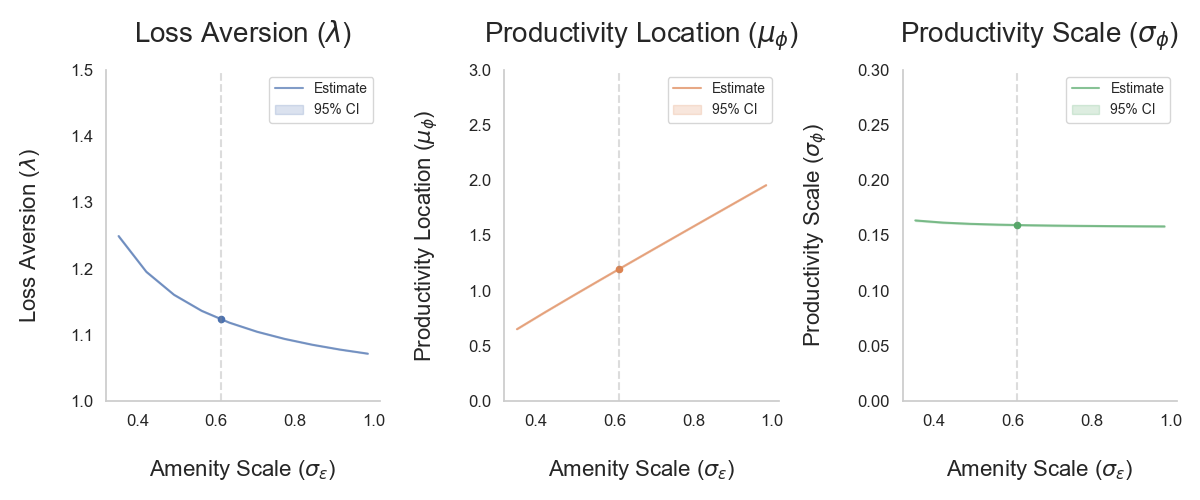}}
    
    \caption{Sensitivity of Parameters to Calibrations for Amenity Scale $\sigma_\epsilon$} \label{fig:calib_params}
    \caption*{\normalsize 
        This is an implementation of \autoref{tab:parameters_main} with different calibrations for the amenity scale parameter $\sigma_\epsilon$ in the minimum distance procedure (discussed in \autoref{subsec:calibrated_lossav}). The left panel plots estimated loss aversion $\hat{\lambda}$ for each calibration of $\sigma_\epsilon$, and the middle/right panels are corresponding figures for location/scale parameters $(\mu_\phi, \sigma_\phi)$ of the productivity distribution (respectively).
    }
\end{figure}
\vspace*{\fill}
\clearpage

\vspace*{\fill}
\begin{figure}[H]

    \makebox[\textwidth][c]{\includegraphics[width=0.95\textwidth]{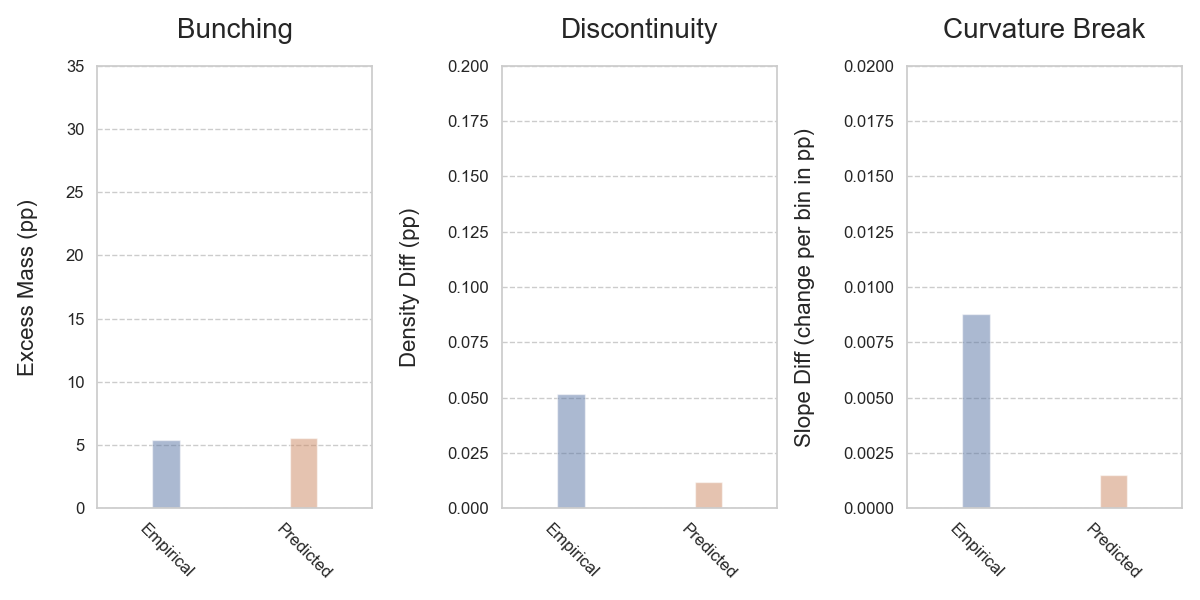}}
    
    \caption{Empirical and Predicted Magnitudes of Anomalies (including the zero bin)} \label{fig:anomaly_comparison_zero}
    \caption*{\normalsize 
        This is an implementation of \autoref{fig:anomaly_comparison} that includes the zero bin in the minimum distance procedure (discussed in \autoref{subsec:fit_anomalies}).
    } 
\end{figure}
\vspace*{\fill}
\clearpage

\vspace*{\fill}
\begin{figure}[H]

    \makebox[\textwidth][c]{%
        \includegraphics[width=0.47\paperwidth]{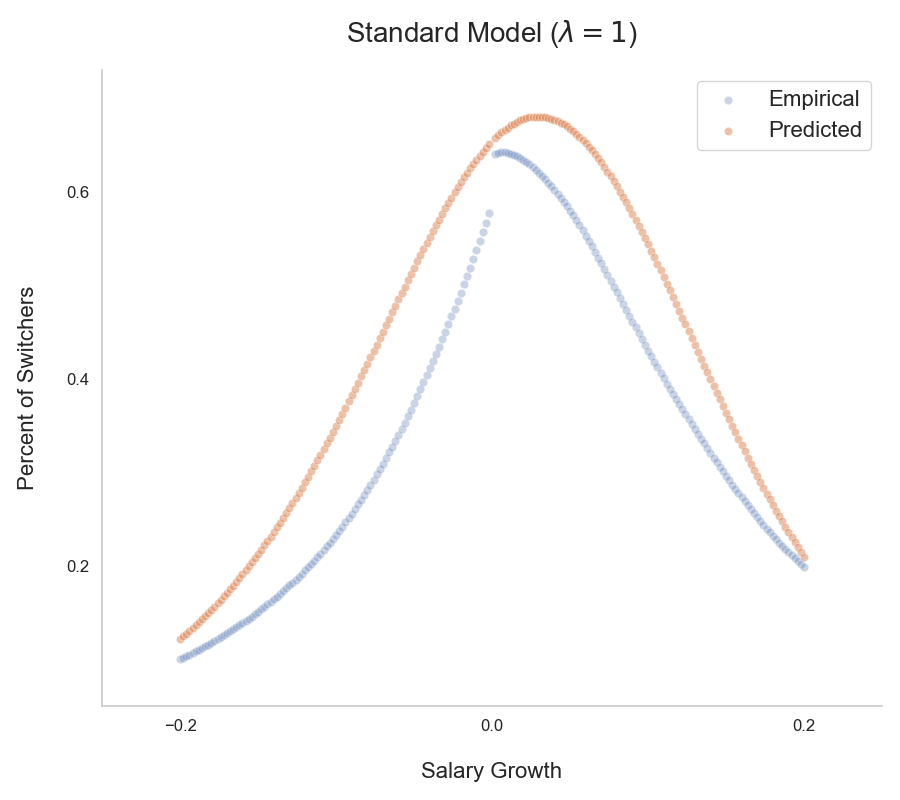}%
        \includegraphics[width=0.47\paperwidth]{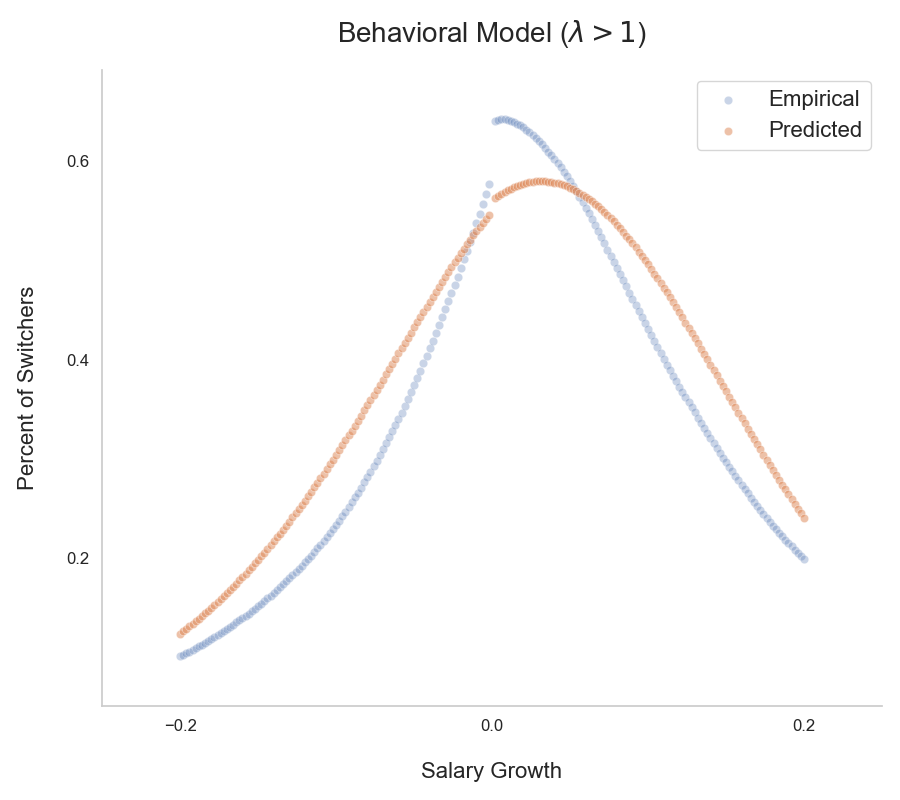}
    }

    \caption{Model Fit to Proportions in Salary Growth Bins (including the zero bin)} \label{fig:model_fit_zero}
    \caption*{\normalsize 
        This is an implementation of \autoref{fig:model_fit} that includes the zero bin in the minimum distance procedure (discussed in \autoref{subsec:fit_anomalies}).
    }
\end{figure}
\vspace*{\fill}
\clearpage

\vspace*{\fill}
\begin{figure}[H]

    \makebox[\textwidth][c]{%
        \includegraphics[width=0.47\paperwidth]{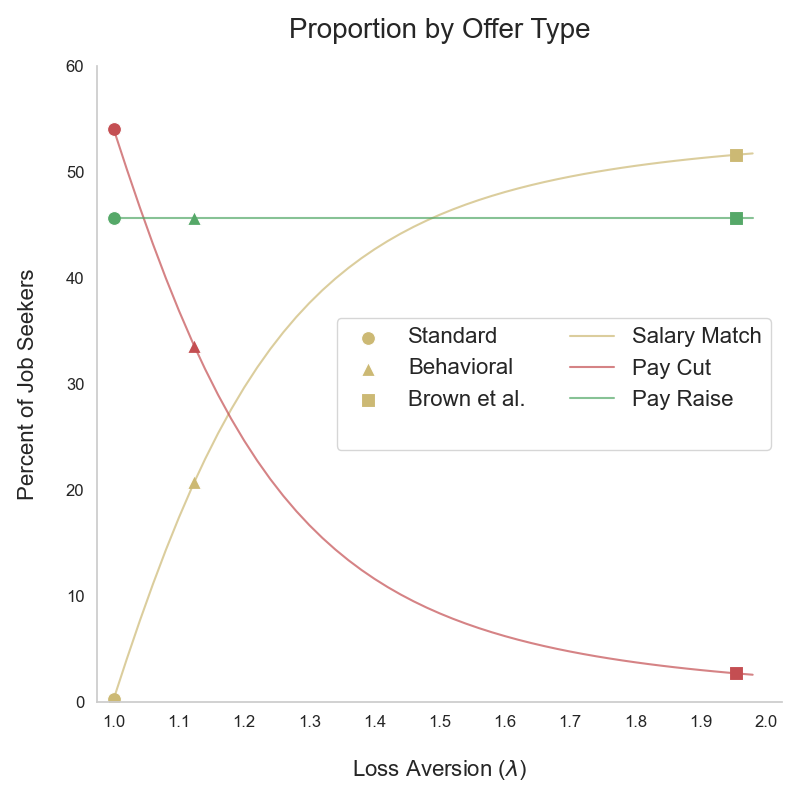}%
        \includegraphics[width=0.47\paperwidth]{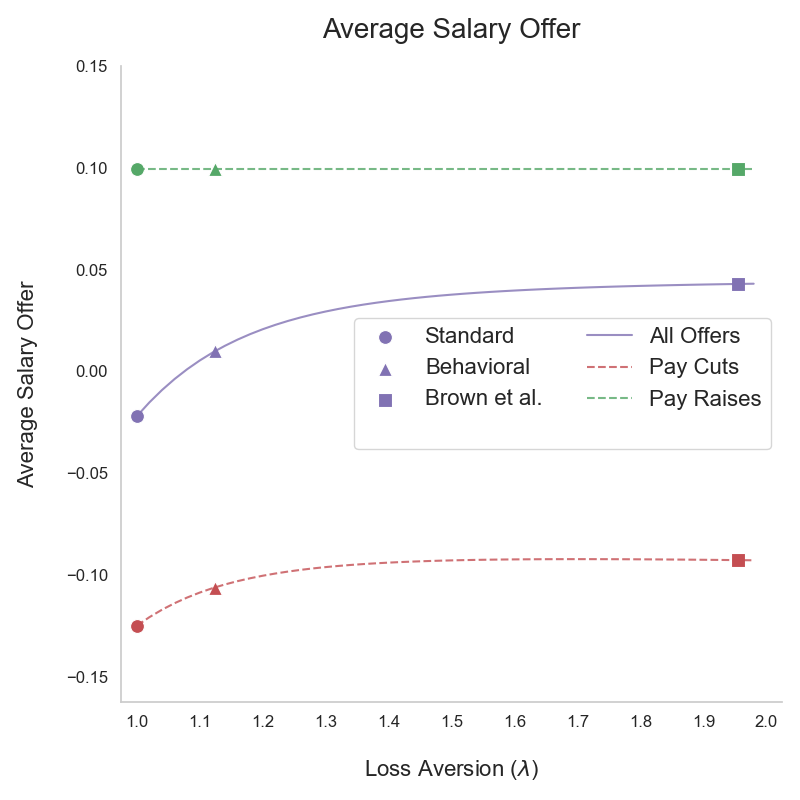}
    }

    \caption{Implications of Loss Aversion on Salary Offers} \label{fig:wage_implications}
    \caption*{\normalsize
        This figure uses the fitted model to predict implications of loss aversion on salary offers received by job seekers (discussed in \autoref{subsec:salary_implications}).
        The left panel plots the predicted proportion of pay cuts (in red), pay raises (in green), and salary matches (in yellow) offered to job seekers for each value of loss aversion, holding other parameters fixed at \autoref{tab:parameters_main}. The right panel is an analogous figure for the average magnitude of pay cuts (in red), pay raises (in green), and all salary offers (in purple). Scatter points correspond to three specific values of loss aversion: $\lambda=1$ for standard preferences (circles), $\lambda=1.123$ for loss aversion in this study (triangles), and $\lambda=1.955$ for average loss aversion across studies in \cite{brown_meta-analysis_2024} (squares). 
    }
\end{figure}
\vspace*{\fill}
\clearpage

\begin{figure}[ht]
    \makebox[\textwidth][c]{\includegraphics[width=0.83\paperwidth]{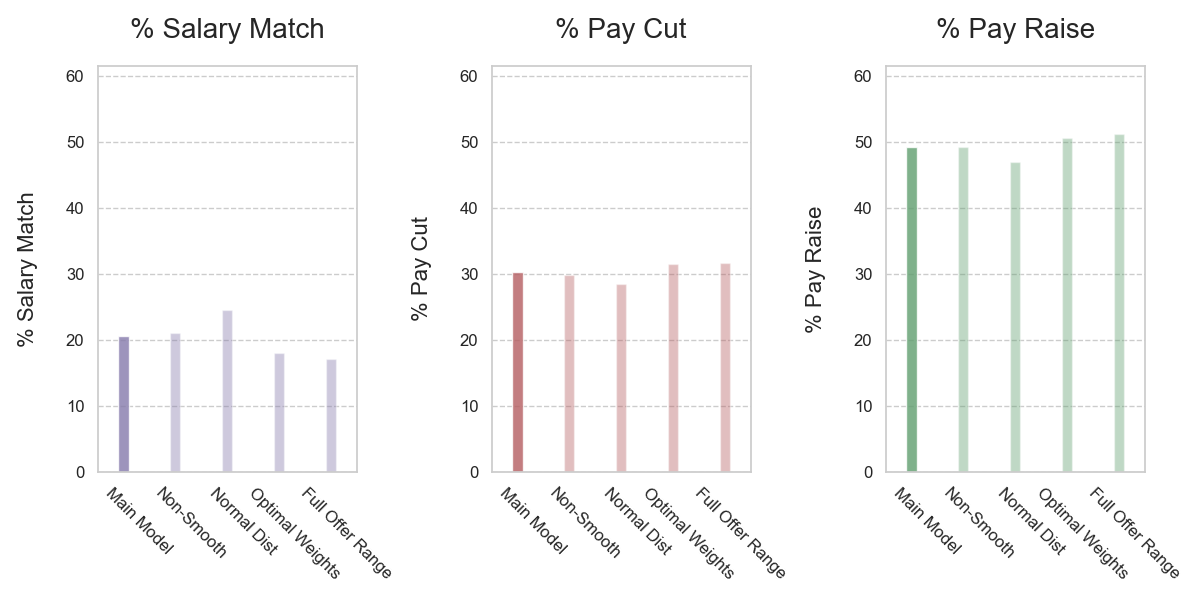}}        
    \vspace{1em}

    \makebox[\textwidth][c]{\includegraphics[width=0.83\paperwidth]{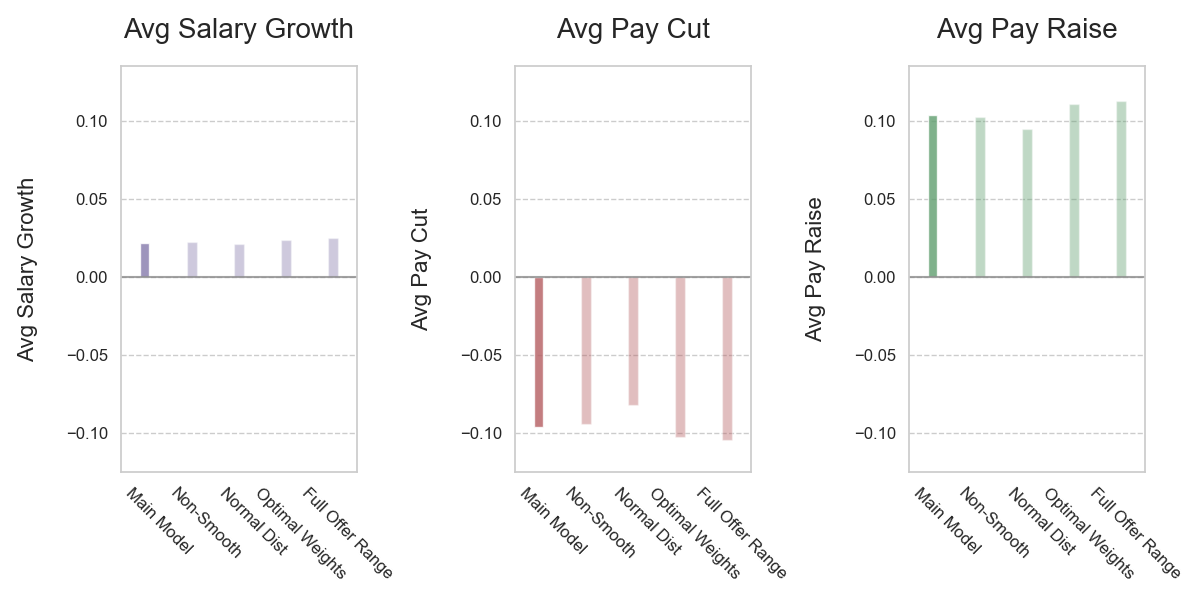}}
    
    \caption{Sensitivity of Salary Impacts to Alternative Specifications} \label{fig:salary_impact_adjust}
    \caption*{\normalsize 
        This is an implementation of \autoref{fig:wage_implications} with adjustments to the minimum distance procedure (discussed in \autoref{subsec:salary_implications}). Top panels are predicted proportions of salary matches (in purple), pay cuts (in red), and pay raises (in green) offered to job seekers. Bottom panels are analogous figures for the average magnitude of pay cuts (in red), pay raises (in green), and all salary offers (in purple). Bar 1 is the primary specification in \autoref{subsec:mindist_procedure}. Bar 2 replaces kernel estimates with raw proportions in the data. Bar 3 parameterizes unobserved heterogeneity with the normal distribution. Bar 4 uses optimal weights for bins. Bar 5 expands the range of salary growth from $\pm 0.2$ to $\pm 1.0$ log points.
    }
\end{figure}
\clearpage

\vspace*{\fill}
\begin{figure}[ht]

    \makebox[\textwidth][c]{\includegraphics[width=0.83\paperwidth]{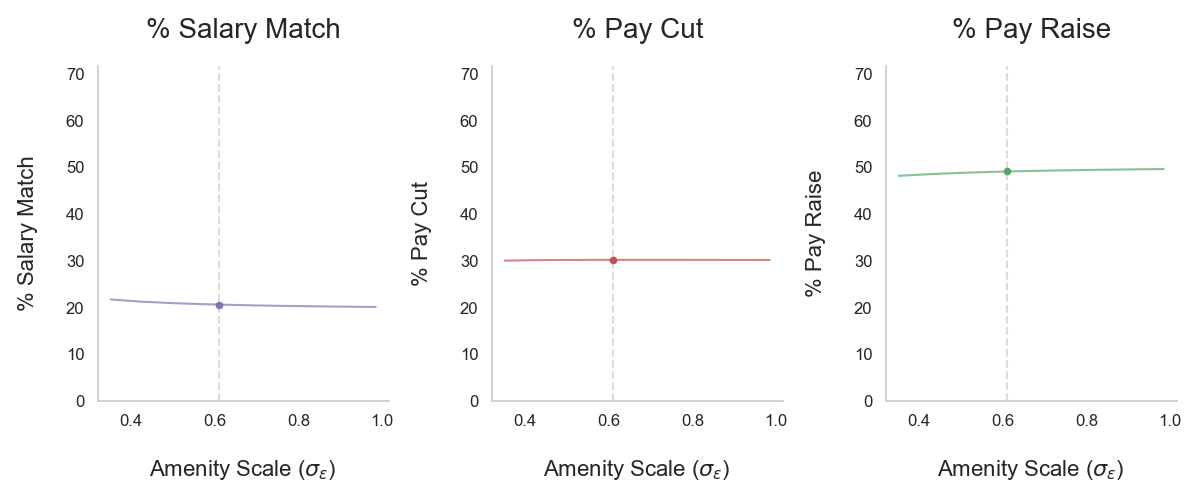}}

    \makebox[\textwidth][c]{\includegraphics[width=0.83\paperwidth]{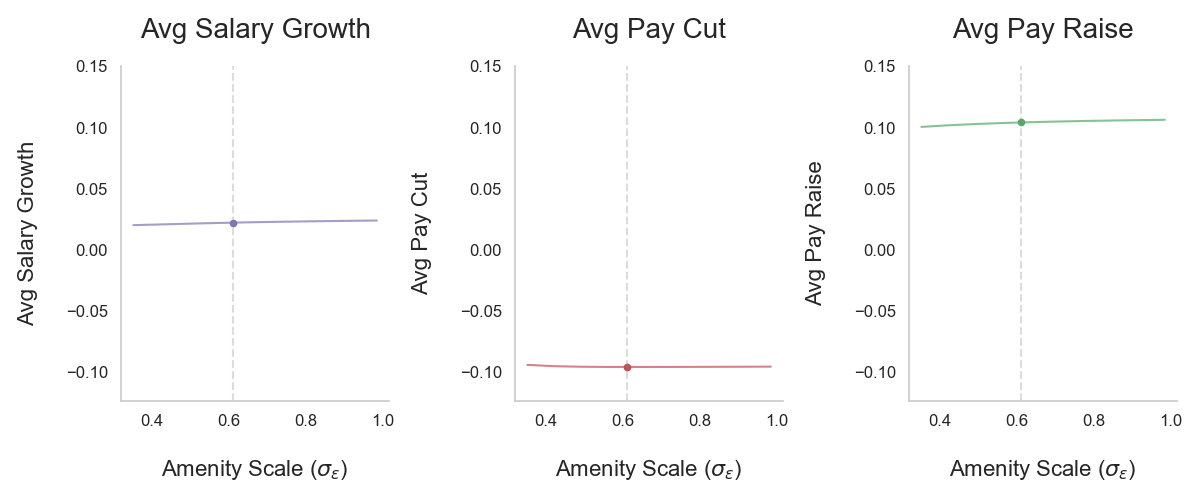}}
    
    \caption{Sensitivity of Salary Impacts to Calibrations for Amenity Scale $\sigma_\epsilon$} \label{fig:salary_impact_calib}
    \caption*{\normalsize 
        This is an implementation of \autoref{fig:wage_implications} with different calibrations for the amenity scale parameter $\sigma_\epsilon$ in the minimum distance procedure (discussed in \autoref{subsec:salary_implications}). Top panels are predicted proportions of salary matches (in purple), pay cuts (in red), and pay raises (in green) offered to job seekers. Bottom panels are analogous figures for the average magnitude of pay cuts (in red), pay raises (in green), and all salary offers (in purple).
    }
\end{figure}
\vspace*{\fill}
\clearpage

\begin{figure}[ht]

    \makebox[\textwidth][c]{\includegraphics[width=0.83\paperwidth]{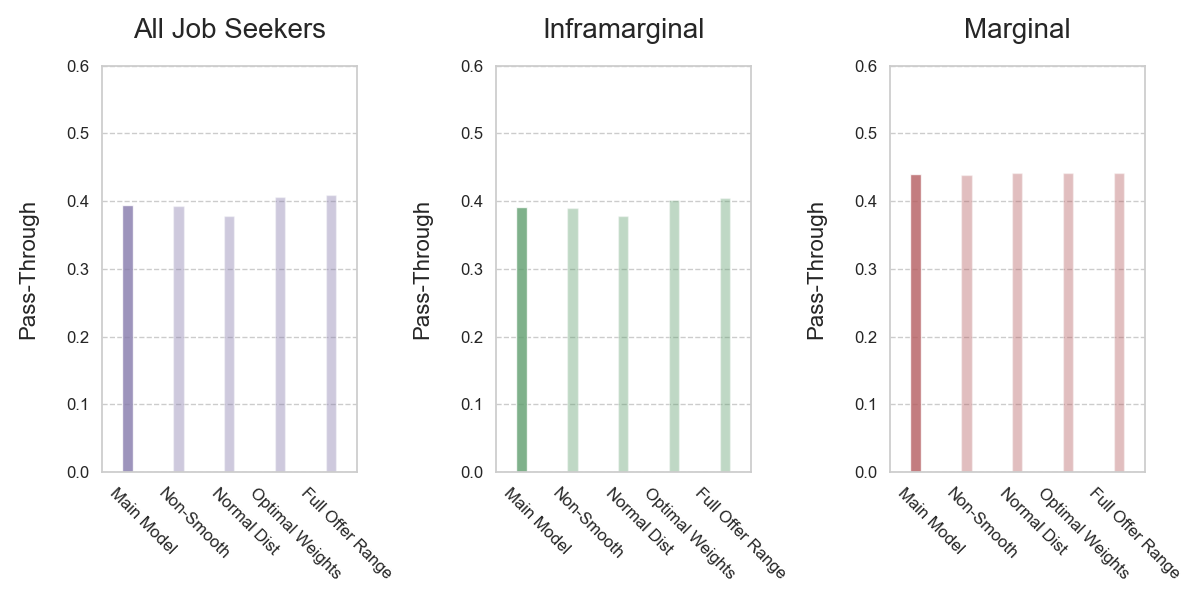}}        
    \vspace{1em}
    
    \makebox[\textwidth][c]{\includegraphics[width=0.83\paperwidth]{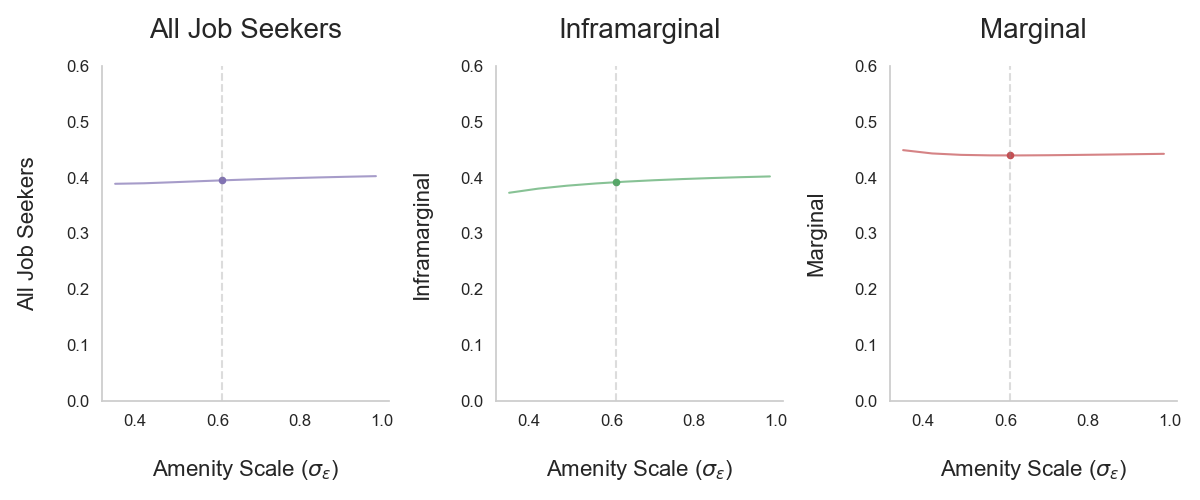}}

    \caption{Sensitivity of Pass-Through to Alternative Specifications and Calibrations for $\sigma_\epsilon$} \label{fig:impact_sensitivity}
    \caption*{\normalsize 
        This is an implementation of \autoref{fig:passthru_lossav} with adjustments to the minimum distance procedure (top panels) and different calibrations for the amenity scale parameter $\sigma_\epsilon$ (bottom panels), which is discussed in \autoref{subsec:pass_through}. Left panels are pass-through for all job seekers, and middle/right panels are pass-through for inframarginal and marginal job seekers (respectively). ``Inframarginal'' refers to job seekers who would accept salary offers regardless of the subsidy (offer is already high enough to be accepted without the subsidy). ``Marginal'' refers to job seekers who would only accept their salary offer if augmented with the subsidy (offer is only high enough with the subsidy). In top panels, Bar 1 is the primary specification in \autoref{subsec:mindist_procedure}. Bar 2 replaces kernel estimates with raw proportions in the data. Bar 3 parameterizes unobserved heterogeneity with the normal distribution. Bar 4 uses optimal weights for bins. Bar 5 expands the range of salary growth from $\pm 0.2$ to $\pm 1.0$ log points.
    }
\end{figure}
\clearpage

\restoregeometry

%% file: 0_tables_apx.tex
\section*{Appendix Tables}
\addcontentsline{toc}{subsection}{Appendix Tables}

\vspace*{\fill}
\begin{table}[H]
    \caption{Sensitivity of Anomalies to Local Polynomial Degree and Bandwidth} \label{tab:anomalies_degbw}
    \begingroup
    \renewcommand{\arraystretch}{2.5}
    \makebox[\textwidth][c]{\small\input{figtable/search/table/anomalies_degbw.tex}}
    \endgroup

    \vspace{24pt}
    \caption*{\normalsize 
        This is an implementation of \autoref{tab:anomalies} with varying levels of polynomial degree and kernel bandwidth (discussed in \autoref{subsec:anomalies}). Standard errors (in parentheses) are bootstrapped with 10,000 iterations. Local-linear polynomials are degree 1, while local-quadratic polynomials are degree 2. RoT is the Rule-of-Thumb bandwidth that minimizes the weighted mean squared error. RoT bandwidths for degree 1 are 0.019 for pay cuts and 0.022 for pay raises, while RoT bandwidths for degree 2 are 0.062 for pay cuts and 0.068 for pay raises.
    }
\end{table}
\vspace*{\fill}
\clearpage

\newgeometry{top=0.1in, bottom=0.8in, left=0.8in, right=0.8in}

\vspace*{\fill}
\begin{table}[H]
    \caption{Sensitivity of Parameters to Alternative Specifications} \label{tab:parameters_adjust}
    \begingroup
    \small
    \renewcommand{\arraystretch}{2.5}
    \makebox[\textwidth][c]{\input{figtable/search/table/parameters_adjust.tex}}
    \endgroup

    \vspace{24pt}
    \caption*{\normalsize 
        This is an implementation of \autoref{tab:parameters_main} with adjustments to the minimum distance procedure (discussed in \autoref{subsec:calibrated_lossav}). Row 1 is the primary specification in \autoref{subsec:mindist_procedure}. Row 2 replaces kernel estimates with raw proportions in the data. Row 3 parameterizes unobserved heterogeneity with the normal distribution. Row 4 uses optimal weights for bins. Row 5 expands the range of salary growth bins from $\pm 0.2$ to $\pm 1.0$ log points.
    }
\end{table}
\vspace*{\fill}
\clearpage

\vspace*{\fill}
\begin{table}[H]
    \caption{Parameter Estimates and Standard Errors (including the zero bin)} \label{tab:parameters_sub}
    \begingroup
    \renewcommand{\arraystretch}{2.5}
    \makebox[\textwidth][c]{\input{figtable/search/table/parameters_sub.tex}}
    \endgroup

    \vspace{24pt}
    \caption*{\normalsize 
        This is an implementation of \autoref{tab:parameters_main} that includes the zero bin in the minimum distance procedure (discussed in \autoref{subsec:fit_anomalies}).
    }
\end{table}
\vspace*{\fill}
\clearpage

%% file: figtable/search/table/anomalies_degbw.tex
\centering
    \begin{tabular}{cc|cccc|cc}
    \toprule[2pt]
    \textbf{Degree} & \textbf{Bandwidth} & \boldmath{{$\hat{b}_2$}} & \boldmath{{$\hat{b}_0$}} & \boldmath{{$\hat{a}_0$}} & \boldmath{{$\hat{a}_2$}} & \textbf{Discontinuity} & \textbf{Curvature Break} \\ \hhline{|--------|}
    \hline \multirow{6}{*}{1} & \shortstack{0.010\\ \ } & \shortstack{0.5855\\(0.0037)} & \shortstack{0.5946\\(0.0045)} & \shortstack{0.6385\\(0.0047)} & \shortstack{0.6392\\(0.0039)} & \shortstack{0.0439\\(0.0066)} & \shortstack{0.0084\\(0.0016)} \\
  & \shortstack{0.015\\ \ } & \shortstack{0.5816\\(0.0031)} & \shortstack{0.5929\\(0.0035)} & \shortstack{0.6334\\(0.0037)} & \shortstack{0.6342\\(0.0033)} & \shortstack{0.0404\\(0.0052)} & \shortstack{0.0105\\(0.0009)} \\
  & \shortstack{0.020\\ \ } & \shortstack{0.5767\\(0.0027)} & \shortstack{0.5871\\(0.0030)} & \shortstack{0.6386\\(0.0032)} & \shortstack{0.6402\\(0.0029)} & \shortstack{0.0515\\(0.0044)} & \shortstack{0.0088\\(0.0006)} \\
  & \shortstack{RoT\\ \ } & \shortstack{0.5772\\(0.0027)} & \shortstack{0.5878\\(0.0030)} & \shortstack{0.6433\\(0.0028)} & \shortstack{0.6445\\(0.0025)} & \shortstack{0.0554\\(0.0042)} & \shortstack{0.0094\\(0.0005)} \\
  & \shortstack{0.025\\ \ } & \shortstack{0.5731\\(0.0024)} & \shortstack{0.5828\\(0.0026)} & \shortstack{0.6483\\(0.0028)} & \shortstack{0.6481\\(0.0026)} & \shortstack{0.0655\\(0.0039)} & \shortstack{0.0099\\(0.0004)} \\
  & \shortstack{0.030\\ \ } & \shortstack{0.5697\\(0.0022)} & \shortstack{0.5791\\(0.0023)} & \shortstack{0.6561\\(0.0025)} & \shortstack{0.6549\\(0.0024)} & \shortstack{0.0770\\(0.0035)} & \shortstack{0.0106\\(0.0003)} \\
\hline \multirow{6}{*}{2} & \shortstack{0.055\\ \ } & \shortstack{0.5778\\(0.0024)} & \shortstack{0.5884\\(0.0026)} & \shortstack{0.6481\\(0.0028)} & \shortstack{0.6485\\(0.0026)} & \shortstack{0.0596\\(0.0038)} & \shortstack{0.0102\\(0.0004)} \\
  & \shortstack{0.060\\ \ } & \shortstack{0.5769\\(0.0022)} & \shortstack{0.5873\\(0.0024)} & \shortstack{0.6530\\(0.0026)} & \shortstack{0.6529\\(0.0025)} & \shortstack{0.0657\\(0.0037)} & \shortstack{0.0105\\(0.0003)} \\
  & \shortstack{0.065\\ \ } & \shortstack{0.5762\\(0.0021)} & \shortstack{0.5865\\(0.0023)} & \shortstack{0.6583\\(0.0025)} & \shortstack{0.6575\\(0.0024)} & \shortstack{0.0719\\(0.0035)} & \shortstack{0.0111\\(0.0003)} \\
  & \shortstack{RoT\\ \ } & \shortstack{0.5766\\(0.0022)} & \shortstack{0.5871\\(0.0024)} & \shortstack{0.6611\\(0.0026)} & \shortstack{0.6599\\(0.0025)} & \shortstack{0.0740\\(0.0037)} & \shortstack{0.0116\\(0.0003)} \\
  & \shortstack{0.070\\ \ } & \shortstack{0.5754\\(0.0021)} & \shortstack{0.5857\\(0.0022)} & \shortstack{0.6637\\(0.0024)} & \shortstack{0.6622\\(0.0023)} & \shortstack{0.0780\\(0.0033)} & \shortstack{0.0117\\(0.0002)} \\
  & \shortstack{0.075\\ \ } & \shortstack{0.5744\\(0.0020)} & \shortstack{0.5846\\(0.0021)} & \shortstack{0.6686\\(0.0023)} & \shortstack{0.6665\\(0.0022)} & \shortstack{0.0840\\(0.0032)} & \shortstack{0.0123\\(0.0002)} \\
\bottomrule[2pt]
\end{tabular}

%% file: figtable/search/table/parameters_adjust.tex
\begin{tabular}{lccc}
\toprule[2pt]
\textbf{Adjustment} & \textbf{\boldmath{{$\lambda$}} (Loss Aversion)} & \textbf{\boldmath{{$\mu_\phi$}} (Productivity Location)} & \textbf{\boldmath{{$\sigma_\phi$}} (Productivity Scale)} \\
\midrule
Main Estimate & \makecell[c]{1.1235 \\ (0.0004)} & \makecell[c]{1.1953 \\ (0.0003)} & \makecell[c]{0.1594 \\ (0.0003)} \\
Smoothed $\rightarrow$ Raw Proportions & \makecell[c]{1.1243 \\ (0.0004)} & \makecell[c]{1.1968 \\ (0.0003)} & \makecell[c]{0.1570 \\ (0.0002)} \\
Logistic $\rightarrow$ Normal Distribution & \makecell[c]{1.1314 \\ (0.0003)} & \makecell[c]{1.3540 \\ (0.0003)} & \makecell[c]{0.2557 \\ (0.0004)} \\
Identity $\rightarrow$ Optimal Weights & \makecell[c]{1.1116 \\ (0.0003)} & \makecell[c]{1.2013 \\ (0.0003)} & \makecell[c]{0.1687 \\ (0.0002)} \\
Center Range $\rightarrow$ Full Range & \makecell[c]{1.1075 \\ (0.0003)} & \makecell[c]{1.2048 \\ (0.0003)} & \makecell[c]{0.1711 \\ (0.0002)} \\
\bottomrule[2pt]
\end{tabular}

%% file: figtable/search/table/parameters_sub.tex
\begin{tabular}{lcc}
\toprule[2pt]
\textbf{Parameter} & \textbf{Standard} & \textbf{Behavioral} \\
\midrule
$\lambda$: Loss Aversion & $\cdot$ & \makecell[c]{1.0338 \\ (0.0001)} \\
$\mu_\phi$: Productivity Location & \makecell[c]{1.2449 \\ (0.0003)} & \makecell[c]{1.2355 \\ (0.0003)} \\
$\sigma_\phi$: Productivity Scale & \makecell[c]{0.1515 \\ (0.0002)} & \makecell[c]{0.1773 \\ (0.0002)} \\
QLR Test & \multicolumn{2}{c}{\makecell[c]{8280.80 \\(CV: 3.84)}}  \\
\bottomrule[2pt]
\end{tabular}

%% file: references.bib
@article{burdett_wage_1998,
	title = {Wage Differentials, Employer Size, and Unemployment},
	volume = {39},
	issn = {00206598},
	url = {https://www.jstor.org/stable/2527292?origin=crossref},
	doi = {10.2307/2527292},
	pages = {257},
	number = {2},
	journaltitle = {International Economic Review},
	shortjournal = {International Economic Review},
	author = {Burdett, Kenneth and Mortensen, Dale T.},
	urldate = {2024-05-22},
	date = {1998-05},
	langid = {english},
}

@article{postelvinay_distribution_2002,
	title = {The Distribution of Earnings in an Equilibrium Search Model with State-Dependent Offers and Counteroffers},
	volume = {43},
	rights = {http://onlinelibrary.wiley.com/{termsAndConditions}\#vor},
	issn = {0020-6598, 1468-2354},
	url = {https://onlinelibrary.wiley.com/doi/10.1111/1468-2354.t01-1-00045},
	doi = {10.1111/1468-2354.t01-1-00045},
	pages = {989--1016},
	number = {4},
	journaltitle = {International Economic Review},
	shortjournal = {Int Economic Review},
	author = {Postel-Vinay, Fabien and Robin, Jean-Marc},
	urldate = {2024-05-22},
	date = {2002-11},
	langid = {english},
}

@article{fongoni_why_nodate,
	title = {Why Wages Don’t Fall in Jobs with Incomplete Contracts},
	year = {forthcoming},
	journaltitle = {Management Science},
	author = {Fongoni, Marco and Schaefer, Daniel and Singleton, Carl},
	langid = {english},
}

@article{benjamin_theory_2015,
	title = {A Theory of Fairness in Labour Markets},
	volume = {66},
	rights = {http://doi.wiley.com/10.1002/tdm\_license\_1.1},
	issn = {13524739},
	url = {http://doi.wiley.com/10.1111/jere.12069},
	doi = {10.1111/jere.12069},
	shorttitle = {A Theory of Fairness in Labour Markets},
	pages = {182--225},
	number = {2},
	journaltitle = {Japanese Economic Review},
	shortjournal = {Japanese Economic Review},
	author = {Benjamin, Daniel J.},
	urldate = {2025-03-04},
	date = {2015-06},
	langid = {english},
}

@article{ahrens_theory_2014,
	title = {A Theory of Wage Adjustment under Loss Aversion},
	volume = {No. 8699},
	journaltitle = {{IZA} Discussion Paper},
	author = {Ahrens, Steffen and Pirschel, Inske and Snower, Dennis},
	date = {2014},
	langid = {english},
}

@article{santana_theory_2024,
	title = {A Theory of Downward Wage Rigidity},
	volume = {Northwestern University},
	journaltitle = {Job Market Paper},
	author = {Santana, Miguel},
	date = {2024},
	langid = {english},
}

@article{koch_can_2021,
	title = {Can Reference Points Explain Wage Rigidity? Experimental Evidence},
	volume = {55},
	issn = {2510-5019, 2510-5027},
	url = {https://labourmarketresearch.springeropen.com/articles/10.1186/s12651-021-00284-2},
	doi = {10.1186/s12651-021-00284-2},
	shorttitle = {Can reference points explain wage rigidity?},
	pages = {5},
	number = {1},
	journaltitle = {Journal for Labour Market Research},
	shortjournal = {J Labour Market Res},
	author = {Koch, Christian},
	urldate = {2025-03-10},
	date = {2021-12},
	langid = {english},
}

@article{hazell_downward_2024,
	title = {Downward Rigidity in the Wage for New Hires},
	journaltitle = {American Economic Review},
	author = {Hazell, Jonathon and Taska, Bledi},
	date = {2024},
	langid = {english},
}

@article{grigsby_aggregate_2021,
	title = {Aggregate Nominal Wage Adjustments: New Evidence from Administrative Payroll Data},
	volume = {111},
	issn = {0002-8282},
	url = {https://pubs.aeaweb.org/doi/10.1257/aer.20190318},
	doi = {10.1257/aer.20190318},
	shorttitle = {Aggregate Nominal Wage Adjustments},
	pages = {428--471},
	number = {2},
	journaltitle = {American Economic Review},
	shortjournal = {American Economic Review},
	author = {Grigsby, John and Hurst, Erik and Yildirmaz, Ahu},
	urldate = {2025-03-10},
	date = {2021-02-01},
	langid = {english},
}

@article{kaur_nominal_2019,
	title = {Nominal Wage Rigidity in Village Labor Markets},
	volume = {109},
	issn = {0002-8282},
	url = {https://pubs.aeaweb.org/doi/10.1257/aer.20141625},
	doi = {10.1257/aer.20141625},
	pages = {3585--3616},
	number = {10},
	journaltitle = {American Economic Review},
	shortjournal = {American Economic Review},
	author = {Kaur, Supreet},
	urldate = {2025-03-10},
	date = {2019-10-01},
	langid = {english},
}

@article{breza_morale_2018,
	title = {The Morale Effects of Pay Inequality},
	volume = {133},
	issn = {0033-5533, 1531-4650},
	url = {https://academic.oup.com/qje/article/133/2/611/4430649},
	doi = {10.1093/qje/qjx041},
	pages = {611--663},
	number = {2},
	journaltitle = {The Quarterly Journal of Economics},
	author = {Breza, Emily and Kaur, Supreet and Shamdasani, Yogita},
	urldate = {2025-03-10},
	date = {2018-05-01},
	langid = {english},
}

@article{campbell_reasons_1997,
	title = {The Reasons for Wage Rigidity: Evidence from a Survey of Firms},
	volume = {112},
	issn = {0033-5533, 1531-4650},
	url = {https://academic.oup.com/qje/article-lookup/doi/10.1162/003355397555343},
	doi = {10.1162/003355397555343},
	shorttitle = {The Reasons for Wage Rigidity},
	pages = {759--789},
	number = {3},
	journaltitle = {The Quarterly Journal of Economics},
	shortjournal = {The Quarterly Journal of Economics},
	author = {Campbell, C. M. and Kamlani, K. S.},
	urldate = {2025-03-10},
	date = {1997-08-01},
	langid = {english},
}

@article{card_inequality_2012,
	title = {Inequality at Work: The Effect of Peer Salaries on Job Satisfaction},
	volume = {102},
	issn = {0002-8282},
	url = {https://pubs.aeaweb.org/doi/10.1257/aer.102.6.2981},
	doi = {10.1257/aer.102.6.2981},
	shorttitle = {Inequality at Work},
	pages = {2981--3003},
	number = {6},
	journaltitle = {American Economic Review},
	shortjournal = {American Economic Review},
	author = {Card, David and Mas, Alexandre and Moretti, Enrico and Saez, Emmanuel},
	urldate = {2025-03-10},
	date = {2012-10-01},
	langid = {english},
}

@article{goette_wage_2007,
	title = {Wage Rigidity: Measurement, Causes and Consequences},
	volume = {117},
	rights = {http://doi.wiley.com/10.1002/tdm\_license\_1.1},
	issn = {0013-0133, 1468-0297},
	url = {https://academic.oup.com/ej/article/117/524/F499-F507/5086540},
	doi = {10.1111/j.1468-0297.2007.02093.x},
	shorttitle = {Wage Rigidity},
	pages = {F499--F507},
	number = {524},
	journaltitle = {The Economic Journal},
	author = {Goette, Lorenz and Sunde, Uwe and Bauer, Thomas},
	urldate = {2025-03-10},
	date = {2007-11-01},
	langid = {english},
}

@article{kleven_bunching_2016,
	title = {Bunching},
	volume = {8},
	issn = {1941-1383, 1941-1391},
	url = {https://www.annualreviews.org/doi/10.1146/annurev-economics-080315-015234},
	doi = {10.1146/annurev-economics-080315-015234},
	pages = {435--464},
	number = {1},
	journaltitle = {Annual Review of Economics},
	shortjournal = {Annu. Rev. Econ.},
	author = {Kleven, Henrik Jacobsen},
	urldate = {2025-03-10},
	date = {2016-10-31},
	langid = {english},
}

@article{bhaskar_wage_1990,
	title = {Wage Relativities and the Natural Range of Unemployment},
	volume = {100},
	issn = {00130133},
	url = {https://academic.oup.com/ej/article/100/400/60-66/5190050},
	doi = {10.2307/2234184},
	pages = {60},
	number = {400},
	journaltitle = {The Economic Journal},
	shortjournal = {The Economic Journal},
	author = {Bhaskar, V.},
	urldate = {2025-03-10},
	date = {1990},
	langid = {english},
}

@article{mcdonald_how_2001,
	title = {How Monetary Policy can have Permanent Real Effects with Only Temporary Nominal Rigidity},
	volume = {48},
	issn = {0036-9292, 1467-9485},
	url = {https://onlinelibrary.wiley.com/doi/10.1111/1467-9485.00213},
	doi = {10.1111/1467-9485.00213},
	pages = {532--546},
	number = {5},
	journaltitle = {Scottish Journal of Political Economy},
	shortjournal = {Scottish J Political Eco},
	author = {{McDonald}, Ian M. and Scully, Hugh},
	urldate = {2025-03-10},
	date = {2001-11},
	langid = {english},
}

@article{saez_taxpayers_2010,
	title = {Do Taxpayers Bunch at Kink Points?},
	volume = {2},
	issn = {1945-7731, 1945-774X},
	url = {https://pubs.aeaweb.org/doi/10.1257/pol.2.3.180},
	doi = {10.1257/pol.2.3.180},
	pages = {180--212},
	number = {3},
	journaltitle = {American Economic Journal: Economic Policy},
	shortjournal = {American Economic Journal: Economic Policy},
	author = {Saez, Emmanuel},
	urldate = {2025-03-10},
	date = {2010-08-01},
	langid = {english},
}

@article{lehmann_non-wage_2025,
	title = {Non-Wage Job Values and Implications for Inequality},
	volume = {16663},
	journaltitle = {{IZA} Discussion Paper},
	author = {Lehmann, Tobias},
	date = {2025},
	langid = {english},
}

@incollection{zimmermann_behavioral_2020,
	location = {Cham},
	title = {Behavioral Job Search},
	isbn = {978-3-319-57365-6},
	url = {http://link.springer.com/10.1007/978-3-319-57365-6_116-1},
	pages = {1--22},
	booktitle = {Handbook of Labor, Human Resources and Population Economics},
	publisher = {Springer International Publishing},
	author = {Cooper, Michael and Kuhn, Peter},
	editor = {Zimmermann, Klaus F.},
	urldate = {2025-03-10},
	date = {2020},
	langid = {english},
	doi = {10.1007/978-3-319-57365-6_116-1},
}

@article{dellavigna_reference-dependent_2017,
	title = {Reference-Dependent Job Search: Evidence from Hungary},
	volume = {132},
	issn = {0033-5533, 1531-4650},
	url = {https://academic.oup.com/qje/article/132/4/1969/3796325},
	doi = {10.1093/qje/qjx015},
	shorttitle = {Reference-Dependent Job Search},
	pages = {1969--2018},
	number = {4},
	journaltitle = {The Quarterly Journal of Economics},
	author = {{DellaVigna}, Stefano and Lindner, Attila and Reizer, Balázs and Schmieder, Johannes F.},
	urldate = {2025-03-10},
	date = {2017-11-01},
	langid = {english},
}

@article{andersen_reference_2022,
	title = {Reference Dependence in the Housing Market},
	volume = {112},
	issn = {0002-8282},
	url = {https://pubs.aeaweb.org/doi/10.1257/aer.20191766},
	doi = {10.1257/aer.20191766},
	pages = {3398--3440},
	number = {10},
	journaltitle = {American Economic Review},
	shortjournal = {American Economic Review},
	author = {Andersen, Steffen and Badarinza, Cristian and Liu, Lu and Marx, Julie and Ramadorai, Tarun},
	urldate = {2025-03-10},
	date = {2022-10-01},
	langid = {english},
}

@article{rees-jones_quantifying_2018,
	title = {Quantifying Loss-Averse Tax Manipulation},
	volume = {85},
	issn = {0034-6527, 1467-937X},
	url = {https://academic.oup.com/restud/article/85/2/1251/3897019},
	doi = {10.1093/restud/rdx038},
	pages = {1251--1278},
	number = {2},
	journaltitle = {The Review of Economic Studies},
	author = {Rees-Jones, Alex},
	urldate = {2025-03-10},
	date = {2018-04-01},
	langid = {english},
}

@article{shafir_money_1997,
	title = {Money Illusion},
	volume = {112},
	pages = {pp. 341--374},
	number = {2},
	journaltitle = {The Quarterly Journal of Economics},
	author = {Shafir, Eldar and Diamond, Peter and Tversky, Amos},
	date = {1997},
}

@article{boheim_great_2011,
	title = {Great Expectations: Past Wages and Unemployment Durations},
	volume = {18},
	rights = {https://www.elsevier.com/tdm/userlicense/1.0/},
	issn = {09275371},
	url = {https://linkinghub.elsevier.com/retrieve/pii/S0927537111000704},
	doi = {10.1016/j.labeco.2011.06.009},
	shorttitle = {Great expectations},
	pages = {778--785},
	number = {6},
	journaltitle = {Labour Economics},
	shortjournal = {Labour Economics},
	author = {Böheim, Renè and Horvath, Gerard Thomas and Winter-Ebmer, Rudolf},
	urldate = {2025-03-10},
	date = {2011-12},
	langid = {english},
}

@article{cahuc_wage_2006,
	title = {Wage Bargaining with On-the-Job Search: Theory and Evidence},
	volume = {74},
	rights = {http://doi.wiley.com/10.1002/tdm\_license\_1.1},
	issn = {0012-9682, 1468-0262},
	url = {http://doi.wiley.com/10.1111/j.1468-0262.2006.00665.x},
	doi = {10.1111/j.1468-0262.2006.00665.x},
	shorttitle = {Wage Bargaining with On-the-Job Search},
	pages = {323--364},
	number = {2},
	journaltitle = {Econometrica},
	shortjournal = {Econometrica},
	author = {Cahuc, Pierre and Postel-Vinay, Fabien and Robin, Jean-Marc},
	urldate = {2025-03-10},
	date = {2006-03},
	langid = {english},
}

@article{jones_loss_2020,
	title = {Loss Aversion and Property Tax Avoidance},
	volume = {No. 3511751},
	issn = {1556-5068},
	url = {https://www.ssrn.com/abstract=3511751},
	doi = {10.2139/ssrn.3511751},
	journaltitle = {SSRN Working Paper},
	author = {Jones, Peter},
	urldate = {2025-03-10},
	date = {2020},
	langid = {english},
}

@article{ehrlich_wage_2024,
	title = {Wage Rigidity and Employment Outcomes: Evidence from Administrative Data},
	volume = {16},
	issn = {1945-7707, 1945-7715},
	url = {https://pubs.aeaweb.org/doi/10.1257/mac.20200125},
	doi = {10.1257/mac.20200125},
	shorttitle = {Wage Rigidity and Employment Outcomes},
	pages = {147--206},
	number = {1},
	journaltitle = {American Economic Journal: Macroeconomics},
	shortjournal = {American Economic Journal: Macroeconomics},
	author = {Ehrlich, Gabriel and Montes, Joshua},
	urldate = {2025-03-10},
	date = {2024-01-01},
	langid = {english},
}

@article{dellavigna_evidence_2022,
	title = {Evidence on Job Search Models from a Survey of Unemployed Workers in Germany},
	volume = {137},
	rights = {https://academic.oup.com/journals/pages/open\_access/funder\_policies/chorus/standard\_publication\_model},
	issn = {0033-5533, 1531-4650},
	url = {https://academic.oup.com/qje/article/137/2/1181/6413570},
	doi = {10.1093/qje/qjab039},
	pages = {1181--1232},
	number = {2},
	journaltitle = {The Quarterly Journal of Economics},
	author = {{DellaVigna}, Stefano and Heining, Jörg and Schmieder, Johannes F and Trenkle, Simon},
	urldate = {2025-03-10},
	date = {2022-04-08},
	langid = {english},
}

@article{barbanchon_gender_2020,
	title = {Gender Differences in Job Search: Trading Off Commute Against Wage},
	volume = {136},
	rights = {http://creativecommons.org/licenses/by-nc-nd/4.0/},
	issn = {0033-5533, 1531-4650},
	url = {https://academic.oup.com/qje/article/136/1/381/5928590},
	doi = {10.1093/qje/qjaa033},
	shorttitle = {Gender Differences in Job Search},
	pages = {381--426},
	number = {1},
	journaltitle = {The Quarterly Journal of Economics},
	author = {Barbanchon, Thomas Le and Rathelot, Roland and Roulet, Alexandra},
	urldate = {2025-03-10},
	date = {2020-12-22},
	langid = {english},
}

@article{blinder_shred_1990,
	title = {A Shred of Evidence on Theories of Wage Stickiness},
	volume = {105},
	issn = {00335533},
	url = {https://academic.oup.com/qje/article-lookup/doi/10.2307/2937882},
	doi = {10.2307/2937882},
	pages = {1003},
	number = {4},
	journaltitle = {The Quarterly Journal of Economics},
	shortjournal = {The Quarterly Journal of Economics},
	author = {Blinder, Alan S. and Choi, Don H.},
	urldate = {2025-03-10},
	date = {1990-11},
}

@article{roussille_bidding_2024,
	title = {Bidding for Talent: A Test of Conduct in a High-Wage Labor Market},
	volume = {No. 16352},
	journaltitle = {{IZA} Discussion Paper},
	author = {Roussille, Nina and Scuderi, Benjamin},
	date = {2024},
}

@article{akerlof_labor_1982,
	title = {Labor Contracts as Partial Gift Exchange},
	volume = {97},
	issn = {00335533},
	url = {https://academic.oup.com/qje/article-lookup/doi/10.2307/1885099},
	doi = {10.2307/1885099},
	pages = {543},
	number = {4},
	journaltitle = {The Quarterly Journal of Economics},
	shortjournal = {The Quarterly Journal of Economics},
	author = {Akerlof, George A.},
	urldate = {2025-03-10},
	date = {1982-11},
}

@article{akerlof_fair_1990,
	title = {The Fair Wage-Effort Hypothesis and Unemployment},
	volume = {105},
	issn = {00335533},
	url = {https://academic.oup.com/qje/article-lookup/doi/10.2307/2937787},
	doi = {10.2307/2937787},
	pages = {255},
	number = {2},
	journaltitle = {The Quarterly Journal of Economics},
	shortjournal = {The Quarterly Journal of Economics},
	author = {Akerlof, George A. and Yellen, Janet L.},
	urldate = {2025-03-10},
	date = {1990-05},
}

@article{dube_monopsony_2020,
	title = {Monopsony and Employer Mis-optimization Explain Why Wages Bunch at Round Numbers},
	volume = {No. 24991},
	url = {http://www.nber.org/papers/w24991.pdf},
	journaltitle = {{NBER} Working Paper},
	author = {Dube, Arindrajit and Manning, Alan and Naidu, Suresh},
	urldate = {2025-03-10},
	date = {2020},
	langid = {english},
	doi = {10.3386/w24991},
}

@article{brown_meta-analysis_2024,
	title = {Meta-analysis of Empirical Estimates of Loss Aversion},
	volume = {62},
	issn = {0022-0515, 2328-8175},
	url = {https://pubs.aeaweb.org/doi/10.1257/jel.20221698},
	doi = {10.1257/jel.20221698},
	pages = {485--516},
	number = {2},
	journaltitle = {Journal of Economic Literature},
	shortjournal = {Journal of Economic Literature},
	author = {Brown, Alexander L. and Imai, Taisuke and Vieider, Ferdinand M. and Camerer, Colin F.},
	urldate = {2025-03-10},
	date = {2024-06-01},
	langid = {english},
}

@article{quach_wage_2025,
	title = {Wage Hysteresis and Entitlement Effects: The Persistent Impacts of a Temporary Overtime Policy},
	rights = {https://academic.oup.com/journals/pages/open\_access/funder\_policies/chorus/standard\_publication\_model},
	issn = {0033-5533, 1531-4650},
	url = {https://academic.oup.com/qje/advance-article/doi/10.1093/qje/qjaf008/7994410},
	doi = {10.1093/qje/qjaf008},
	shorttitle = {Wage Hysteresis and Entitlement Effects},
	pages = {qjaf008},
	journaltitle = {The Quarterly Journal of Economics},
	author = {Quach, Simon},
	urldate = {2025-03-10},
	date = {2025-01-31},
	langid = {english},
}

@article{krueger_contribution_2016,
	title = {A Contribution to the Empirics of Reservation Wages},
	volume = {8},
	issn = {1945-7731, 1945-774X},
	url = {https://pubs.aeaweb.org/doi/10.1257/pol.20140211},
	doi = {10.1257/pol.20140211},
	pages = {142--179},
	number = {1},
	journaltitle = {American Economic Journal: Economic Policy},
	shortjournal = {American Economic Journal: Economic Policy},
	author = {Krueger, Alan B. and Mueller, Andreas I.},
	urldate = {2025-03-10},
	date = {2016-02-01},
	langid = {english},
}

@article{faia_cost_2024,
	title = {The Cost of Wage Rigidity},
	volume = {91},
	rights = {https://creativecommons.org/licenses/by/4.0/},
	issn = {0034-6527, 1467-937X},
	url = {https://academic.oup.com/restud/article/91/1/301/7041131},
	doi = {10.1093/restud/rdad020},
	pages = {301--339},
	number = {1},
	journaltitle = {Review of Economic Studies},
	author = {Faia, Ester and Pezone, Vincenzo},
	urldate = {2025-03-10},
	date = {2024-01-12},
	langid = {english},
}

@article{card_does_1997,
	title = {Does Inflation 'Grease the Wheels of the Labor Market'?},
	pages = {p. 71 -- 122},
	journaltitle = {Reducing Inflation: Motivation and Strategy ({NBER})},
	author = {Card, David and Hyslop, Dean},
	date = {1997},
	langid = {english},
}

@article{nickell_nominal_2003,
	title = {Nominal Wage Rigidity and the Rate of Inflation},
	volume = {113},
	rights = {http://doi.wiley.com/10.1002/tdm\_license\_1.1},
	issn = {0013-0133, 1468-0297},
	url = {https://academic.oup.com/ej/article/113/490/762-781/5079618},
	doi = {10.1111/1468-0297.t01-1-00161},
	pages = {762--781},
	number = {490},
	journaltitle = {The Economic Journal},
	author = {Nickell, Stephen and Quintini, Glenda},
	urldate = {2025-03-13},
	date = {2003-10-01},
	langid = {english},
}

@article{hart_contracts_2008,
	title = {Contracts as Reference Points},
	volume = {123},
	issn = {0033-5533, 1531-4650},
	url = {https://academic.oup.com/qje/article-lookup/doi/10.1162/qjec.2008.123.1.1},
	doi = {10.1162/qjec.2008.123.1.1},
	pages = {1--48},
	number = {1},
	journaltitle = {Quarterly Journal of Economics},
	shortjournal = {Quarterly Journal of Economics},
	author = {Hart, Oliver and Moore, John},
	urldate = {2025-03-13},
	date = {2008-02},
	langid = {english},
}

@article{farber_is_2005,
	title = {Is Tomorrow Another Day? The Labor Supply of New York City Cabdrivers},
	volume = {113},
	issn = {0022-3808, 1537-534X},
	url = {https://www.journals.uchicago.edu/doi/10.1086/426040},
	doi = {10.1086/426040},
	shorttitle = {Is Tomorrow Another Day?},
	pages = {46--82},
	number = {1},
	journaltitle = {Journal of Political Economy},
	shortjournal = {Journal of Political Economy},
	author = {Farber, Henry S.},
	urldate = {2025-03-13},
	date = {2005-02},
	langid = {english},
}

@article{farber_why_2015,
	title = {Why you Can’t Find a Taxi in the Rain and Other Labor Supply Lessons from Cab Drivers},
	volume = {130},
	issn = {1531-4650, 0033-5533},
	url = {https://academic.oup.com/qje/article/130/4/1975/1916077},
	doi = {10.1093/qje/qjv026},
	pages = {1975--2026},
	number = {4},
	journaltitle = {The Quarterly Journal of Economics},
	author = {Farber, Henry S.},
	urldate = {2025-03-13},
	date = {2015-11-01},
	langid = {english},
}

@article{crawford_new_2011,
	title = {New York City Cab Drivers' Labor Supply Revisited: Reference-Dependent Preferences with Rational-Expectations Targets for Hours and Income},
	volume = {101},
	issn = {0002-8282},
	url = {https://pubs.aeaweb.org/doi/10.1257/aer.101.5.1912},
	doi = {10.1257/aer.101.5.1912},
	shorttitle = {New York City Cab Drivers' Labor Supply Revisited},
	pages = {1912--1932},
	number = {5},
	journaltitle = {American Economic Review},
	shortjournal = {American Economic Review},
	author = {Crawford, Vincent P and Meng, Juanjuan},
	urldate = {2025-03-13},
	date = {2011-08-01},
	langid = {english},
}

@article{thakral_daily_2021,
	title = {Daily Labor Supply and Adaptive Reference Points},
	volume = {111},
	issn = {0002-8282},
	url = {https://pubs.aeaweb.org/doi/10.1257/aer.20170768},
	doi = {10.1257/aer.20170768},
	pages = {2417--2443},
	number = {8},
	journaltitle = {American Economic Review},
	shortjournal = {American Economic Review},
	author = {Thakral, Neil and Tô, Linh T.},
	urldate = {2025-03-13},
	date = {2021-08-01},
	langid = {english},
}

@article{dickens_how_2007,
	title = {How Wages Change: Micro Evidence from the International Wage Flexibility Project},
	volume = {21},
	issn = {0895-3309},
	url = {https://pubs.aeaweb.org/doi/10.1257/jep.21.2.195},
	doi = {10.1257/jep.21.2.195},
	shorttitle = {How Wages Change},
	pages = {195--214},
	number = {2},
	journaltitle = {Journal of Economic Perspectives},
	shortjournal = {Journal of Economic Perspectives},
	author = {Dickens, William T and Goette, Lorenz and Groshen, Erica L and Holden, Steinar and Messina, Julian and Schweitzer, Mark E and Turunen, Jarkko and Ward, Melanie E},
	urldate = {2025-03-13},
	date = {2007-04-01},
	langid = {english},
}

@article{kahn_evidence_1997,
	title = {Evidence of Nominal Wage Stickiness from Microdata},
	volume = {87},
	issn = {00028282},
	url = {http://www.jstor.org/stable/2951337},
	pages = {993--1008},
	number = {5},
	journaltitle = {The American Economic Review},
	author = {Kahn, Shulamit},
	urldate = {2025-03-13},
	date = {1997},
}

@article{shapiro_equilibrium_1984,
	title = {Equilibrium Unemployment as a Worker Discipline Device},
	volume = {74},
	issn = {00028282, 19447981},
	url = {http://www.jstor.org/stable/1804018},
	pages = {433--444},
	number = {3},
	journaltitle = {The American Economic Review},
	author = {Shapiro, Carl and Stiglitz, Joseph E.},
	urldate = {2025-03-13},
	date = {1984},
}

@article{kahneman_fairness_1986,
	title = {Fairness as a Constraint on Profit Seeking: Entitlements in the Market},
	volume = {76},
	issn = {00028282},
	url = {http://www.jstor.org/stable/1806070},
	pages = {728--741},
	number = {4},
	journaltitle = {The American Economic Review},
	author = {Kahneman, Daniel and Knetsch, Jack L. and Thaler, Richard},
	urldate = {2025-03-13},
	date = {1986},
}

@book{bewley_why_1999,
	title = {Why Wages Don't Fall during a Recession},
	isbn = {978-0-674-95241-6},
	url = {http://www.jstor.org/stable/j.ctv1pncnkx},
	publisher = {Harvard University Press},
	author = {Bewley, Truman F.},
	urldate = {2025-03-13},
	date = {1999},
	doi = {10.2307/j.ctv1pncnkx},
}

@article{kujansuu_fairness_2024,
	title = {The Fairness of Wage Cuts},
	volume = {No. 4879770},
	url = {https://www.ssrn.com/abstract=4879770},
	doi = {10.2139/ssrn.4879770},
	journaltitle = {Social Science Research Network ({SSRN})},
	author = {Kujansuu, Essi},
	urldate = {2025-03-13},
	date = {2024},
}

@article{mas_pay_2006,
	title = {Pay, Reference Points, and Police Performance},
	volume = {121},
	issn = {0033-5533},
	url = {https://doi.org/10.1162/qjec.121.3.783},
	doi = {10.1162/qjec.121.3.783},
	pages = {783--821},
	number = {3},
	journaltitle = {The Quarterly Journal of Economics},
	shortjournal = {The Quarterly Journal of Economics},
	author = {Mas, Alexandre},
	urldate = {2025-03-13},
	date = {2006-08-01},
}

@article{cohn_social_2014,
	title = {Social Comparison and Effort Provision: Evidence from a Field Experiment},
	volume = {12},
	issn = {1542-4766},
	url = {https://doi.org/10.1111/jeea.12079},
	doi = {10.1111/jeea.12079},
	pages = {877--898},
	number = {4},
	journaltitle = {Journal of the European Economic Association},
	shortjournal = {Journal of the European Economic Association},
	author = {Cohn, Alain and Fehr, Ernst and Herrmann, Benedikt and Schneider, Frédéric},
	urldate = {2025-03-13},
	date = {2014-08-01},
}

@article{eliaz_reference_2014,
	title = {Reference Dependence and Labor Market Fluctuations},
	volume = {28},
	issn = {08893365, 15372642},
	url = {http://www.jstor.org/stable/10.1086/674596},
	doi = {10.1086/674596},
	pages = {159--200},
	number = {1},
	journaltitle = {{NBER} Macroeconomics Annual},
	author = {Eliaz, Kfir and Spiegler, Ran},
	urldate = {2025-03-13},
	date = {2014},
}

@article{doerrenberg_asymmetric_2023,
	title = {Asymmetric Labor-Supply Responses to Wage Changes: Experimental Evidence from an Online Labor Market},
	volume = {81},
	issn = {0927-5371},
	url = {https://www.sciencedirect.com/science/article/pii/S0927537122001956},
	doi = {10.1016/j.labeco.2022.102305},
	pages = {102305},
	journaltitle = {Labour Economics},
	shortjournal = {Labour Economics},
	author = {Doerrenberg, Philipp and Duncan, Denvil and Löffler, Max},
	date = {2023-04-01},
}

@article{reyes_coarse_2024,
	title = {Coarse Wage-Setting and Behavioral Firms},
	issn = {0034-6535},
	url = {https://doi.org/10.1162/rest_a_01470},
	doi = {10.1162/rest_a_01470},
	pages = {1--41},
	journaltitle = {The Review of Economics and Statistics},
	shortjournal = {The Review of Economics and Statistics},
	author = {Reyes, Germán},
	urldate = {2025-03-13},
	date = {2024-06-14},
}

@book{pissarides_equilibrium_2000,
	title = {Equilibrium Unemployment Theory},
	publisher = {{MIT} press},
	author = {Pissarides, Christopher},
	date = {2000},
}

@article{koszegi_reference-dependent_2007,
	title = {Reference-Dependent Risk Attitudes},
	volume = {97},
	issn = {00028282},
	url = {http://www.jstor.org/stable/30034084},
	pages = {1047--1073},
	number = {4},
	journaltitle = {The American Economic Review},
	author = {Kőszegi, Botond and Rabin, Matthew},
	urldate = {2025-03-14},
	date = {2007},
}

@article{koszegi_model_2006,
	title = {A Model of Reference-Dependent Preferences},
	volume = {121},
	issn = {0033-5533},
	url = {https://doi.org/10.1093/qje/121.4.1133},
	doi = {10.1093/qje/121.4.1133},
	pages = {1133--1165},
	number = {4},
	journaltitle = {The Quarterly Journal of Economics},
	shortjournal = {The Quarterly Journal of Economics},
	author = {Kőszegi, Botond and Rabin, Matthew},
	urldate = {2025-03-14},
	date = {2006-11-01},
}

@article{kahneman_prospect_1979,
	title = {Prospect Theory: An Analysis of Decision under Risk},
	volume = {47},
	issn = {00129682, 14680262},
	url = {http://www.jstor.org/stable/1914185},
	doi = {10.2307/1914185},
	pages = {263--291},
	number = {2},
	journaltitle = {Econometrica},
	author = {Kahneman, Daniel and Tversky, Amos},
	urldate = {2025-03-14},
	date = {1979},
}

@article{dhami_loss_2025,
	title = {Loss Aversion and Tax Evasion: Theory and Evidence},
	journaltitle = {Arthaniti: Journal of Economic Theory and Practice},
	author = {Dhami, Sanjit and Hajimoladarvish, Narges and Mamidi, Pavan},
	date = {2025},
	langid = {english},
}

@article{fehr_workers_2007,
	title = {Do Workers Work More if Wages Are High? Evidence from a Randomized Field Experiment},
	volume = {97},
	url = {https://www.aeaweb.org/articles?id=10.1257/aer.97.1.298},
	doi = {10.1257/aer.97.1.298},
	pages = {298--317},
	number = {1},
	journaltitle = {American Economic Review},
	shortjournal = {American Economic Review},
	author = {Fehr, Ernst and Goette, Lorenz},
	date = {2007},
}

@article{camerer_labor_1997,
	title = {Labor Supply of New York City Cabdrivers: One Day at a Time},
	volume = {112},
	issn = {0033-5533},
	url = {https://doi.org/10.1162/003355397555244},
	doi = {10.1162/003355397555244},
	pages = {407--441},
	number = {2},
	journaltitle = {The Quarterly Journal of Economics},
	shortjournal = {The Quarterly Journal of Economics},
	author = {Camerer, Colin and Babcock, Linda and Loewenstein, George and Thaler, Richard},
	urldate = {2025-03-14},
	date = {1997-05-01},
}

@article{genesove_loss_2001,
	title = {Loss Aversion and Seller Behavior: Evidence from the Housing Market},
	volume = {116},
	issn = {00335533, 15314650},
	url = {http://www.jstor.org/stable/2696458},
	pages = {1233--1260},
	number = {4},
	journaltitle = {The Quarterly Journal of Economics},
	author = {Genesove, David and Mayer, Christopher},
	urldate = {2025-03-14},
	date = {2001},
}

@article{barberis_prospect_2001,
	title = {Prospect Theory and Asset Prices},
	volume = {116},
	issn = {00335533, 15314650},
	url = {http://www.jstor.org/stable/2696442},
	pages = {1--53},
	number = {1},
	journaltitle = {The Quarterly Journal of Economics},
	author = {Barberis, Nicholas and Huang, Ming and Santos, Tano},
	urldate = {2025-03-14},
	date = {2001},
}

@article{benartzi_myopic_1995,
	title = {Myopic Loss Aversion and the Equity Premium Puzzle},
	volume = {110},
	issn = {00335533, 15314650},
	url = {http://www.jstor.org/stable/2118511},
	doi = {10.2307/2118511},
	pages = {73--92},
	number = {1},
	journaltitle = {The Quarterly Journal of Economics},
	author = {Benartzi, Shlomo and Thaler, Richard H.},
	urldate = {2025-03-14},
	date = {1995},
}

@article{baker_effect_2012,
	title = {The effect of reference point prices on mergers and acquisitions},
	volume = {106},
	issn = {0304-405X},
	url = {https://www.sciencedirect.com/science/article/pii/S0304405X12000712},
	doi = {10.1016/j.jfineco.2012.04.010},
	pages = {49--71},
	number = {1},
	journaltitle = {Journal of Financial Economics},
	shortjournal = {Journal of Financial Economics},
	author = {Baker, Malcolm and Pan, Xin and Wurgler, Jeffrey},
	date = {2012-10-01},
}

@article{pope_is_2011,
	title = {Is Tiger Woods Loss Averse? Persistent Bias in the Face of Experience, Competition, and High Stakes},
	volume = {101},
	url = {https://www.aeaweb.org/articles?id=10.1257/aer.101.1.129},
	doi = {10.1257/aer.101.1.129},
	pages = {129--57},
	number = {1},
	journaltitle = {American Economic Review},
	shortjournal = {American Economic Review},
	author = {Pope, Devin G. and Schweitzer, Maurice E.},
	date = {2011},
}

@article{allen_reference-dependent_2017,
	title = {Reference-Dependent Preferences: Evidence from Marathon Runners},
	volume = {63},
	issn = {0025-1909},
	url = {https://doi.org/10.1287/mnsc.2015.2417},
	doi = {10.1287/mnsc.2015.2417},
	pages = {1657--1672},
	number = {6},
	journaltitle = {Management Science},
	shortjournal = {Management Science},
	author = {Allen, Eric J. and Dechow, Patricia M. and Pope, Devin G. and Wu, George},
	urldate = {2025-03-14},
	date = {2017-06-01},
}

@article{fu_social_2019,
	title = {Social comparisons in job search},
	volume = {168},
	issn = {01672681},
	url = {https://linkinghub.elsevier.com/retrieve/pii/S016726811930321X},
	doi = {10.1016/j.jebo.2019.10.013},
	pages = {338--361},
	journaltitle = {Journal of Economic Behavior \& Organization},
	shortjournal = {Journal of Economic Behavior \& Organization},
	author = {Fu, Jingcheng and Sefton, Martin and Upward, Richard},
	urldate = {2025-03-17},
	date = {2019-12},
	langid = {english},
}

@article{sorkin_ranking_2018,
	title = {Ranking Firms Using Revealed Preference},
	volume = {133},
	rights = {http://academic.oup.com/journals/pages/about\_us/legal/notices},
	issn = {0033-5533, 1531-4650},
	url = {https://academic.oup.com/qje/article/133/3/1331/4813638},
	doi = {10.1093/qje/qjy001},
	pages = {1331--1393},
	number = {3},
	journaltitle = {The Quarterly Journal of Economics},
	author = {Sorkin, Isaac},
	urldate = {2025-03-25},
	date = {2018-08-01},
	langid = {english},
}

@article{bursztyn_misperceived_2020,
	title = {Misperceived Social Norms: Women Working Outside the Home in Saudi Arabia},
	volume = {110},
	issn = {0002-8282},
	url = {https://pubs.aeaweb.org/doi/10.1257/aer.20180975},
	doi = {10.1257/aer.20180975},
	shorttitle = {Misperceived Social Norms},
	pages = {2997--3029},
	number = {10},
	journaltitle = {American Economic Review},
	shortjournal = {American Economic Review},
	author = {Bursztyn, Leonardo and González, Alessandra L. and Yanagizawa-Drott, David},
	urldate = {2025-03-28},
	date = {2020-10-01},
	langid = {english},
}
